\documentclass[12pt]{article}
\usepackage{amsmath}
\allowdisplaybreaks

\begin{document}

\author{C. Bizdadea\thanks{%
e-mail address: bizdadea@central.ucv.ro}, E. M. Cioroianu\thanks{%
e-mail address: manache@central.ucv.ro}, \and S. O. Saliu\thanks{%
e-mail address: osaliu@central.ucv.ro}, S. C. S\u{a}raru\thanks{%
e-mail address: scsararu@central.ucv.ro}, M. Iordache \\
Faculty of Physics, University of Craiova\\
13 A. I. Cuza Str., Craiova 200585, Romania}
\title{Four-dimensional couplings among BF and massless Rarita-Schwinger
theories: a BRST cohomological approach}
\date{}
\maketitle

\begin{abstract}
The local and manifestly covariant Lagrangian interactions in four
spacetime dimensions that can be added to a free\ model that
describes a massless Rarita-Schwinger theory and an Abelian BF
theory are constructed by means of deforming the solution to the
master equation on behalf of specific cohomological techniques.

\end{abstract}

\section{Introduction\label{intro}}

Topological field theories~\cite{birmingham91}--\cite{labastida97} are
important in view of the fact that certain interacting, non-Abelian versions
are related to a Poisson structure algebra~\cite{stroblspec} present in
various versions of Poisson sigma models~\cite{psmikeda94}--\cite%
{psmcattaneo2001}, which are known to be useful at the study of
two-dimensional gravity~\cite{grav2teit83}--\cite{grav2grumvassil02} (for a
detailed approach, see~\cite{grav2strobl00}). It is well known that pure
three-dimensional gravity is just a BF theory. Moreover, in higher
dimensions general relativity and supergravity in Ashtekar formalism may
also be formulated as topological BF theories with some extra constraints
\cite{ezawa}--\cite{ling}. In view of these results, it is important to know
the self-interactions in BF theories as well as the couplings between BF
models and other theories. This problem has been considered in literature in
relation with self-interactions in various classes of BF models~\cite%
{defBFizawa2000}--\cite{defBFijmpajuvi06} and couplings to other (matter or
gauge) fields~\cite{defBFepjc}--\cite{defBFjhep06} by using the powerful
BRST cohomological reformulation of the problem of constructing consistent
interactions within the Lagrangian~\cite{deflag} or the Hamiltonian~\cite%
{defham} setting. Other aspects concerning interacting, topological BF
models can be found in~\cite{otherBFikeda02} and \cite{otherBFikedaizawa04}.

The scope of this paper is to investigate the consistent interactions that
can be added to a free, Abelian gauge theory consisting of a BF model and a
massless Rarita-Schwinger field in $D=4$. This matter is addressed by means
of the deformation of the solution to the master equation from the
BRST-antifield formalism~\cite{deflag}. Under the hypotheses of smooth,
local, Lorentz covariant, and Poincar\'{e} invariant interactions,
supplemented with the requirement on the preservation of the number of
derivatives on each field with respect to the free theory, we obtain the
most general form of the theory that describes the cross-couplings between a
BF model and a massless spin-$3/2$ field. The resulting interacting model is
accurately formulated in terms of a gauge theory with gauge transformations
that close according to an open algebra (the commutators among the deformed
gauge transformations only close on the stationary surface of deformed field
equations), which are on-shell, second-order reducible. An interesting
feature of the coupled theory is the appearance of certain similarities with
the gauge symmetries from the gravitini sector of $N=1$, $D=4$ conformal
SUGRA at the level of local $Q$-supersymmetry and $U\left( 1\right) $ gauge
symmetry.

\section{Free model: Lagrangian formulation and BRST symmetry\label{free}}

We start from a free\ four-dimensional theory whose Lagrangian action is
written as the sum between the action for a massless Rarita-Schwinger field
and the action for a topological BF theory involving one scalar field, two
one-forms and one two-form
\begin{eqnarray}
S_{0}[\psi _{\mu },A^{\mu },H^{\mu },\varphi ,B^{\mu \nu }] &=&\int
d^{4}x\left( -\frac{\mathrm{i}}{2}\bar{\psi}_{\mu }\gamma ^{\mu \nu \rho
}\partial _{\nu }\psi _{\rho }+H_{\mu }\partial ^{\mu }\varphi +\frac{1}{2}%
B^{\mu \nu }\partial _{\lbrack \mu }A_{\nu ]}\right)  \notag \\
&\equiv &\int d^{4}x\left( \mathcal{L}_{0}^{\mathrm{RS}}+\mathcal{L}_{0}^{%
\mathrm{BF}}\right) .  \label{bfa1}
\end{eqnarray}%
We work with a Minkowski-flat metric tensor of `mostly minus' signature $%
\sigma ^{\mu \nu }=\sigma _{\mu \nu }=\left( +---\right) $ and employ the
Majorana representation of the Clifford algebra%
\begin{equation}
\gamma _{\mu }\gamma _{\nu }+\gamma _{\nu }\gamma _{\mu }=2\sigma _{\mu \nu }%
\mathbf{1}.  \label{conv1}
\end{equation}%
This means that all the $\gamma $-matrices are purely imaginary, with $%
\gamma _{0}$ Hermitian and $\gamma _{i}$ anti-Hermitian. The charge
conjugation matrix in this representation is given by%
\begin{equation}
\mathcal{C}=-\gamma ^{0},  \label{conv2}
\end{equation}%
while the Dirac and the charge conjugation operations are respectively
defined by the expressions
\begin{equation}
\bar{\psi}\equiv \psi ^{\dag }\gamma ^{0},\qquad \psi ^{c}\equiv \left(
\mathcal{C}\psi \right) ^{\top }.  \label{conv3}
\end{equation}%
In the above we denoted by $\dag $ and $\top $ the operations of Hermitian
conjugation and transposition, respectively. The Rarita-Schwinger field $%
\psi _{\mu }$ is a Majorana vector spinor%
\begin{equation}
\bar{\psi}_{\mu }=\psi _{\mu }^{c}.  \label{conv4}
\end{equation}%
For definiteness, we take a basis in the vector space of $4\times 4$ complex
matrices of the form%
\begin{equation}
\left\{ \mathbf{1},\gamma _{\mu },\gamma _{\mu \nu },\gamma _{\mu \nu \rho
},\gamma _{5}\right\} ,  \label{b2}
\end{equation}%
where the generic notation $\gamma _{\mu _{1}\cdots \mu _{k}}$ means the
(normalized) antisymmetrical product of $k$ $\gamma $-matrices%
\begin{equation}
\gamma _{\mu _{1}\cdots \mu _{k}}=\frac{1}{k!}\sum\limits_{\sigma \in
S_{k}}\left( -\right) ^{\sigma }\gamma _{\mu _{\sigma (1)}}\cdots \gamma
_{\mu _{\sigma (k)}}.  \label{conv5}
\end{equation}%
$S_{k}$ and $\left( -\right) ^{\sigma }$ denote the set of permutations of $%
\left\{ 1,\ldots ,k\right\} $ and the signature of the permutation $\sigma $%
, respectively. Finally, the matrix $\gamma _{5}$ is defined in the standard
manner as $\gamma _{5}=\mathrm{i}\gamma ^{0}\gamma ^{1}\gamma ^{2}\gamma
^{3} $. Also, it is useful to recall the four-dimensional duality relations
among the various matrices $\gamma _{\mu _{1}\cdots \mu _{k}}$, namely
\begin{eqnarray}
\gamma ^{\mu \nu \rho \lambda } &=&\mathrm{i}\varepsilon ^{\mu \nu \rho
\lambda }\gamma _{5},\qquad \gamma ^{\mu \nu \rho }=-\mathrm{i}\varepsilon
^{\mu \nu \rho \lambda }\gamma _{\lambda }\gamma _{5},  \label{d1} \\
\gamma ^{\mu \nu } &=&-\frac{\mathrm{i}}{2}\varepsilon ^{\mu \nu \rho
\lambda }\gamma _{\rho \lambda }\gamma _{5},\qquad \gamma ^{\mu }=\frac{%
\mathrm{i}}{6}\varepsilon ^{\mu \nu \rho \lambda }\gamma _{\nu \rho \lambda
}\gamma _{5},  \label{d2}
\end{eqnarray}%
where we used the convention $\varepsilon _{0123}=-\varepsilon ^{0123}=1$
for the four-dimensional Levi-Civita symbol. In the chosen representation of
algebra (\ref{conv1}), the $\gamma $-matrices exhibit the following
symmetry/antisymmetry properties%
\begin{eqnarray}
\left( \gamma ^{0}\gamma ^{\mu }\right) ^{\top } &=&\gamma ^{0}\gamma ^{\mu
},\qquad \left( \gamma ^{0}\gamma ^{\mu \nu }\right) ^{\top }=\gamma
^{0}\gamma ^{\mu \nu },  \label{sym1} \\
\left( \gamma ^{0}\gamma ^{\mu \nu \rho }\right) ^{\top } &=&-\gamma
^{0}\gamma ^{\mu \nu \rho },\qquad \left( \gamma ^{0}\gamma _{5}\right)
^{\top }=-\gamma ^{0}\gamma _{5}.  \label{sym2}
\end{eqnarray}%
Spinor-like indices will be denoted everywhere by Latin capital letters,
such that%
\begin{equation*}
\psi _{\mu }\equiv \left( \psi _{\mu }^{A}\right) _{A=\overline{1,4}}\equiv
\left(
\begin{array}{c}
\psi _{\mu }^{1} \\
\psi _{\mu }^{2} \\
\psi _{\mu }^{3} \\
\psi _{\mu }^{4}%
\end{array}%
\right) .
\end{equation*}

Action (\ref{bfa1}) is found invariant under the gauge transformations
\begin{eqnarray}
\delta _{\epsilon }A^{\mu } &=&\partial ^{\mu }\epsilon ,\qquad \delta
_{\epsilon }H^{\mu }=2\partial _{\nu }\epsilon ^{\mu \nu },\qquad \delta
_{\epsilon }B^{\mu \nu }=-3\partial _{\rho }\epsilon ^{\mu \nu \rho },
\label{bfa2.1a} \\
\delta _{\epsilon }\varphi &=&0,\qquad \delta _{\epsilon }\psi _{\mu
}=\partial _{\mu }\chi ,  \label{bfa2.1b}
\end{eqnarray}%
where the gauge parameters $\epsilon $, $\epsilon ^{\mu \nu }$, and $%
\epsilon ^{\mu \nu \rho }$ are bosonic, with $\epsilon ^{\mu \nu }$ and $%
\epsilon ^{\mu \nu \rho }$ completely antisymmetric. In addition, the gauge
parameter $\chi $ is a Majorana spinor%
\begin{equation*}
\chi \equiv \left( \chi ^{A}\right) _{A=\overline{1,4}}\equiv \left(
\begin{array}{c}
\chi ^{1} \\
\chi ^{2} \\
\chi ^{3} \\
\chi ^{4}%
\end{array}%
\right) .
\end{equation*}%
From (\ref{bfa2.1a}) and (\ref{bfa2.1b}) we read the nonvanishing gauge
generators of the fields, written in De Witt condensed notations, as
\begin{eqnarray}
(Z_{(A)}^{\mu }) &=&\partial ^{\mu },\qquad (Z_{(H)}^{\mu })_{\alpha \beta
}=-\partial _{\left[ \alpha \right. }\delta _{\left. \beta \right] }^{\mu },
\label{bfa2a} \\
(Z_{(B)}^{\mu \nu })_{\alpha \beta \gamma } &=&-\frac{1}{2}\partial _{\left[
\alpha \right. }\delta _{\beta }^{\mu }\delta _{\left. \gamma \right] }^{\nu
},\qquad (Z_{(\psi )}^{A\mu })_{B}=\delta _{B}^{A}\partial ^{\mu },
\label{bfa2b}
\end{eqnarray}%
where we put an extra lower index ($(A)$, $(H)$, etc.) in order to indicate
the field to which a certain gauge generator is associated with. Everywhere
in this paper we use the convention that the symbol $\left[ \alpha \beta
\ldots \gamma \right] $ signifies the operation of complete antisymmetry
with respect to the (Lorentz) indices between brackets, with the conventions
that the minimum number of terms is always used and the result is never
divided by the number of terms. The above gauge transformations are Abelian
and off-shell, second-order reducible. More precisely, the gauge generators
of the one-form $H^{\mu }$ are second-order reducible, with the first- and
respectively second-order reducibility functions
\begin{equation}
(Z_{1}^{\alpha \beta })_{\mu ^{\prime }\nu ^{\prime }\rho ^{\prime }}=-\frac{%
1}{2}\partial _{\left[ \mu ^{\prime }\right. }\delta _{\nu ^{\prime
}}^{\alpha }\delta _{\left. \rho ^{\prime }\right] }^{\beta },\qquad
(Z_{2}^{\mu ^{\prime }\nu ^{\prime }\rho ^{\prime }})_{\alpha ^{\prime
}\beta ^{\prime }\gamma ^{\prime }\delta ^{\prime }}=-\frac{1}{6}\partial _{%
\left[ \alpha ^{\prime }\right. }\delta _{\beta ^{\prime }}^{\mu ^{\prime
}}\delta _{\gamma ^{\prime }}^{\nu ^{\prime }}\delta _{\left. \delta
^{\prime }\right] }^{\rho ^{\prime }},  \label{bfa4}
\end{equation}%
while the gauge generators of the two-form $B^{\mu \nu }$ are first-order
reducible, with the reducibility functions
\begin{equation}
(Z_{1}^{\alpha \beta \gamma })_{\mu ^{\prime }\nu ^{\prime }\rho ^{\prime
}\lambda ^{\prime }}=-\frac{1}{6}\partial _{\left[ \mu ^{\prime }\right.
}\delta _{\nu ^{\prime }}^{\alpha }\delta _{\rho ^{\prime }}^{\beta }\delta
_{\left. \lambda ^{\prime }\right] }^{\gamma },  \label{bfa5}
\end{equation}%
such that the concrete form of the first- and second-order reducibility
relations are expressed by
\begin{equation}
(Z_{(H)}^{\mu })_{\alpha \beta }(Z_{1}^{\alpha \beta })_{\mu ^{\prime }\nu
^{\prime }\rho ^{\prime }}=0,\qquad (Z_{(B)}^{\mu \nu })_{\alpha \beta
\gamma }(Z_{1}^{\alpha \beta \gamma })_{\mu ^{\prime }\nu ^{\prime }\rho
^{\prime }\lambda ^{\prime }}=0  \label{bfa5a}
\end{equation}%
and
\begin{equation}
(Z_{1}^{\alpha \beta })_{\mu ^{\prime }\nu ^{\prime }\rho ^{\prime
}}(Z_{2}^{\mu ^{\prime }\nu ^{\prime }\rho ^{\prime }})_{\alpha ^{\prime
}\beta ^{\prime }\gamma ^{\prime }\delta ^{\prime }}=0,  \label{bfa5b}
\end{equation}%
respectively. We observe that the theory described by action (\ref{bfa1}) is
a usual linear gauge theory (its field equations are linear in the fields
and first-order in their spacetime derivatives), whose generating set of
gauge transformations is second-order reducible, such that we can define in
a consistent manner its Cauchy order, which is found equal to four.

In order to construct the BRST symmetry of this free\ theory, we introduce
the field/ghost and antifield spectra
\begin{eqnarray}
\Phi ^{\alpha _{0}} &=&\left( A^{\mu },H^{\mu },\varphi ,B^{\mu \nu },\psi
_{\mu }\right) ,\qquad \Phi _{\alpha _{0}}^{\ast }=\left( A_{\mu }^{\ast
},H_{\mu }^{\ast },\varphi ^{\ast },B_{\mu \nu }^{\ast },\psi ^{\ast \mu
}\right) ,  \label{bfa6} \\
\eta ^{\alpha _{1}} &=&\left( \eta ,C^{\mu \nu },\eta ^{\mu \nu \rho },\xi
\right) ,\qquad \eta _{\alpha _{1}}^{\ast }=\left( \eta ^{\ast },C_{\mu \nu
}^{\ast },\eta _{\mu \nu \rho }^{\ast },\xi ^{\ast }\right) ,  \label{bfa7}
\\
\eta ^{\alpha _{2}} &=&\left( C^{\mu \nu \rho },\eta ^{\mu \nu \rho \lambda
}\right) ,\qquad \eta _{\alpha _{2}}^{\ast }=\left( C_{\mu \nu \rho }^{\ast
},\eta _{\mu \nu \rho \lambda }^{\ast }\right) ,  \label{bfa8} \\
\eta ^{\alpha _{3}} &=&C^{\mu \nu \rho \lambda },\qquad \eta _{\alpha
_{3}}^{\ast }=C_{\mu \nu \rho \lambda }^{\ast }.  \label{bfa9}
\end{eqnarray}%
The fermionic ghosts $\left( \eta ,C^{\mu \nu },\eta ^{\mu \nu \rho }\right)
$ respectively correspond to the bosonic gauge parameters $\left( \epsilon
,\epsilon ^{\mu \nu },\epsilon ^{\mu \nu \rho }\right) $, the bosonic ghosts
for ghosts $\eta ^{\alpha _{2}}$ are due to the first-order reducibility
relations (\ref{bfa5a}), while the fermionic ghosts for ghosts for ghosts $%
\eta ^{\alpha _{3}}$ are required by the second-order reducibility relations
(\ref{bfa5b}). In addition, the ghost $\xi $, associated with the gauge
parameter $\chi $, is a bosonic spinor of purely imaginary components. The
star variables represent the antifields of the corresponding fields/ghosts.
Their Grassmann parities are obtained via the usual rule
\begin{equation*}
\varepsilon \left( \chi _{\Delta }^{\ast }\right) =\left( \varepsilon \left(
\chi ^{\Delta }\right) +1\right) \mathrm{mod}\;2,
\end{equation*}%
where we employed the notations
\begin{equation}
\chi ^{\Delta }=\left( \Phi ^{\alpha _{0}},\eta ^{\alpha _{1}},\eta ^{\alpha
_{2}},\eta ^{\alpha _{3}}\right) ,\qquad \chi _{\Delta }^{\ast }=\left( \Phi
_{\alpha _{0}}^{\ast },\eta _{\alpha _{1}}^{\ast },\eta _{\alpha _{2}}^{\ast
},\eta _{\alpha _{3}}^{\ast }\right) .  \label{notat}
\end{equation}

Since both the gauge generators and the reducibility functions are
field-independent, it follows that the BRST differential simply reduces to
\begin{equation}
s=\delta +\gamma ,  \label{sfree}
\end{equation}%
where $\delta $ is the Koszul-Tate differential and $\gamma $ means the
exterior longitudinal derivative. The Koszul-Tate differential is graded in
terms of the antighost number ($\mathrm{agh}$, $\mathrm{agh}\left( \delta
\right) =-1$, $\mathrm{agh}\left( \gamma \right) =0$) and enforces a
resolution of the algebra of smooth functions defined on the stationary
surface of field equations for action (\ref{bfa1}), $C^{\infty }\left(
\Sigma \right) $, $\Sigma :\delta S_{0}/\delta \Phi ^{\alpha _{0}}=0$. The
exterior longitudinal derivative is graded in terms of the pure ghost number
($\mathrm{pgh}$, $\mathrm{pgh}\left( \gamma \right) =1$, $\mathrm{pgh}\left(
\delta \right) =0$) and is correlated with the original gauge symmetry via
its cohomology at pure ghost number zero computed in $C^{\infty }\left(
\Sigma \right) $, which is isomorphic to the algebra of physical observables
for the free theory. These two degrees of the generators (\ref{bfa6})--(\ref%
{bfa9}) from the BRST complex are valued like
\begin{eqnarray}
\mathrm{pgh}\left( \Phi ^{\alpha _{0}}\right) &=&0,\qquad \mathrm{pgh}\left(
\eta ^{\alpha _{1}}\right) =1,  \label{bfa10} \\
\mathrm{pgh}\left( \eta ^{\alpha _{2}}\right) &=&2,\qquad \mathrm{pgh}\left(
\eta ^{\alpha _{3}}\right) =3,  \label{bfa10a} \\
\mathrm{pgh}\left( \Phi _{\alpha _{0}}^{\ast }\right) &=&\mathrm{pgh}\left(
\eta _{\alpha _{1}}^{\ast }\right) =\mathrm{pgh}\left( \eta _{\alpha
_{2}}^{\ast }\right) =\mathrm{pgh}\left( \eta _{\alpha _{3}}^{\ast }\right)
=0,  \label{bfa12} \\
\mathrm{agh}\left( \Phi ^{\alpha _{0}}\right) &=&\mathrm{agh}\left( \eta
^{\alpha _{1}}\right) =\mathrm{agh}\left( \eta ^{\alpha _{2}}\right) =%
\mathrm{agh}\left( \eta ^{\alpha _{3}}\right) =0,  \label{bfa11} \\
\mathrm{agh}\left( \Phi _{\alpha _{0}}^{\ast }\right) &=&1,\qquad \mathrm{agh%
}\left( \eta _{\alpha _{1}}^{\ast }\right) =2,  \label{bfa13} \\
\mathrm{agh}\left( \eta _{\alpha _{2}}^{\ast }\right) &=&3,\qquad \mathrm{agh%
}\left( \eta _{\alpha _{3}}^{\ast }\right) =4,  \label{bfa13a}
\end{eqnarray}%
where the (right) actions of $\delta $ and $\gamma $ on them read as%
\begin{gather}
\delta \Phi ^{\alpha _{0}}=\delta \eta ^{\alpha _{1}}=\delta \eta ^{\alpha
_{2}}=\delta \eta ^{\alpha _{3}}=0,  \label{bfa15} \\
\delta A_{\mu }^{\ast }=\partial ^{\nu }B_{\nu \mu },\qquad \delta H_{\mu
}^{\ast }=-\partial _{\mu }\varphi ,  \label{bfa16} \\
\delta \varphi ^{\ast }=\partial ^{\mu }H_{\mu },\qquad \delta B_{\mu \nu
}^{\ast }=-\frac{1}{2}\partial _{\lbrack \mu }A_{\nu ]},  \label{bfa16.1} \\
\delta \psi ^{\ast \mu }=-\mathrm{i}\partial _{\nu }\bar{\psi}
_{\rho }\gamma ^{\mu \nu \rho },\qquad \delta \eta ^{\ast
}=-\partial ^{\mu }A_{\mu }^{\ast },\qquad \delta C_{\mu \nu }^{\ast
}=\partial _{\lbrack \mu }H_{\nu ]}^{\ast
},  \label{bfa17} \\
\delta \eta _{\mu \nu \rho }^{\ast }=\partial _{\lbrack \mu }B_{\nu \rho
]}^{\ast },\qquad \delta \xi ^{\ast }=\partial _{\mu }\psi ^{\ast \mu
},\qquad \delta C_{\mu \nu \rho }^{\ast }=-\partial _{\lbrack \mu }C_{\nu
\rho ]}^{\ast },  \label{bfa18} \\
\delta \eta _{\mu \nu \rho \lambda }^{\ast }=-\partial _{\lbrack \mu }\eta
_{\nu \rho \lambda ]}^{\ast },\qquad \delta C_{\mu \nu \rho \lambda }^{\ast
}=\partial _{\lbrack \mu }C_{\nu \rho \lambda ]}^{\ast },  \label{bfa19} \\
\gamma \Phi _{\alpha _{0}}^{\ast }=\gamma \eta _{\alpha _{1}}^{\ast }=\gamma
\eta _{\alpha _{2}}^{\ast }=\gamma \eta _{\alpha _{3}}^{\ast }=0,
\label{bfa20} \\
\gamma A^{\mu }=\partial ^{\mu }\eta ,\qquad \gamma H^{\mu }=2\partial _{\nu
}C^{\mu \nu },\qquad \gamma B^{\mu \nu }=-3\partial _{\rho }\eta ^{\mu \nu
\rho },  \label{bfa21} \\
\gamma \varphi =0,\qquad \gamma \psi _{\mu }=\partial _{\mu }\xi ,
\label{bfa21.1} \\
\gamma \eta =\gamma \xi =0,\qquad \gamma C^{\mu \nu }=-3\partial _{\rho
}C^{\mu \nu \rho },\qquad \gamma \eta ^{\mu \nu \rho }=4\partial _{\lambda
}\eta ^{\mu \nu \rho \lambda },  \label{bfa22} \\
\gamma C^{\mu \nu \rho }=4\partial _{\lambda }C^{\mu \nu \rho \lambda
},\qquad \gamma \eta ^{\mu \nu \rho \lambda }=\gamma C^{\mu \nu \rho \lambda
}=0.  \label{bfa23}
\end{gather}

The overall degree of the BRST complex is named ghost number ($\mathrm{gh}$)
and is defined like the difference between the pure ghost number and the
antighost number, such that $\mathrm{gh}\left( s\right) =\mathrm{gh}\left(
\delta \right) =\mathrm{gh}\left( \gamma \right) =1$. The BRST differential
is known to have a canonical action in a structure named antibracket and
denoted by the symbol $\left( ,\right) $ ($s\cdot =\left( \cdot ,\bar{S}%
\right) $), which is obtained by decreeing the fields/ghosts respectively
conjugated to the corresponding antifields. The generator of the BRST
symmetry is a bosonic functional of ghost number zero ($\mathrm{gh}\left(
\bar{S}\right) =0$, $\varepsilon \left( \bar{S}\right) =0$), which is
solution to the classical master equation $\left( \bar{S},\bar{S}\right) =0$%
. In the case of the free\ theory under discussion, the solution to the
master equation takes the form
\begin{eqnarray}
\bar{S} &=&S_{0}+\int d^{4}x\left( A_{\mu }^{\ast }\partial ^{\mu }\eta
+2H_{\mu }^{\ast }\partial _{\nu }C^{\mu \nu }-3B_{\mu \nu }^{\ast }\partial
_{\rho }\eta ^{\mu \nu \rho }\right.  \notag \\
&&\left. +\psi ^{\ast \mu }\partial _{\mu }\xi -3C_{\mu \nu }^{\ast
}\partial _{\rho }C^{\mu \nu \rho }+4\eta _{\mu \nu \rho }^{\ast }\partial
_{\lambda }\eta ^{\mu \nu \rho \lambda }+4C_{\mu \nu \rho }^{\ast }\partial
_{\lambda }C^{\mu \nu \rho \lambda }\right) .  \label{bfa14}
\end{eqnarray}%
The solution to the master equation encodes all the information on the gauge
structure of a given theory. We remark that in our case the solution (\ref%
{bfa14}) breaks into terms with antighost numbers ranging from zero to
three. The piece with antighost number zero is nothing but the Lagrangian
action (\ref{bfa1}), while the elements of antighost number one include the
gauge generators (\ref{bfa2a})--(\ref{bfa2b}). If the gauge algebra were
non-Abelian, then there would appear at least terms linear in the antighost
number two antifields and quadratic in the pure ghost number one ghosts. The
absence of such terms in our case reflects that the gauge transformations
are Abelian. The terms from (\ref{bfa14}) of higher antighost number give us
information on the reducibility functions (\ref{bfa4}) and (\ref{bfa5}). If
the reducibility relations held on-shell, then the solution of the master
equation would contain components linear in the ghosts for ghosts (ghosts of
pure ghost number strictly greater than one) and quadratic in the various
antifields. Such pieces are not present in (\ref{bfa14}), since the
reducibility relations hold off-shell. Other possible components in the
solution to the master equation offer information on the higher-order
structure functions related to the tensor gauge structure of the theory.
There are no such terms in (\ref{bfa14}), as a consequence of the fact that
all higher-order structure functions vanish for this (free) model.

\section{Deformation of the master equation --- brief review\label{defgen}}

We begin with a \textquotedblleft free\textquotedblright\ gauge theory,
described by a Lagrangian action $S_{0}\left[ \Phi ^{\alpha _{0}}\right] $,
invariant under some gauge transformations
\begin{equation}
\delta _{\epsilon }\Phi ^{\alpha _{0}}=Z_{\;\;\alpha _{1}}^{\alpha
_{0}}\epsilon ^{\alpha _{1}},\qquad \frac{\delta S_{0}}{\delta \Phi ^{\alpha
_{0}}}Z_{\;\;\alpha _{1}}^{\alpha _{0}}=0,  \label{bfa2.1}
\end{equation}%
and consider the problem of constructing consistent interactions among the
fields $\Phi ^{\alpha _{0}}$ such that the couplings preserve the field
spectrum and the original number of gauge symmetries. This matter is
addressed by means of reformulating the problem of constructing consistent
interactions as a deformation problem of the solution to the master equation
corresponding to the \textquotedblleft free\textquotedblright\ theory \cite%
{deflag}, \cite{17and5}. Such a reformulation is possible due to the fact
that the solution to the master equation contains all the information on the
gauge structure of the theory. If a consistent interacting gauge theory can
be constructed, then the solution $\bar{S}$ to the master equation
associated with the \textquotedblleft free\textquotedblright\ theory, $%
\left( \bar{S},\bar{S}\right) =0$, can be deformed into a solution $S$,
\begin{equation}
\bar{S}\rightarrow S=\bar{S}+\lambda S_{1}+\lambda ^{2}S_{2}+\cdots =\bar{S}%
+\lambda \int d^{D}x\,a+\lambda ^{2}\int d^{D}x\,b+\cdots ,  \label{bfa2.2}
\end{equation}%
of the master equation for the deformed theory
\begin{equation}
\left( S,S\right) =0,  \label{bfa2.3}
\end{equation}%
such that both the ghost and antifield spectra of the initial theory are
preserved. The symbol $\left( ,\right) $ denotes the antibracket. The
equation (\ref{bfa2.3}) splits, according to the various orders in the
coupling constant (or deformation parameter) $\lambda $, into
\begin{eqnarray}
\left( \bar{S},\bar{S}\right)  &=&0,  \label{bfa2.4} \\
2\left( S_{1},\bar{S}\right)  &=&0,  \label{bfa2.5} \\
2\left( S_{2},\bar{S}\right) +\left( S_{1},S_{1}\right)  &=&0,
\label{bfa2.6} \\
\left( S_{3},\bar{S}\right) +\left( S_{1},S_{2}\right)  &=&0,  \label{bfa2.7}
\\
&&\vdots   \notag
\end{eqnarray}

Equation (\ref{bfa2.4}) is fulfilled by hypothesis. The next equation
requires that the first-order deformation of the solution to the master
equation, $S_{1}$, is a co-cycle of the \textquotedblleft
free\textquotedblright\ BRST differential $s\cdot =\left( \cdot ,\bar{S}%
\right) $. However, only cohomologically nontrivial solutions to (\ref%
{bfa2.5}) should be taken into account, because the BRST-exact ones can be
eliminated by a (possibly nonlinear) field redefinition. This means that $%
S_{1}$ pertains to the ghost number zero cohomological space of $s$, $%
H^{0}\left( s\right) $, which is generically nonempty due to its isomorphism
to the space of physical observables of the \textquotedblleft
free\textquotedblright\ theory. It has been shown in \cite{deflag}, \cite%
{17and5} (by means of the triviality of the antibracket map in the
cohomology of the BRST differential) that there are no obstructions in
finding solutions to the remaining equations ((\ref{bfa2.6})--(\ref{bfa2.7}%
), etc.). However, the resulting interactions may be nonlocal, and there
might even appear obstructions if one insists on their locality. As it will
be seen below, this is not the case here since all the interactions in the
case of the model under study turn out to be local.

\section{Consistent interactions between a massless Rarita-Schwinger field
and a topological BF theory\label{inter}}

In this section we determine the consistent interactions that can be added
to the free\ theory (\ref{bfa1}) that describes a massless Rarita-Schwinger
field plus a topological BF model in four spacetime dimensions. This is done
by solving the Lagrangian deformation equations ((\ref{bfa2.5})--(\ref%
{bfa2.7}), etc.) via specific cohomological BRST techniques. The interacting
theory and its gauge structure are deduced from the analysis of the deformed
solution to the master equation that is consistent to all orders in the
deformation parameter. For obvious reasons, we consider only smooth, local,
Lorentz covariant, and Poincar\'{e} invariant deformations (i.e., we do not
allow explicit dependence on the spacetime coordinates). In the meantime we
require that the maximum number of derivatives allowed to enter the
interaction vertices is equal to one, i.e. the maximum number of derivatives
from the free Lagrangian. The smoothness of deformations refers to the fact
that the deformed solution to the master equation, (\ref{bfa2.2}), is smooth
in the coupling constant $\lambda $ and reduces to the original solution, (%
\ref{bfa14}), in the free limit ($\lambda =0$).

\subsection{Standard material: basic cohomologies\label{stand}}

If we make the notation $S_{1}=\int d^{4}x\,a$, with $a$ a local function,
then equation (\ref{bfa2.5}), which we have seen that controls the
first-order deformation, takes the local form
\begin{equation}
sa=\partial _{\mu }m^{\mu },\qquad \mathrm{gh}\left( a\right) =0,\qquad
\varepsilon \left( a\right) =0,  \label{3.1}
\end{equation}%
for some local $m^{\mu }$. It shows that the nonintegrated density of the
first-order deformation pertains to the local cohomology of $s$ in ghost
number zero, $a\in H^{0}\left( s|d\right) $, where $d$ denotes the exterior
spacetime differential. The solution to (\ref{3.1}) is unique up to $s$%
-exact pieces plus divergences
\begin{equation}
a\rightarrow a+sb+\partial _{\mu }n^{\mu },\qquad \mathrm{gh}\left( b\right)
=-1,\qquad \varepsilon \left( b\right) =1.  \label{3.1a}
\end{equation}%
At the same time, if the general solution to (\ref{3.1}) is found to be
completely trivial, $a=sb+\partial _{\mu }n^{\mu }$, then it can be made to
vanish, $a=0$.

In order to analyze equation (\ref{3.1}) we develop $a$ according to the
antighost number
\begin{equation}
a=\sum\limits_{i=0}^{I}a_{i},\qquad \mathrm{agh}\left( a_{i}\right)
=i,\qquad \mathrm{gh}\left( a_{i}\right) =0,\qquad \varepsilon \left(
a_{i}\right) =0,  \label{3.2}
\end{equation}%
and assume, without loss of generality, that the above decomposition stops
at some finite value of $I$. This can be shown, for instance, like in~\cite%
{gen2} (Section 3), under the sole assumption that the interacting
Lagrangian at the first order in the coupling constant, $a_{0}$, has a
finite, but otherwise arbitrary derivative order. Inserting decomposition (%
\ref{3.2}) into equation (\ref{3.1}) and projecting it on the various values
of the antighost number, we obtain the tower of equations
\begin{eqnarray}
\gamma a_{I} &=&\partial _{\mu }\overset{\left( I\right) }{m}^{\mu },
\label{3.3} \\
\delta a_{I}+\gamma a_{I-1} &=&\partial _{\mu }\overset{\left( I-1\right) }{m%
}^{\mu },  \label{3.4} \\
\delta a_{i}+\gamma a_{i-1} &=&\partial _{\mu }\overset{\left( i-1\right) }{m%
}^{\mu },\qquad 1\leq i\leq I-1,  \label{3.5}
\end{eqnarray}%
where $\left( \overset{\left( i\right) }{m}^{\mu }\right) _{i=\overline{0,I}%
} $ are some local currents with $\mathrm{agh}\left( \overset{\left(
i\right) }{m}^{\mu }\right) =i$. Equation (\ref{3.3}) can be replaced in
strictly positive values of the antighost number by
\begin{equation}
\gamma a_{I}=0,\qquad I>0.  \label{3.6}
\end{equation}%
Due to the second-order nilpotency of $\gamma $ ($\gamma ^{2}=0$), the
solution to (\ref{3.6}) is clearly unique up to $\gamma $-exact
contributions
\begin{equation}
a_{I}\rightarrow a_{I}+\gamma b_{I},\qquad \mathrm{agh}\left( b_{I}\right)
=I,\qquad \mathrm{pgh}\left( b_{I}\right) =I-1,\qquad \varepsilon \left(
b_{I}\right) =1.  \label{r68}
\end{equation}%
Meanwhile, if it turns out that $a_{I}$ exclusively reduces to $\gamma $%
-exact terms, $a_{I}=\gamma b_{I}$, then it can be made to vanish, $a_{I}=0$%
. In other words, the nontriviality of the first-order deformation $a$ is
translated at its highest antighost number component into the requirement
that $a_{I}\in H^{I}\left( \gamma \right) $, where $H^{I}\left( \gamma
\right) $ denotes the cohomology of the exterior longitudinal derivative $%
\gamma $ in pure ghost number equal to $I$. So, in order to solve equation (%
\ref{3.1}) (equivalent with (\ref{3.6}) and (\ref{3.4})--(\ref{3.5})), we
need to compute the cohomology of $\gamma $, $H\left( \gamma \right) $, and,
as it will be made clear below, also the local homology of $\delta $, $%
H\left( \delta |d\right) $.

On behalf of definitions (\ref{bfa20})--(\ref{bfa23}) it is simple to see
that $H\left( \gamma \right) $ is spanned by
\begin{equation}
F_{\bar{A}}=\left( \varphi ,\partial _{\lbrack \mu }A_{\nu ]},\partial ^{\mu
}H_{\mu },\partial _{\mu }B^{\mu \nu },\partial _{\lbrack \mu }\psi _{\nu
]}\right) ,  \label{3.7}
\end{equation}%
the antifields $\chi _{\Delta }^{\ast }$, all of their spacetime derivatives
as well as by the undifferentiated ghosts
\begin{equation}
\eta ^{\bar{\Upsilon}}=\left( \eta ,\xi ,\eta ^{\mu \nu \rho \lambda
},C_{\mu \nu \rho \lambda }\right) .  \label{notat1}
\end{equation}%
(The derivatives of the ghosts $\eta ^{\bar{\Upsilon}}$ are removed from $%
H\left( \gamma \right) $ since they are $\gamma $-exact, in agreement with
the first relation from (\ref{bfa21}), the last formula in (\ref{bfa21.1}),
the fourth equation in (\ref{bfa22}), and the first definition from (\ref%
{bfa23}).) If we denote by $e^{M}\left( \eta ^{\bar{\Upsilon}}\right) $ the
elements with pure ghost number $M$ of a basis in the space of the
polynomials in the ghosts (\ref{notat1}), then it follows that the general
solution to equation (\ref{3.6}) takes the form
\begin{equation}
a_{I}=\alpha _{I}\left( \left[ F_{\bar{A}}\right] ,\left[ \chi _{\Delta
}^{\ast }\right] \right) e^{I}\left( \eta ^{\bar{\Upsilon}}\right) ,
\label{3.8}
\end{equation}%
where $\mathrm{agh}\left( \alpha _{I}\right) =I$ and $\mathrm{pgh}\left(
e^{I}\right) =I$. The notation $f([q])$ means that $f$ depends on $q$ and
its spacetime derivatives up to a finite order. The objects $\alpha _{I}$
(obviously nontrivial in $H^{0}\left( \gamma \right) $) will be called
\textquotedblleft invariant polynomials\textquotedblright . The result that
we can replace equation (\ref{3.3}) with the less obvious one (\ref{3.6}) is
a nice consequence of the fact that the cohomology of the exterior spacetime
differential is trivial in the space of invariant polynomials in strictly
positive antighost numbers.

Inserting (\ref{3.8}) in (\ref{3.4}) we obtain that a necessary (but not
sufficient) condition for the existence of (nontrivial) solutions $a_{I-1}$
is that the invariant polynomials $\alpha _{I}$ are (nontrivial) objects
from the local cohomology of Koszul-Tate differential $H\left( \delta
|d\right) $ in antighost number $I>0$ and in pure ghost number zero,
\begin{equation}
\delta \alpha _{I}=\partial _{\mu }\overset{\left( I-1\right) }{j}^{\mu
},\qquad \mathrm{agh}\left( \overset{\left( I-1\right) }{j}^{\mu }\right)
=I-1,\qquad \mathrm{pgh}\left( \overset{\left( I-1\right) }{j}^{\mu }\right)
=0.  \label{3.10a}
\end{equation}%
We recall that the local cohomology $H\left( \delta |d\right) $ is
completely trivial in both strictly positive antighost \textit{and} pure
ghost numbers (for instance, see~\cite{gen1}, Theorem 5.4, and~\cite{gen2}
), so from now on it is understood that by $H\left( \delta |d\right) $ we
mean the local cohomology of $\delta $ at pure ghost number zero. Using the
fact that the free BF model under study is a linear gauge theory of Cauchy
order equal to four and the general result from~\cite{gen1,gen2}, according
to which the local cohomology of the Koszul-Tate differential is trivial in
antighost numbers strictly greater than its Cauchy order, we can state that
\begin{equation}
H_{J}\left( \delta |d\right) =0\qquad \mathrm{for\,all}\qquad J>4,
\label{3.11}
\end{equation}%
where $H_{J}\left( \delta |d\right) $ represents the local cohomology of the
Koszul-Tate differential in antighost number $J$. Moreover, if the invariant
polynomial $\alpha _{J}$, with \textrm{agh}$\left( \alpha _{J}\right) =J\geq
4$, is trivial in $H_{J}\left( \delta |d\right) $, then it can be taken to
be trivial also in $H_{J}^{\mathrm{inv}}\left( \delta |d\right) $:%
\begin{equation}
\left( \alpha _{J}=\delta b_{J+1}+\partial _{\mu }\overset{(J)}{c}^{\mu },%
\mathrm{agh}\left( \alpha _{J}\right) =J\geq 4\right) \Rightarrow \alpha
_{J}=\delta \beta _{J+1}+\partial _{\mu }\overset{(J)}{\gamma }^{\mu },
\label{3.12ax}
\end{equation}%
with both $\beta _{J+1}$ and $\overset{(J)}{\gamma }^{\mu }$ invariant
polynomials. Here, $H_{J}^{\mathrm{inv}}\left( \delta |d\right) $ denotes
the invariant characteristic cohomology in antighost number $J$ (the local
cohomology of the Koszul-Tate differential in the space of invariant
polynomials). (An element of $H_{I}^{\mathrm{inv}}\left( \delta |d\right) $
is defined via an equation like (\ref{3.10a}), but with the corresponding
current an invariant polynomial.). This result together with (\ref{3.11})
ensures that the entire invariant characteristic cohomology in antighost
numbers strictly greater than four is trivial
\begin{equation}
H_{J}^{\mathrm{inv}}\left( \delta |d\right) =0\qquad \mathrm{for\, all}%
\qquad J>4.  \label{3.12x}
\end{equation}

The nontrivial representatives of $H_{J}(\delta |d)$ and $H_{J}^{\mathrm{inv}%
}(\delta |d)$ for $J\geq 2$ depend neither on $\left( \partial _{\lbrack \mu
}A_{\nu ]},\partial ^{\mu }H_{\mu },\partial _{\mu }B^{\mu \nu },\partial
_{\lbrack \mu }\psi _{\nu ]}\right) $ nor on the spacetime derivatives of $%
F_{\bar{A}}$ defined in (\ref{3.7}), but only on the undifferentiated scalar
field $\varphi $. With the help of relations (\ref{bfa15})--(\ref{bfa19}),
it can be shown that $H_{4}^{\mathrm{inv}}\left( \delta |d\right) $ is
generated by the elements
\begin{eqnarray}
\left( P\left( W\right) \right) ^{\mu \nu \rho \lambda } &=&\frac{dW}{%
d\varphi }C^{\ast \mu \nu \rho \lambda }+\frac{d^{2}W}{d\varphi ^{2}}\left(
H^{\ast \lbrack \mu }C^{\ast \nu \rho \lambda ]}+C^{\ast \lbrack \mu \nu
}C^{\ast \rho \lambda ]}\right)  \notag \\
&&+\frac{d^{3}W}{d\varphi ^{3}}H^{\ast \lbrack \mu }H^{\ast \nu }C^{\ast
\rho \lambda ]}+\frac{d^{4}W}{d\varphi ^{4}}H^{\ast \mu }H^{\ast \nu
}H^{\ast \rho }H^{\ast \lambda },  \label{3.13}
\end{eqnarray}%
where $W=W\left( \varphi \right) $ is an arbitrary, smooth function
depending only on the undifferentiated scalar field $\varphi $. Indeed,
direct computation yields
\begin{equation}
\delta \left( P\left( W\right) \right) ^{\mu \nu \rho \lambda }=\partial
^{\lbrack \mu }\left( P\left( W\right) \right) ^{\nu \rho \lambda ]},\qquad
\mathrm{agh}\left( \left( P\left( W\right) \right) ^{\nu \rho \lambda
}\right) =3,  \label{3.13a}
\end{equation}%
where we made the notation
\begin{equation}
\left( P\left( W\right) \right) ^{\mu \nu \rho }=\frac{dW}{d\varphi }C^{\ast
\mu \nu \rho }+\frac{d^{2}W}{d\varphi ^{2}}H^{\ast \lbrack \mu }C^{\ast \nu
\rho ]}+\frac{d^{3}W}{d\varphi ^{3}}H^{\ast \mu }H^{\ast \nu }H^{\ast \rho }.
\label{3.14}
\end{equation}%
It is clear that $\left( P\left( W\right) \right) ^{\mu \nu \rho }$ is an
invariant polynomial. By applying the operator $\delta $ on it, we have that
\begin{equation}
\delta \left( P\left( W\right) \right) ^{\mu \nu \rho }=-\partial ^{\lbrack
\mu }\left( P\left( W\right) \right) ^{\nu \rho ]},\qquad \mathrm{agh}\left(
\left( P\left( W\right) \right) ^{\nu \rho }\right) =2,  \label{3.14a}
\end{equation}%
where we employed the convention
\begin{equation}
\left( P\left( W\right) \right) ^{\mu \nu }=\frac{dW}{d\varphi }C^{\ast \mu
\nu }+\frac{d^{2}W}{d\varphi ^{2}}H^{\ast \mu }H^{\ast \nu }.  \label{3.15}
\end{equation}%
Since $\left( P\left( W\right) \right) ^{\mu \nu }$ is also an invariant
polynomial, from (\ref{3.14a}) it follows that $\left( P\left( W\right)
\right) ^{\mu \nu \rho }$ belongs to $H_{3}^{\mathrm{inv}}\left( \delta
|d\right) $. Moreover, further calculations produce%
\begin{equation}
\delta \left( P\left( W\right) \right) ^{\mu \nu }=\partial ^{\lbrack \mu
}\left( P\left( W\right) \right) ^{\nu ]},\qquad \mathrm{agh}\left( \left(
P\left( W\right) \right) ^{\nu }\right) =1,  \label{3.15a}
\end{equation}%
with
\begin{equation}
\left( P\left( W\right) \right) ^{\mu }=\frac{dW}{d\varphi }H^{\ast \mu }.
\label{3.16}
\end{equation}%
Due to the fact that $\left( P\left( W\right) \right) ^{\mu }$ is an
invariant polynomial, we deduce that $\left( P\left( W\right) \right) ^{\mu
\nu }$ pertains to $H_{2}^{\mathrm{inv}}\left( \delta |d\right) $. Using
again the actions of $\delta $ on the BRST generators, it can be proved that
$H_{3}^{\mathrm{inv}}\left( \delta |d\right) $ is spanned, beside the
elements $\left( P\left( W\right) \right) ^{\mu \nu \rho }$ given in (\ref%
{3.14}), also by the undifferentiated antifields $\eta _{\mu \nu \rho
\lambda }^{\ast }$ (according to the first definition from (\ref{bfa19})).
Putting together the above results we can state that $H_{2}^{\mathrm{inv}%
}\left( \delta |d\right) $ is spanned by $\left( P\left( W\right) \right)
^{\mu \nu }$ listed in (\ref{3.15}) and the undifferentiated antifields $%
\eta ^{\ast }$, $\eta _{\mu \nu \rho }^{\ast }$, and $\xi ^{\ast }$ (in
agreement with the second definition in (\ref{bfa17}), the first formula
from (\ref{bfa18}), and the second relation in (\ref{bfa18})). The above
results are synthesized in the following array%
\begin{equation}
\begin{array}{cc}
\mathrm{agh} &
\begin{array}{l}
\mathrm{nontrivial\;representatives} \\
\mathrm{spanning\;}H_{J}\left( \delta |d\right) \;\mathrm{and}\;H_{J}^{%
\mathrm{inv}}\left( \delta |d\right)%
\end{array}
\\
J>4 & \mathrm{none} \\
J=4 & \left( P\left( W\right) \right) ^{\mu \nu \rho \lambda } \\
J=3 & \eta _{\mu \nu \rho \lambda }^{\ast },\left( P\left( W\right) \right)
^{\mu \nu \rho } \\
J=2 & \eta _{\mu \nu \rho }^{\ast },\left( P\left( W\right) \right) ^{\mu
\nu },\eta ^{\ast },\xi ^{\ast }%
\end{array}%
.  \label{hinvdeltamodd}
\end{equation}

In contrast to the spaces $\left( H_{J}(\delta \vert d)\right) _{J\geq 2}$
and $\left( H_{J}^{\mathrm{inv}}(\delta \vert d)\right) _{J\geq 2}$, which
are finite-dimensional, the cohomology $H_{1}(\delta \vert d)$ (known to be
related to global symmetries and ordinary conservation laws) is
infinite-dimensional since the theory is free. Fortunately, it will not be
needed in the sequel.

The previous results on $H(\delta |d)$ and $H^{\mathrm{inv}}(\delta |d)$ in
strictly positive antighost numbers are important because they control the
obstructions to removing the antifields from the first-order deformation.
More precisely, we can successively eliminate all the pieces of antighost
number strictly greater that four from the nonintegrated density of the
first-order deformation by adding solely trivial terms, so we can take,
without loss of nontrivial objects, the condition $I\leq 4$ into (\ref{3.2}%
). In addition, the last representative is of the form (\ref{3.8}), where
the invariant polynomial necessarily is a nontrivial object from $H_{4}^{%
\mathrm{inv}}(\delta |d)$.

\subsection{First-order deformation\label{first}}

In the case $I=4$ the nonintegrated density of the first-order deformation
(see (\ref{3.2})) becomes
\begin{equation}
a=a_{0}+a_{1}+a_{2}+a_{3}+a_{4}.  \label{fo1}
\end{equation}%
We can further decompose $a$ in a natural manner as a sum between two kinds
of deformations
\begin{equation}
a=a^{\mathrm{BF}}+a^{\mathrm{int}},  \label{fo2}
\end{equation}%
where $a^{\mathrm{BF}}$ contains only fields/ghosts/antifields from the BF
sector and $a^{\mathrm{int}}$ describes the cross-interactions between the
two theories. Strictly speaking, we should have added to (\ref{fo2}) a
component $a^{\mathrm{RS}}$ that involves only the Rarita-Schwinger sector.
As it will be seen in the end of this subsection, $a^{\mathrm{RS}}$ will
automatically be included into $a^{\mathrm{int}}$. The piece $a^{\mathrm{BF}%
} $ is completely known and satisfies (separately) an equation of the type (%
\ref{3.1}). It admits a decomposition similar to (\ref{fo1})
\begin{equation}
a^{\mathrm{BF}}=a_{0}^{\mathrm{BF}}+a_{1}^{\mathrm{BF}}+a_{2}^{\mathrm{BF}%
}+a_{3}^{\mathrm{BF}}+a_{4}^{\mathrm{BF}},  \label{descBF}
\end{equation}%
where
\begin{equation}
a_{4}^{\mathrm{BF}}=\left( P\left( W\right) \right) ^{\mu \nu \rho \lambda
}\eta C_{\mu \nu \rho \lambda }+\tfrac{1}{2}\varepsilon _{\mu \nu \rho
\lambda }\left( P\left( M\right) \right) ^{\mu \nu \rho \lambda }\eta
_{\alpha \beta \gamma \delta }\eta ^{\alpha \beta \gamma \delta },
\label{a4}
\end{equation}%
\begin{eqnarray}
a_{3}^{\mathrm{BF}} &=&\left( P\left( W\right) \right) ^{\mu \nu \rho
}\left( -\eta C_{\mu \nu \rho }+4A^{\lambda }C_{\mu \nu \rho \lambda }\right)
\notag \\
&&+2\left( 6\left( P\left( W\right) \right) ^{\mu \nu }B^{\ast \rho \lambda
}+4\left( P\left( W\right) \right) ^{\mu }\eta ^{\ast \nu \rho \lambda
}+W\eta ^{\ast \mu \nu \rho \lambda }\right) C_{\mu \nu \rho \lambda }
\notag \\
&&-\varepsilon _{\mu \nu \rho \lambda }\left( P\left( M\right) \right)
_{\alpha \beta \gamma }\eta ^{\alpha \beta \gamma }\eta ^{\mu \nu \rho
\lambda },  \label{a3}
\end{eqnarray}%
\begin{eqnarray}
a_{2}^{\mathrm{BF}} &=&\left( P\left( W\right) \right) ^{\mu \nu }\left(
\eta C_{\mu \nu }-3A^{\rho }C_{\mu \nu \rho }\right) -2\left( 3\left(
P\left( W\right) \right) ^{\mu }B^{\ast \nu \rho }\right.  \notag \\
&&\left. +W\eta ^{\ast \mu \nu \rho }\right) C_{\mu \nu \rho }+\tfrac{9}{2}%
\varepsilon ^{\mu \nu \rho \lambda }\left( P\left( M\right) \right) _{\mu
\nu }\eta _{\rho \alpha \beta }\eta _{\lambda }^{\;\;\;\alpha \beta }  \notag
\\
&&+\varepsilon _{\mu \nu \rho \lambda }\left( 2\left( P\left( M\right)
\right) _{\alpha }A^{\ast \alpha }-2M\eta ^{\ast }+\left( P\left( M\right)
\right) _{\alpha \beta }B^{\alpha \beta }\right) \eta _{b}^{\mu \nu \rho
\lambda },  \label{a2}
\end{eqnarray}%
\begin{eqnarray}
a_{1}^{\mathrm{BF}} &=&\left( P\left( W\right) \right) ^{\mu }\left( -\eta
H_{\mu }+2A^{\nu }C_{\mu \nu }\right) +W\left( 2B_{\mu \nu }^{\ast }C^{\mu
\nu }+\varphi ^{\ast }\eta \right)  \notag \\
&&+2\varepsilon _{\nu \rho \sigma \lambda }\left( \left( P\left( M\right)
\right) _{\mu }B^{\mu \nu }-MA^{\ast \nu }\right) \eta ^{\rho \sigma \lambda
},  \label{a1}
\end{eqnarray}%
\begin{equation}
a_{0}^{\mathrm{BF}}=-WA^{\mu }H_{\mu }+\tfrac{1}{2}\varepsilon ^{\mu \nu
\rho \lambda }MB_{\mu \nu }B_{\rho \lambda }.  \label{a0}
\end{equation}%
In (\ref{a4})--(\ref{a1}) the quantities denoted by $\left( P\left( W\right)
\right) ^{\mu _{1}\ldots \mu _{k}}$ and $\left( P\left( M\right) \right)
^{\mu _{1}\ldots \mu _{k}}$ read as in (\ref{3.13}), (\ref{3.14}), (\ref%
{3.15}), and (\ref{3.16}) for $k=4$, $k=3$, $k=2$, and $k=1$ respectively,
modulo the successive replacement of $W\left( \varphi \right) $ with the
real smooth functions $W\left( \varphi \right) $ and $M\left( \varphi
\right) $, respectively.

Due to the fact that $a^{\mathrm{BF}}$ and $a^{\mathrm{int}}$ involve
different types of fields and because $a^{\mathrm{BF}}$ satisfies
individually an equation of the type (\ref{3.1}), it follows that $a^{%
\mathrm{int}}$ is subject to the equation
\begin{equation}
sa^{\mathrm{int}}=\partial ^{\mu }m_{\mu }^{\mathrm{int}},  \label{fo3}
\end{equation}%
for some local current $m_{\mu }^{\mathrm{int}}$. In the sequel we determine
the general solution to (\ref{fo3}) that complies with all the hypotheses
mentioned in the beginning of the previous subsection.

In agreement with (\ref{fo1}), the solution to the equation $sa^{\mathrm{int}%
}=\partial ^{\mu }m_{\mu }^{\mathrm{int}}$ can be decomposed as
\begin{equation}
a^{\mathrm{int}}=a_{0}^{\mathrm{int}}+a_{1}^{\mathrm{int}}+a_{2}^{\mathrm{int%
}}+a_{3}^{\mathrm{int}}+a_{4}^{\mathrm{int}},  \label{fo5}
\end{equation}%
where the components on the right-hand side of (\ref{fo5}) are subject to
the equations
\begin{eqnarray}
\gamma a_{4}^{\mathrm{int}} &=&0,  \label{fo6a} \\
\delta a_{k}^{\mathrm{int}}+\gamma a_{k-1}^{\mathrm{int}} &=&\partial ^{\mu }%
\overset{(k-1)}{m}_{\mu }^{\mathrm{int}},\qquad k=\overline{1,4}.
\label{fo6b}
\end{eqnarray}%
The piece $a_{4}^{\mathrm{int}}$ as solution to equation (\ref{fo6a}) has
the general form expressed by (\ref{3.8}) for $I=4$, with $\alpha _{4}$ from
$H_{4}^{\mathrm{inv}}(\delta |d)$ and $e^{4}$ spanned by
\begin{equation}
\left\{ \xi ^{A}\xi ^{B}\xi ^{C}\xi ^{D},\xi ^{A}\xi ^{B}\xi ^{C}\eta ,\xi
^{A}\xi ^{B}\eta ^{\mu \nu \rho \lambda },\eta C_{\mu \nu \rho \lambda
},\eta ^{\mu \nu \rho \lambda }\eta ^{\alpha \beta \gamma \delta }\right\} .
\label{fo7}
\end{equation}%
Taking into account the result that the general representative of $H_{4}^{%
\mathrm{inv}}(\delta |d)$ is given by (\ref{3.13}) and recalling that $%
a_{4}^{\mathrm{int}}$ should mix the BF and the massless spin-$3/2$ sectors
(in order to provide cross-couplings), it follows that the eligible
representatives of $e^{4}$ from (\ref{fo7}) allowed to enter $a_{4}^{\mathrm{%
int}}$ are those elements containing at least one ghost of the type $\xi
^{A} $. Recalling the symmetry properties (\ref{sym1}) and (\ref{sym2}) of
the $\gamma $-matrices, we deduce the general solution of equation (\ref%
{fo6a}) under the form
\begin{eqnarray}
a_{4}^{\mathrm{int}} &=&\frac{1}{4\cdot 4!}\varepsilon _{\mu \nu \rho
\lambda }\left[ \left( P\left( U_{1}\right) \right) ^{\mu \nu \rho \lambda
}\left( \bar{\xi}\gamma _{\alpha }\xi \right) \bar{\xi}\gamma ^{\alpha }\xi +%
\bar{\xi}\gamma _{\alpha \beta }\xi \left( \left( P\left( U_{2}\right)
\right) ^{\mu \nu \rho \lambda }\bar{\xi}\gamma ^{\alpha \beta }\xi \right.
\right.  \notag \\
&&\left. \left. +2\left( P\left( U_{3}\right) \right) ^{\mu \nu \rho \lambda
}\bar{\xi}\gamma ^{\alpha \beta }\gamma _{5}\xi \right) \right] ,
\label{ai4}
\end{eqnarray}%
where $U_{1}$, $U_{2}$, and $U_{3}$ are smooth functions depending only on
the undifferentiated scalar field $\varphi $ ($\left( P\left( U_{i}\right)
\right) ^{\mu \nu \rho \lambda }$, with $i=1,2,3$, read as in (\ref{3.13}),
but with the function $W$ replaced by $U_{i}$). Introducing (\ref{ai4}) in
equation (\ref{fo6b}) for $k=4$ and employing definitions (\ref{bfa15})--(%
\ref{bfa23}), we determine the component of antighost number three from (\ref%
{fo5}) as%
\begin{eqnarray}
a_{3}^{\mathrm{int}} &=&\frac{1}{3!}\varepsilon _{\mu \nu \rho \lambda }%
\left[ \left( P\left( U_{1}\right) \right) ^{\nu \rho \lambda }\left( \bar{%
\xi}\gamma _{\alpha }\xi \right) \bar{\xi}\gamma ^{\alpha }\psi ^{\mu }-\bar{%
\xi}\gamma _{\alpha \beta }\psi ^{\mu }\left( \left( P\left( U_{2}\right)
\right) ^{\nu \rho \lambda }\bar{\xi}\gamma ^{\alpha \beta }\xi \right.
\right.  \notag \\
&&\left. \left. +\left( P\left( U_{3}\right) \right) ^{\nu \rho \lambda }%
\bar{\xi}\gamma ^{\alpha \beta }\gamma _{5}\xi \right) +\left( P\left(
U_{3}\right) \right) ^{\nu \rho \lambda }\left( \bar{\xi}\gamma _{\alpha
\beta }\xi \right) \bar{\xi}\gamma ^{\alpha \beta }\gamma _{5}\psi ^{\mu }%
\right] +\bar{a}_{3}^{\mathrm{int}},  \label{ai3}
\end{eqnarray}%
where the objects $\left( \left( P\left( U_{i}\right) \right) ^{\mu \nu \rho
}\right) _{i=1,2,3}$ are of the form (\ref{3.14}), up to replacing the
function $W$ with $U_{i}$, and $\bar{a}_{3}^{\mathrm{int}}$ is the general
solution of the `homogeneous' equation $\gamma \bar{a}_{3}^{\mathrm{int}%
}=\partial _{\mu }u_{3}^{\mu }$, which, according to the discussion from the
previous subsection, can be replaced with%
\begin{equation}
\gamma \bar{a}_{3}^{\mathrm{int}}=0.  \label{om3}
\end{equation}%
This means that we can always take $\bar{a}_{3}^{\mathrm{int}}$ as a
nontrivial object of $H\left( \gamma \right) $. At this stage it is useful
to decompose $\bar{a}_{3}^{\mathrm{int}}$ into
\begin{equation}
\bar{a}_{3}^{\mathrm{int}}=\hat{a}_{3}^{\mathrm{int}}+\tilde{a}_{3}^{\mathrm{%
int}}.  \label{cai36a}
\end{equation}%
The first piece, $\hat{a}_{3}^{\mathrm{int}}$, denotes the component of the
solution to (\ref{om3}) required by the consistency of $a_{3}^{\mathrm{int}}$
in antighost number two (ensures that (\ref{fo6b}) possesses solutions for $%
k=3$ with respect to the terms from $a_{3}^{\mathrm{int}}$ containing the
functions $U_{i}$) and $\tilde{a}_{3}^{\mathrm{int}}$ signifies the part of
the solution to (\ref{om3}) that is independently consistent in antighost
number two%
\begin{equation}
\delta \tilde{a}_{3}^{\mathrm{int}}=-\gamma \tilde{c}_{2}+\partial _{\mu }%
\tilde{m}_{2}^{\mu }.  \label{cai36b}
\end{equation}

By means of definitions (\ref{bfa15})--(\ref{bfa23}) and recalling the
decomposition (\ref{cai36a}) one infers (by direct computation) that
\begin{equation}
\delta a_{3}^{\mathrm{int}}=\delta \hat{a}_{3}^{\mathrm{int}}+\gamma
c_{2}+\partial _{\mu }j_{2}^{\mu }+\chi _{2},  \label{cai31}
\end{equation}%
where we made the notations%
\begin{eqnarray}
c_{2} &=&-\tilde{c}_{2}+\frac{1}{2}\varepsilon _{\mu \nu \rho \lambda
}\left\{ \left( P\left( U_{1}\right) \right) ^{\rho \lambda }\left[ -\left(
\bar{\xi}\gamma _{\alpha }\psi ^{\mu }\right) \bar{\xi}\gamma ^{\alpha }\psi
^{\nu }+\frac{1}{2}\left( \bar{\xi}\gamma _{\alpha }\xi \right) \bar{\psi}%
^{\mu }\gamma ^{\alpha }\psi ^{\nu }\right] \right.  \notag \\
&&+\left( P\left( U_{2}\right) \right) ^{\rho \lambda }\left[ -\left( \bar{%
\xi}\gamma _{\alpha \beta }\psi ^{\mu }\right) \bar{\xi}\gamma ^{\alpha
\beta }\psi ^{\nu }+\frac{1}{2}\left( \bar{\xi}\gamma _{\alpha \beta }\xi
\right) \bar{\psi}^{\mu }\gamma ^{\alpha \beta }\psi ^{\nu }\right]  \notag
\\
&&+\frac{1}{2}\left( P\left( U_{3}\right) \right) ^{\rho \lambda }\left[
\left( \bar{\psi}^{\mu }\gamma _{\alpha \beta }\psi ^{\nu }\right) \bar{\xi}%
\gamma ^{\alpha \beta }\gamma _{5}\xi +\left( \bar{\xi}\gamma _{\alpha \beta
}\xi \right) \bar{\psi}^{\mu }\gamma ^{\alpha \beta }\gamma _{5}\psi ^{\nu
}\right.  \notag \\
&&\left. \left. -4\left( \bar{\xi}\gamma _{\alpha \beta }\psi ^{\mu }\right)
\bar{\xi}\gamma ^{\alpha \beta }\gamma _{5}\psi ^{\nu }\right] \right\} ,
\label{cai32}
\end{eqnarray}%
\begin{eqnarray}
j_{2}^{\mu } &=&\tilde{m}_{2}^{\mu }-\frac{1}{2}\varepsilon ^{\mu \nu \rho
\lambda }\left\{ \left( P\left( U_{1}\right) \right) _{\nu \rho }\left( \bar{%
\xi}\gamma _{\alpha }\xi \right) \bar{\xi}\gamma ^{\alpha }\psi _{\lambda
}+\left( P\left( U_{2}\right) \right) _{\nu \rho }\left( \bar{\xi}\gamma
_{\alpha \beta }\xi \right) \bar{\xi}\gamma ^{\alpha \beta }\psi _{\lambda
}\right.  \notag \\
&&\left. +\left( P\left( U_{3}\right) \right) _{\nu \rho }\left[ \left( \bar{%
\xi}\gamma _{\alpha \beta }\psi _{\lambda }\right) \bar{\xi}\gamma ^{\alpha
\beta }\gamma _{5}\xi +\left( \bar{\xi}\gamma _{\alpha \beta }\xi \right)
\bar{\xi}\gamma ^{\alpha \beta }\gamma _{5}\psi _{\lambda }\right] \right\} ,
\label{cai33}
\end{eqnarray}%
\begin{eqnarray}
\chi _{2} &=&\frac{1}{4}\varepsilon ^{\mu \nu \rho \lambda }\left\{ \left(
P\left( U_{1}\right) \right) _{\mu \nu }\left( \bar{\xi}\gamma _{\alpha }\xi
\right) \bar{\xi}\gamma ^{\alpha }\partial _{\lbrack \rho }\psi _{\lambda
]}+\left( P\left( U_{2}\right) \right) _{\mu \nu }\left( \bar{\xi}\gamma
_{\alpha \beta }\xi \right) \bar{\xi}\gamma ^{\alpha \beta }\partial
_{\lbrack \rho }\psi _{\lambda ]}\right.  \notag \\
&&\left. +\left( P\left( U_{3}\right) \right) _{\mu \nu }\left[ \left( \bar{%
\xi}\gamma ^{\alpha \beta }\gamma _{5}\xi \right) \bar{\xi}\gamma _{\alpha
\beta }\partial _{\lbrack \rho }\psi _{\lambda ]}+\left( \bar{\xi}\gamma
_{\alpha \beta }\xi \right) \bar{\xi}\gamma ^{\alpha \beta }\gamma
_{5}\partial _{\lbrack \rho }\psi _{\lambda ]}\right] \right\} .
\label{cai34}
\end{eqnarray}%
Comparing (\ref{cai31}) with (\ref{fo6b}) for $k=3$, it follows that the
existence of $a_{2}^{\mathrm{int}}$ is ensured if and only if $\chi _{2}$
satisfies the equation%
\begin{equation}
\chi _{2}=-\delta \hat{a}_{3}^{\mathrm{int}}+\gamma \hat{c}_{2}+\partial
_{\mu }\hat{\jmath}_{2}^{\mu },  \label{cai35}
\end{equation}%
where%
\begin{equation}
\hat{c}_{2}=-\left( a_{2}^{\mathrm{int}}+c_{2}\right) ,\qquad \hat{\jmath}%
_{2}^{\mu }=\overset{(2)}{m}^{\mathrm{int~}\mu }-j_{2}^{\mu }.  \label{cai36}
\end{equation}%
We will show that (\ref{cai35}) cannot hold unless $\chi _{2}=0$. In view of
this, we assume equation (\ref{cai35}) is valid. By taking its
Euler-Lagrange (EL) derivatives with respect to $C_{\mu \nu }^{\ast }$ we
infer%
\begin{equation}
\frac{\delta \chi _{2}}{\delta C_{\mu \nu }^{\ast }}=-\frac{\delta \left(
\delta \hat{a}_{3}^{\mathrm{int}}\right) }{\delta C_{\mu \nu }^{\ast }}%
+\gamma \left( \frac{\delta \hat{c}_{2}}{\delta C_{\mu \nu }^{\ast }}\right)
.  \label{cai37}
\end{equation}%
Direct computation based on (\ref{cai34}) leads to%
\begin{eqnarray}
\frac{\delta \chi _{2}}{\delta C_{\mu \nu }^{\ast }} &=&\frac{1}{4}%
\varepsilon ^{\mu \nu \rho \lambda }\left\{ \frac{dU_{1}}{d\varphi }\left(
\bar{\xi}\gamma _{\alpha }\xi \right) \bar{\xi}\gamma ^{\alpha }\partial
_{\lbrack \rho }\psi _{\lambda ]}+\frac{dU_{2}}{d\varphi }\left( \bar{\xi}%
\gamma _{\alpha \beta }\xi \right) \bar{\xi}\gamma ^{\alpha \beta }\partial
_{\lbrack \rho }\psi _{\lambda ]}\right.  \notag \\
&&\left. +\frac{dU_{3}}{d\varphi }\left[ \left( \bar{\xi}\gamma ^{\alpha
\beta }\gamma _{5}\xi \right) \bar{\xi}\gamma _{\alpha \beta }\partial
_{\lbrack \rho }\psi _{\lambda ]}+\left( \bar{\xi}\gamma _{\alpha \beta }\xi
\right) \bar{\xi}\gamma ^{\alpha \beta }\gamma _{5}\partial _{\lbrack \rho
}\psi _{\lambda ]}\right] \right\} .  \label{cai38}
\end{eqnarray}%
It is easy to see that the right-hand side of (\ref{cai38}) is a nontrivial
object from $H\left( \gamma \right) $, so relation (\ref{cai37}) implies%
\begin{equation}
\frac{\delta \chi _{2}}{\delta C_{\mu \nu }^{\ast }}=-\frac{\delta \left(
\delta \hat{a}_{3}^{\mathrm{int}}\right) }{\delta C_{\mu \nu }^{\ast }}%
,\qquad \gamma \left( \frac{\delta \hat{c}_{2}}{\delta C_{\mu \nu }^{\ast }}%
\right) =0.  \label{cai39}
\end{equation}%
Due to (\ref{cai38}), the former formula in (\ref{cai39}) cannot take place.
This is because $\delta \hat{a}_{3}^{\mathrm{int}}$ comprises two spacetime
derivatives, while $\frac{\delta \chi _{2}}{\delta C_{\mu \nu }^{\ast }}$
has only one. Indeed, $\hat{a}_{3}^{\mathrm{int}}$ is expressed by (\ref{3.8}%
) for $I=3$ and is also simultaneously linear in $C_{\mu \nu \rho }^{\ast }$
and $\partial _{\lbrack \mu }\psi _{\nu ]}$ (the linearity in $\partial
_{\lbrack \mu }\psi _{\nu ]}$ is imposed by the linearity of the right-hand
side of (\ref{cai38})). Taking into account these observations, we get that $%
\delta \hat{a}_{3}^{\mathrm{int}}$ is linear in both $\partial _{\lbrack \mu
}C_{\nu \rho ]}^{\ast }$ and $\partial _{\lbrack \mu }\psi _{\nu ]}$, so it
displays precisely two spacetime derivatives. As a consequence, the former
equality from (\ref{cai39}) does not hold, so neither do formulas (\ref%
{cai37}) or (\ref{cai35}). In conclusion, $\chi _{2}$ must vanish, which
further implies that $\left( U_{i}\left( \varphi \right) \right) _{i=1,2,3}$
must be constant, and thus $a_{4}^{\mathrm{int}}$ itself vanishes.

Since the decomposition (\ref{fo5}) cannot stop at antighost number four, we
pass to the next possibility, namely that $a^{\mathrm{int}}$ ends at
antighost number three:%
\begin{equation}
a^{\mathrm{int}}=a_{0}^{\mathrm{int}}+a_{1}^{\mathrm{int}}+a_{2}^{\mathrm{int%
}}+a_{3}^{\mathrm{int}},  \label{fo8}
\end{equation}%
where the components on the right-hand side of (\ref{fo8}) are subject to
the equations
\begin{eqnarray}
\gamma a_{3}^{\mathrm{int}} &=&0,  \label{fo9a} \\
\delta a_{k}^{\mathrm{int}}+\gamma a_{k-1}^{\mathrm{int}} &=&\partial ^{\mu }%
\overset{(k-1)}{m}_{\mu }^{\mathrm{int}},\qquad k=\overline{1,3}.
\label{fo9b}
\end{eqnarray}%
The piece $a_{3}^{\mathrm{int}}$ as solution to equation (\ref{fo9a}) has
the general form expressed by (\ref{3.8}) for $I=3$, with $\alpha _{3}$ from
$H_{3}^{\mathrm{inv}}(\delta |d)$ and $e^{3}$ spanned by%
\begin{equation}
\left\{ \xi ^{A}\xi ^{B}\xi ^{C},\xi ^{A}\xi ^{B}\eta ,\xi ^{A}\eta ^{\mu
\nu \rho \lambda },\eta \eta ^{\mu \nu \rho \lambda }\right\} .  \label{fo10}
\end{equation}%
Given the spinor-like behavior of some of the elements (\ref{fo10}) and also
the general expressions of the generators of $H_{3}^{\mathrm{inv}}(\delta
|d) $ (see (\ref{hinvdeltamodd}) for $J=3$), the general, real solution to
equation (\ref{fo9a}) reads as%
\begin{equation}
a_{3}^{\mathrm{int}}=-\frac{\mathrm{i}}{12}\varepsilon ^{\mu \nu \rho
\lambda }\left( P\left( U_{4}\right) \right) _{\mu \nu \rho }\bar{\xi}\gamma
_{\lambda }\xi \eta ,  \label{fo11}
\end{equation}%
where $U_{4}=U_{4}\left( \varphi \right) $ is a smooth function on the
(undifferentiated) scalar field $\varphi $ and $\left( P\left( U_{4}\right)
\right) _{\mu \nu \rho }$ follows from (\ref{3.14}) with $W$ replaced by $%
U_{4}$. Making use of the latter set of duality relations from (\ref{d1}), (%
\ref{fo11}) can be written as%
\begin{equation}
a_{3}^{\mathrm{int}}=\frac{1}{12}\left( P\left( U_{4}\right) \right) _{\mu
\nu \rho }\bar{\xi}\gamma ^{\mu \nu \rho }\gamma _{5}\xi \eta .  \label{fo12}
\end{equation}%
Substituting (\ref{fo12}) into (\ref{fo9b}) for $k=3$ and recalling
definitions (\ref{bfa15})--(\ref{bfa23}), we identify the component of
antighost number two from the first-order deformation as%
\begin{equation}
a_{2}^{\mathrm{int}}=\frac{1}{4}\left( P\left( U_{4}\right) \right) _{\mu
\nu }\left( \bar{\xi}\gamma ^{\mu \nu \rho }\gamma _{5}\xi A_{\rho }-2\bar{%
\xi}\gamma ^{\mu \nu \rho }\gamma _{5}\psi _{\rho }\eta \right) +\bar{a}%
_{2}^{\mathrm{int}},  \label{fo13}
\end{equation}%
where the quantity $\left( P\left( U_{4}\right) \right) _{\mu \nu }$ is of
the type (\ref{3.15}) and $\bar{a}_{2}^{\mathrm{int}}$ is, like in the
above, the general solution to the `homogeneous' equation $\gamma \bar{a}%
_{2}^{\mathrm{int}}=\partial _{\mu }u_{2}^{\mu }$, which, according to our
discussion from the previous subsection, can be safely replaced with
\begin{equation}
\gamma \bar{a}_{2}^{\mathrm{int}}=0.  \label{om2}
\end{equation}
Just like before (see (\ref{cai36a})), we decompose $\bar{a}_{2}^{\mathrm{int%
}}$ into%
\begin{equation}
\bar{a}_{2}^{\mathrm{int}}=\hat{a}_{2}^{\mathrm{int}}+\tilde{a}_{2}^{\mathrm{%
int}},  \label{ai2}
\end{equation}%
where $\hat{a}_{2}^{\mathrm{int}}$ is the solution to (\ref{om2}) necessary
for the consistency of $a_{2}^{\mathrm{int}}$ in antighost number one (for
the existence of solutions $a_{1}^{\mathrm{int}}$ to (\ref{fo9b}) for $k=2$
with respect to the terms from $a_{2}^{\mathrm{int}}$ containing the
function $U_{4}$) and $\tilde{a}_{2}^{\mathrm{int}}$ is the solution to (\ref%
{om2}) that is independently consistent in antighost number one%
\begin{equation}
\delta \tilde{a}_{2}^{\mathrm{int}}=-\gamma \tilde{c}_{1}+\partial _{\mu }%
\tilde{m}_{1}^{\mu }.  \label{ai21}
\end{equation}%
With the help of definitions (\ref{bfa15})--(\ref{bfa23}), we get%
\begin{eqnarray}
\delta a_{2}^{\mathrm{int}} &=&\delta \left[ \hat{a}_{2}^{\mathrm{int}}-%
\frac{1}{2}\left( P\left( U_{4}\right) \right) _{\mu }\bar{\xi}\gamma ^{\mu
\nu \rho }\gamma _{5}\xi B_{\nu \rho }^{\ast }-\frac{1}{6}U_{4}\bar{\xi}%
\gamma ^{\mu \nu \rho }\gamma _{5}\xi \eta _{\mu \nu \rho }^{\ast }\right.
\notag \\
&&\left. -\mathrm{i}\left( P\left( U_{4}\right) \right) _{\mu }\psi ^{\ast
\mu }\gamma _{5}\xi \eta +\mathrm{i}U_{4}\xi ^{\ast }\gamma _{5}\xi \eta %
\right]  \notag \\
&&+\gamma c_{1}+\partial _{\mu }j_{1}^{\mu },  \label{ai22}
\end{eqnarray}%
where we made the notations
\begin{eqnarray}
c_{1} &=&-\tilde{c}_{1}+\mathrm{i}U_{4}\left( \psi ^{\ast \mu }\gamma
_{5}\psi _{\mu }\eta -\psi ^{\ast \mu }\gamma _{5}\xi A_{\mu }-\mathrm{i}%
\bar{\xi}\gamma ^{\mu \nu \rho }\gamma _{5}\psi _{\mu }B_{\nu \rho }^{\ast
}\right)  \notag \\
&&+\frac{1}{2}\left( P\left( U_{4}\right) \right) _{\mu }\left( 2\bar{\xi}%
\gamma ^{\mu \nu \rho }\gamma _{5}\psi _{\nu }A_{\rho }+\bar{\psi}_{\nu
}\gamma ^{\mu \nu \rho }\gamma _{5}\psi _{\rho }\eta \right) ,  \label{ai23}
\end{eqnarray}%
\begin{eqnarray}
j_{1}^{\mu } &=&\tilde{m}_{1}^{\mu }+\frac{1}{2}\left( P\left( U_{4}\right)
\right) _{\nu }\left( \bar{\xi}\gamma ^{\mu \nu \rho }\gamma _{5}\xi A_{\rho
}-2\bar{\xi}\gamma ^{\mu \nu \rho }\gamma _{5}\psi _{\rho }\eta \right)
\notag \\
&&+\frac{1}{2}U_{4}\left( \bar{\xi}\gamma ^{\mu \nu \rho }\gamma _{5}\xi
B_{\nu \rho }^{\ast }+2\mathrm{i}\psi ^{\ast \mu }\gamma _{5}\xi \eta
\right) ,  \label{ai24}
\end{eqnarray}%
and $\left( P\left( U_{4}\right) \right) _{\mu }$ reads as in (\ref{3.16})
modulo the replacement $W\rightarrow U_{4}$. Comparing (\ref{ai22}) with (%
\ref{fo9b}) for $k=2$, we infer that
\begin{eqnarray}
\hat{a}_{2}^{\mathrm{int}} &=&\frac{1}{2}\left( P\left( U_{4}\right) \right)
_{\mu }\bar{\xi}\gamma ^{\mu \nu \rho }\gamma _{5}\xi B_{\nu \rho }^{\ast }+%
\frac{1}{6}U_{4}\bar{\xi}\gamma ^{\mu \nu \rho }\gamma _{5}\xi \eta _{\mu
\nu \rho }^{\ast }  \notag \\
&&+\mathrm{i}\left( P\left( U_{4}\right) \right) _{\mu }\psi ^{\ast \mu
}\gamma _{5}\xi \eta -\mathrm{i}U_{4}\xi ^{\ast }\gamma _{5}\xi \eta ,
\label{ai25}
\end{eqnarray}%
\begin{eqnarray}
a_{1}^{\mathrm{int}} &=&\tilde{c}_{1}-\mathrm{i}U_{4}\left( \psi ^{\ast \mu
}\gamma _{5}\psi _{\mu }\eta -\psi ^{\ast \mu }\gamma _{5}\xi A_{\mu }-%
\mathrm{i}\bar{\xi}\gamma ^{\mu \nu \rho }\gamma _{5}\psi _{\mu }B_{\nu \rho
}^{\ast }\right)  \notag \\
&&-\frac{1}{2}\left( P\left( U_{4}\right) \right) _{\mu }\left( 2\bar{\xi}%
\gamma ^{\mu \nu \rho }\gamma _{5}\psi _{\nu }A_{\rho }+\bar{\psi}_{\nu
}\gamma ^{\mu \nu \rho }\gamma _{5}\psi _{\rho }\eta \right) +\bar{a}_{1}^{%
\mathrm{int}},  \label{ai26}
\end{eqnarray}%
where $\bar{a}_{1}^{\mathrm{int}}$ is the general solution to the
`homogeneous' equation $\gamma \bar{a}_{1}^{\mathrm{int}}=\partial _{\mu
}u_{1}^{\mu }$, which can again be replaced with%
\begin{equation}
\gamma \bar{a}_{1}^{\mathrm{int}}=0.  \label{om1}
\end{equation}

Regarding the component $\tilde{a}_{2}^{\mathrm{int}}$ of the equation (\ref%
{om2}), it can be represented like in (\ref{3.8}) for $I=2$, with $\alpha
_{2}$ from $H_{2}^{\mathrm{inv}}(\delta |d)$ and $e^{2}$ spanned by
\begin{equation*}
\left\{ \xi ^{A}\xi ^{B},\xi ^{A}\eta ,\eta ^{\mu \nu \rho \lambda }\right\}
.
\end{equation*}%
The most general representatives of $H_{2}^{\mathrm{inv}}(\delta |d)$ are
listed in (\ref{hinvdeltamodd}) for $J=2$, so $\tilde{a}_{2}^{\mathrm{int}}$
is generally expressed by%
\begin{equation}
\tilde{a}_{2}^{\mathrm{int}}=\frac{\mathrm{i}}{4}\left( P\left( U_{5}\right)
\right) _{\mu \nu }\bar{\xi}\gamma ^{\mu \nu }\gamma _{5}\xi -\frac{1}{4}%
\left( P\left( U_{6}\right) \right) _{\mu \nu }\bar{\xi}\gamma ^{\mu \nu
}\xi ,  \label{ai27}
\end{equation}%
where $U_{5}\left( \varphi \right) $ and $U_{6}\left( \varphi \right) $ are
some real, smooth (but otherwise arbitrary) functions depending only on $%
\varphi $. Inserting (\ref{ai25}) and (\ref{ai27}) in (\ref{fo13}), we
arrive at%
\begin{eqnarray}
a_{2}^{\mathrm{int}} &=&\frac{1}{4}\left( P\left( U_{4}\right) \right) _{\mu
\nu }\left( \bar{\xi}\gamma ^{\mu \nu \rho }\gamma _{5}\xi A_{\rho }-2\bar{%
\xi}\gamma ^{\mu \nu \rho }\gamma _{5}\psi _{\rho }\eta \right)   \notag \\
&&+\frac{1}{2}\left( P\left( U_{4}\right) \right) _{\mu }\bar{\xi}\gamma
^{\mu \nu \rho }\gamma _{5}\xi B_{\nu \rho }^{\ast }+\frac{1}{6}U_{4}\bar{\xi%
}\gamma ^{\mu \nu \rho }\gamma _{5}\xi \eta _{\mu \nu \rho }^{\ast }  \notag
\\
&&+\mathrm{i}\left( P\left( U_{4}\right) \right) _{\mu }\psi ^{\ast \mu
}\gamma _{5}\xi \eta -\mathrm{i}U_{4}\xi ^{\ast }\gamma _{5}\xi \eta   \notag
\\
&&+\frac{\mathrm{i}}{4}\left( P\left( U_{5}\right) \right) _{\mu \nu }\bar{%
\xi}\gamma ^{\mu \nu }\gamma _{5}\xi -\frac{1}{4}\left( P\left( U_{6}\right)
\right) _{\mu \nu }\bar{\xi}\gamma ^{\mu \nu }\xi .  \label{fo14}
\end{eqnarray}%
Applying $\delta $ on $\tilde{a}_{2}^{\mathrm{int}}$ given in (\ref{ai27})
and using definitions (\ref{bfa15})--(\ref{bfa23}), we obtain the concrete
expression of the object $\tilde{c}_{1}$ involved in (\ref{ai21}) of the form%
\begin{equation}
\tilde{c}_{1}=-\mathrm{i}\left( P\left( U_{5}\right) \right) _{\mu }\bar{\xi}%
\gamma ^{\mu \nu }\gamma _{5}\psi _{\nu }+\left( P\left( U_{6}\right)
\right) _{\mu }\bar{\xi}\gamma ^{\mu \nu }\psi _{\nu },  \label{ai28}
\end{equation}%
which further inserted in (\ref{ai26}) allows us to write the component of
antighost number one from the first-order deformation as%
\begin{eqnarray}
a_{1}^{\mathrm{int}} &=&-\mathrm{i}U_{4}\left( \psi ^{\ast \mu }\gamma
_{5}\psi _{\mu }\eta -\psi ^{\ast \mu }\gamma _{5}\xi A_{\mu }-\mathrm{i}%
\bar{\xi}\gamma ^{\mu \nu \rho }\gamma _{5}\psi _{\mu }B_{\nu \rho }^{\ast
}\right)   \notag \\
&&-\frac{1}{2}\left( P\left( U_{4}\right) \right) _{\mu }\left( 2\bar{\xi}%
\gamma ^{\mu \nu \rho }\gamma _{5}\psi _{\nu }A_{\rho }+\bar{\psi}_{\nu
}\gamma ^{\mu \nu \rho }\gamma _{5}\psi _{\rho }\eta \right)   \notag \\
&&-\mathrm{i}\left( P\left( U_{5}\right) \right) _{\mu }\bar{\xi}\gamma
^{\mu \nu }\gamma _{5}\psi _{\nu }+\left( P\left( U_{6}\right) \right) _{\mu
}\bar{\xi}\gamma ^{\mu \nu }\psi _{\nu }+\bar{a}_{1}^{\mathrm{int}}.
\label{fo15}
\end{eqnarray}%
We split again $\bar{a}_{1}^{\mathrm{int}}$ into%
\begin{equation}
\bar{a}_{1}^{\mathrm{int}}=\hat{a}_{1}^{\mathrm{int}}+\tilde{a}_{1}^{\mathrm{%
int}},  \label{ai11}
\end{equation}%
with $\hat{a}_{1}^{\mathrm{int}}$ the solution to (\ref{om1}) ensuring the
consistency of $a_{1}^{\mathrm{int}}$ in antighost number zero (i.e., the
existence of solutions $a_{0}^{\mathrm{int}}$ to equation (\ref{fo9b}) for $%
k=1$ for the pieces from $a_{1}^{\mathrm{int}}$ containing the functions $%
U_{m}$, with $m=4,5,6$) and $\tilde{a}_{1}^{\mathrm{int}}$ the solution to (%
\ref{om1}) that is independently consistent at antighost number zero%
\begin{equation}
\delta \tilde{a}_{1}^{\mathrm{int}}=-\gamma \tilde{c}_{0}+\partial _{\mu }%
\tilde{m}_{0}^{\mu }.  \label{ai12}
\end{equation}%
By means of definitions (\ref{bfa15})--(\ref{bfa23}), straightforward
computation produces%
\begin{equation}
\delta a_{1}^{\mathrm{int}}=\delta \left[ \hat{a}_{1}^{\mathrm{int}}+\frac{1%
}{2}U_{5}\psi ^{\ast \mu }\gamma _{\mu }\gamma _{5}\xi +\frac{\mathrm{i}}{2}%
U_{6}\psi ^{\ast \mu }\gamma _{\mu }\xi \right] +\gamma c_{0}+\partial _{\mu
}j_{0}^{\mu },  \label{ai13}
\end{equation}%
where
\begin{equation}
c_{0}=-\tilde{c}_{0}+\frac{1}{2}U_{4}\bar{\psi}_{\mu }\gamma ^{\mu \nu \rho
}\gamma _{5}\psi _{\nu }A_{\rho }+\frac{1}{2}\left( \mathrm{i}U_{5}\bar{\psi}%
_{\mu }\gamma ^{\mu \nu }\gamma _{5}\psi _{\nu }-U_{6}\bar{\psi}_{\mu
}\gamma ^{\mu \nu }\psi _{\nu }\right) ,  \label{ai14}
\end{equation}%
\begin{eqnarray}
j_{0}^{\mu } &=&\tilde{m}_{0}^{\mu }-U_{4}\left( \frac{1}{2}\bar{\psi}_{\nu
}\gamma ^{\mu \nu \rho }\gamma _{5}\psi _{\rho }\eta +\bar{\xi}\gamma ^{\mu
\nu \rho }\gamma _{5}\psi _{\nu }A_{\rho }\right)   \notag \\
&&-\mathrm{i}U_{5}\bar{\xi}\gamma ^{\mu \nu }\gamma _{5}\psi _{\nu }+U_{6}%
\bar{\xi}\gamma ^{\mu \nu }\psi _{\nu }.  \label{ai15}
\end{eqnarray}%
Comparing (\ref{ai13}) with (\ref{fo9b}) for $k=1$, we deduce%
\begin{equation}
\hat{a}_{1}^{\mathrm{int}}=-\frac{1}{2}\left( U_{5}\psi ^{\ast \mu }\gamma
_{\mu }\gamma _{5}\xi +\mathrm{i}U_{6}\psi ^{\ast \mu }\gamma _{\mu }\xi
\right) ,  \label{ai16}
\end{equation}%
\begin{equation}
a_{0}^{\mathrm{int}}=\tilde{c}_{0}-\frac{1}{2}\left( U_{4}\bar{\psi}_{\mu
}\gamma ^{\mu \nu \rho }\gamma _{5}\psi _{\nu }A_{\rho }+\mathrm{i}U_{5}\bar{%
\psi}_{\mu }\gamma ^{\mu \nu }\gamma _{5}\psi _{\nu }-U_{6}\bar{\psi}_{\mu
}\gamma ^{\mu \nu }\psi _{\nu }\right) +\bar{a}_{0}^{\mathrm{int}},
\label{ai17}
\end{equation}%
where $\bar{a}_{0}^{\mathrm{int}}$ is the general solution to the
`homogeneous' equation $\gamma \bar{a}_{0}^{\mathrm{int}}=\partial _{\mu
}u_{0}^{\mu }$ (it \emph{cannot} be replaced any longer with the simpler
equation, corresponding to $u_{0}^{\mu }=0$). The component $\tilde{a}_{1}^{%
\mathrm{int}}$, which is a solution to (\ref{om1}) that is independently
consistent at antighost number zero, i.e. satisfies equation (\ref{ai12}),
can be taken to vanish%
\begin{equation}
\tilde{a}_{1}^{\mathrm{int}}=0,  \label{om159}
\end{equation}%
since it produces only trivial deformations, as it will be shown in Appendix %
\ref{formula1}.

Injecting (\ref{ai16}) and (\ref{om159}) in (\ref{ai11}), and the resulting
expression in (\ref{fo15}), we complete the component of antighost number
one from the first-order deformation as%
\begin{eqnarray}
a_{1}^{\mathrm{int}} &=&-\mathrm{i}U_{4}\left( \psi ^{\ast \mu }\gamma
_{5}\psi _{\mu }\eta -\psi ^{\ast \mu }\gamma _{5}\xi A_{\mu }-\mathrm{i}%
\bar{\xi}\gamma ^{\mu \nu \rho }\gamma _{5}\psi _{\mu }B_{\nu \rho }^{\ast
}\right)   \notag \\
&&-\frac{1}{2}\left( P\left( U_{4}\right) \right) _{\mu }\left( 2\bar{\xi}%
\gamma ^{\mu \nu \rho }\gamma _{5}\psi _{\nu }A_{\rho }+\bar{\psi}_{\nu
}\gamma ^{\mu \nu \rho }\gamma _{5}\psi _{\rho }\eta \right)   \notag \\
&&-\mathrm{i}\left( P\left( U_{5}\right) \right) _{\mu }\bar{\xi}\gamma
^{\mu \nu }\gamma _{5}\psi _{\nu }+\left( P\left( U_{6}\right) \right) _{\mu
}\bar{\xi}\gamma ^{\mu \nu }\psi _{\nu }  \notag \\
&&-\frac{1}{2}\left( U_{5}\psi ^{\ast \mu }\gamma _{\mu }\gamma _{5}\xi +%
\mathrm{i}U_{6}\psi ^{\ast \mu }\gamma _{\mu }\xi \right) .  \label{fo16}
\end{eqnarray}%
We mention that (\ref{om159}) also implies%
\begin{equation}
\tilde{c}_{0}=0  \label{om160}
\end{equation}%
in (\ref{ai14}) and (\ref{ai17}), such that the element of antighost number
zero of the first-order deformation (the interacting Lagrangian at order one
in the coupling constant) is%
\begin{equation}
a_{0}^{\mathrm{int}}=-\frac{1}{2}\left( U_{4}\bar{\psi}_{\mu }\gamma ^{\mu
\nu \rho }\gamma _{5}\psi _{\nu }A_{\rho }+\mathrm{i}U_{5}\bar{\psi}_{\mu
}\gamma ^{\mu \nu }\gamma _{5}\psi _{\nu }-U_{6}\bar{\psi}_{\mu }\gamma
^{\mu \nu }\psi _{\nu }\right) +\bar{a}_{0}^{\mathrm{int}},  \label{fo17}
\end{equation}%
where the piece $\bar{a}_{0}^{\mathrm{int}}$ is subject to the homogenous
equation%
\begin{equation}
\gamma \bar{a}_{0}^{\mathrm{int}}=\partial _{\mu }u_{0}^{\mu }.  \label{om0}
\end{equation}%
The general solution to (\ref{om0}) can be written as%
\begin{equation}
\bar{a}_{0}^{\mathrm{int}}=\bar{a}_{0}^{\prime \mathrm{int}}+\bar{a}%
_{0}^{\prime \prime \mathrm{int}},  \label{om01}
\end{equation}%
with $\bar{a}_{0}^{\prime \mathrm{int}}$ and $\bar{a}_{0}^{\prime \prime
\mathrm{int}}$ solutions to%
\begin{eqnarray}
\gamma \bar{a}_{0}^{\prime \mathrm{int}} &=&0,  \label{om02a} \\
\gamma \bar{a}_{0}^{\prime \prime \mathrm{int}} &=&\partial _{\mu
}u_{0}^{\mu },  \label{om02b}
\end{eqnarray}%
and $u_{0}^{\mu }$ a nonvanishing current. We recall the main properties of $%
\bar{a}_{0}^{\mathrm{int}}$: it should mix the Rarita-Schwinger spinors with
the BF fields (in order to provide cross-couplings) and contain at most one
spacetime derivative (like the original Lagrangian). In agreement with our
result (\ref{3.8}) for $I=0$, the component $\bar{a}_{0}^{\prime \mathrm{int}%
}$ is of the type%
\begin{equation}
\bar{a}_{0}^{\prime \mathrm{int}}=\bar{a}_{0}^{\prime \mathrm{int}}\left( %
\left[ F_{\bar{A}}\right] \right) ,  \label{om03}
\end{equation}%
where $F_{\bar{A}}$ are given in (\ref{3.7}). Due to the above mentioned
main properties, it is easy to see that (\ref{om03}) contains at least two
derivatives. Indeed, it is forced to contain gauge-invariant objects
depending on the Rarita-Schwinger spinors, so it should be quadratic in $%
\partial _{\lbrack \mu }\psi _{\nu ]}$ (in order to render a bosonic
quantity), which contradicts the derivative-order assumption, such that we
must set%
\begin{equation}
\bar{a}_{0}^{\prime \mathrm{int}}=0.  \label{om04}
\end{equation}%
In the meanwhile, as it will be shown in Appendix \ref{formula2}, the
solution to equation (\ref{om02b}) can be taken as trivial%
\begin{equation}
\bar{a}_{0}^{\prime \prime \mathrm{int}}=0,  \label{om034}
\end{equation}%
which, together with (\ref{om04}), leads to the conclusion that the general
solution to the homogeneous equation in antighost number zero, (\ref{om0}),
that complies with all the working hypotheses is also trivial%
\begin{equation}
\bar{a}_{0}^{\mathrm{int}}=0.  \label{om035}
\end{equation}%
In this manner we also completed the component of antighost number zero from
the first-order deformation, which follows from (\ref{fo17}) with $\bar{a}%
_{0}^{\mathrm{int}}$ as in (\ref{om035}):%
\begin{equation}
a_{0}^{\mathrm{int}}=-\frac{1}{2}\left( U_{4}\bar{\psi}_{\mu }\gamma ^{\mu
\nu \rho }\gamma _{5}\psi _{\nu }A_{\rho }+\mathrm{i}U_{5}\bar{\psi}_{\mu
}\gamma ^{\mu \nu }\gamma _{5}\psi _{\nu }-U_{6}\bar{\psi}_{\mu }\gamma
^{\mu \nu }\psi _{\nu }\right) .  \label{fo18}
\end{equation}

Putting together the BF piece (\ref{descBF}) (whose various terms are listed
in (\ref{a4})--(\ref{a0})) with the interacting one (\ref{fo8}) (having the
components (\ref{fo11}), (\ref{fo14}), (\ref{fo16}), and (\ref{fo18})) via
relation (\ref{fo2}) and renaming the functions $\left( U_{m}\left( \varphi
\right) \right) _{m=4,5,6}$ by%
\begin{equation}
U_{4}\left( \varphi \right) \equiv U_{1}\left( \varphi \right) ,\qquad
U_{5}\left( \varphi \right) \equiv U_{2}\left( \varphi \right) ,\qquad
U_{6}\left( \varphi \right) \equiv U_{3}\left( \varphi \right) ,
\label{renot}
\end{equation}%
we can state that the general expression of the first-order deformation of
the solution to the master equation for a four-dimensional BF model and a
massless Rarita-Schwinger field is given by:%
\begin{eqnarray}
&&S_{1}=\int d^{4}x\left[ A_{\mu }\left( -W\left( \varphi \right) H^{\mu
}\right) +\frac{1}{2}M\left( \varphi \right) \varepsilon _{\alpha \beta
\gamma \delta }B^{\alpha \beta }B^{\gamma \delta }\right.  \notag \\
&&-\frac{1}{2}\left( U_{1}\left( \varphi \right) \bar{\psi}_{\mu }\gamma
^{\mu \nu \rho }\gamma _{5}\psi _{\nu }A_{\rho }+\mathrm{i}U_{2}\left(
\varphi \right) \bar{\psi}_{\mu }\gamma ^{\mu \nu }\gamma _{5}\psi _{\nu
}-U_{3}\left( \varphi \right) \bar{\psi}_{\mu }\gamma ^{\mu \nu }\psi _{\nu
}\right)  \notag \\
&&+\frac{dW}{d\varphi }H^{\ast \mu }\left( -\eta H_{\mu }+2A^{\nu }C_{\mu
\nu }\right) +W\left( 2B_{\mu \nu }^{\ast }C^{\mu \nu }+\varphi ^{\ast }\eta
\right)  \notag \\
&&+2\varepsilon _{\nu \rho \sigma \lambda }\left( \frac{dM}{d\varphi }H_{\mu
}^{\ast }B^{\mu \nu }-MA^{\ast \nu }\right) \eta ^{\rho \sigma \lambda }-%
\mathrm{i}U_{1}\left( \psi ^{\ast \mu }\gamma _{5}\psi _{\mu }\eta -\psi
^{\ast \mu }\gamma _{5}\xi A_{\mu }\right.  \notag \\
&&\left. -\mathrm{i}\bar{\xi}\gamma ^{\mu \nu \rho }\gamma _{5}\psi _{\mu
}B_{\nu \rho }^{\ast }\right) -\frac{1}{2}\frac{dU_{1}}{d\varphi }H_{\mu
}^{\ast }\left( 2\bar{\xi}\gamma ^{\mu \nu \rho }\gamma _{5}\psi _{\nu
}A_{\rho }+\bar{\psi}_{\nu }\gamma ^{\mu \nu \rho }\gamma _{5}\psi _{\rho
}\eta \right)  \notag \\
&&-\mathrm{i}\frac{dU_{2}}{d\varphi }H_{\mu }^{\ast }\bar{\xi}\gamma ^{\mu
\nu }\gamma _{5}\psi _{\nu }+\frac{dU_{3}}{d\varphi }H_{\mu }^{\ast }\bar{\xi%
}\gamma ^{\mu \nu }\psi _{\nu }-\frac{1}{2}\left( U_{2}\psi ^{\ast \mu
}\gamma _{\mu }\gamma _{5}\xi \right.  \notag \\
&&\left. +\mathrm{i}U_{3}\psi ^{\ast \mu }\gamma _{\mu }\xi \right) +\left(
P\left( W\right) \right) ^{\mu \nu }\left( \eta C_{\mu \nu }-3A^{\rho
}C_{\mu \nu \rho }\right) -2\left( 3\left( P\left( W\right) \right) ^{\mu
}B^{\ast \nu \rho }\right.  \notag \\
&&\left. +W\eta ^{\ast \mu \nu \rho }\right) C_{\mu \nu \rho }+\tfrac{9}{2}%
\varepsilon ^{\mu \nu \rho \lambda }\left( P\left( M\right) \right) _{\mu
\nu }\eta _{\rho \alpha \beta }\eta _{\lambda }^{\;\;\;\alpha \beta }  \notag
\\
&&+\varepsilon _{\mu \nu \rho \lambda }\left( 2\left( P\left( M\right)
\right) _{\alpha }A^{\ast \alpha }-2M\eta ^{\ast }+\left( P\left( M\right)
\right) _{\alpha \beta }B^{\alpha \beta }\right) \eta _{b}^{\mu \nu \rho
\lambda }  \notag \\
&&+\frac{1}{4}\left( P\left( U_{1}\right) \right) _{\mu \nu }\left( \bar{\xi}%
\gamma ^{\mu \nu \rho }\gamma _{5}\xi A_{\rho }-2\bar{\xi}\gamma ^{\mu \nu
\rho }\gamma _{5}\psi _{\rho }\eta \right)  \notag \\
&&+\frac{dU_{1}}{d\varphi }H_{\mu }^{\ast }\left( \frac{1}{2}\bar{\xi}\gamma
^{\mu \nu \rho }\gamma _{5}\xi B_{\nu \rho }^{\ast }+\mathrm{i}\psi ^{\ast
\mu }\gamma _{5}\xi \eta \right) +\frac{1}{6}U_{1}\bar{\xi}\gamma ^{\mu \nu
\rho }\gamma _{5}\xi \eta _{\mu \nu \rho }^{\ast }  \notag \\
&&-\mathrm{i}U_{1}\xi ^{\ast }\gamma _{5}\xi \eta +\frac{\mathrm{i}}{4}%
\left( P\left( U_{2}\right) \right) _{\mu \nu }\bar{\xi}\gamma ^{\mu \nu
}\gamma _{5}\xi -\frac{1}{4}\left( P\left( U_{3}\right) \right) _{\mu \nu }%
\bar{\xi}\gamma ^{\mu \nu }\xi  \notag \\
&&+\left( P\left( W\right) \right) ^{\mu \nu \rho }\left( -\eta C_{\mu \nu
\rho }+4A^{\lambda }C_{\mu \nu \rho \lambda }\right) -\varepsilon _{\mu \nu
\rho \lambda }\left( P\left( M\right) \right) _{\alpha \beta \gamma }\eta
^{\alpha \beta \gamma }\eta ^{\mu \nu \rho \lambda }  \notag \\
&&+2\left( 6\left( P\left( W\right) \right) ^{\mu \nu }B^{\ast \rho \lambda
}+4\left( P\left( W\right) \right) ^{\mu }\eta ^{\ast \nu \rho \lambda
}+W\eta ^{\ast \mu \nu \rho \lambda }\right) C_{\mu \nu \rho \lambda }
\notag \\
&&+\frac{1}{12}\left( P\left( U_{1}\right) \right) _{\mu \nu \rho }\bar{\xi}%
\gamma ^{\mu \nu \rho }\gamma _{5}\xi \eta +\left( P\left( W\right) \right)
^{\mu \nu \rho \lambda }\eta C_{\mu \nu \rho \lambda }  \notag \\
&&\left. +\tfrac{1}{2}\varepsilon _{\mu \nu \rho \lambda }\left( P\left(
M\right) \right) ^{\mu \nu \rho \lambda }\eta _{\alpha \beta \gamma \delta
}\eta ^{\alpha \beta \gamma \delta }\right] .  \label{bfa44}
\end{eqnarray}%
It is important to notice that it complies with all the working hypotheses,
including the derivative order assumption, and is parameterized by five
smooth, real functions of the undifferentiated scalar field, namely $W$, $M$%
, and $\left( U_{i}\right) _{i=1,2,3}$, which are otherwise arbitrary. The
functional $S_{1}$ is by construction a $s$-cocycle of ghost number zero,
such that $\bar{S}+\lambda S_{1}$ is solution to the master equation (\ref%
{bfa2.3}) to order one in $\lambda $. We will see in the next section that
the consistency of $S_{1}$ at order two in the coupling constant will
restrict these five functions of $\varphi $ to satisfy several equations.

\subsection{Higher-order deformations\label{higher}}

Next, we investigate the equations that control the higher-order
deformations. The second-order deformation is governed by equation (\ref%
{bfa2.6}). Making use of (\ref{bfa44}), the second term in the left
hand-side of (\ref{bfa2.6}) takes the concrete form
\begin{eqnarray}
&&\frac{1}{2}\left( S_{1},S_{1}\right) =\int d^{4}x\left[ \varepsilon _{\mu
\nu \rho \lambda }\sum\limits_{a=0}^{4}\left( T_{a}^{\mu \nu \rho \lambda }%
\frac{d^{a}X}{d\varphi ^{a}}+U_{a}^{\mu \nu \rho \lambda }\frac{d^{a}Y}{%
d\varphi ^{a}}\right) \right.  \notag \\
&&+\sum\limits_{a=0}^{2}\left( Q_{a}^{(1)}\frac{d^{a}Z_{1}}{d\varphi ^{a}}%
+Q_{a}^{(2)}\frac{d^{a}Z_{2}}{d\varphi ^{a}}\right)
+\sum\limits_{a=0}^{1}Q_{a}^{(3)}\frac{d^{a}Z_{3}}{d\varphi ^{a}}  \notag \\
&&\left. +\sum\limits_{a=0}^{3}Q_{a}^{(4)}\frac{d^{a}Z_{4}}{d\varphi ^{a}}%
\right]  \label{bfa45}
\end{eqnarray}%
where we used the notations
\begin{equation}
T_{0}^{\mu \nu \rho \lambda }=4A^{\ast \mu }C^{\nu \rho \lambda }+B^{\mu \nu
}C^{\rho \lambda }+H^{\mu }\eta ^{\nu \rho \lambda }-2\eta ^{\ast }C^{\mu
\nu \rho \lambda }-\varphi ^{\ast }\eta ^{\mu \nu \rho \lambda },
\label{ide1}
\end{equation}%
\begin{eqnarray}
&&T_{1}^{\mu \nu \rho \lambda }=\left( H_{\alpha }^{\ast }H^{\alpha
}+C_{\alpha \beta }^{\ast }C^{\alpha \beta }+C_{\alpha \beta \gamma }^{\ast
}C^{\alpha \beta \gamma }+C_{\alpha \beta \gamma \delta }^{\ast }C^{\alpha
\beta \gamma \delta }\right) \eta ^{\mu \nu \rho \lambda }  \notag \\
&&+\left( 2H_{\alpha }^{\ast }A^{\ast \alpha }+C_{\alpha \beta }^{\ast
}B^{\alpha \beta }-C_{\alpha \beta \gamma }^{\ast }\eta ^{\alpha \beta
\gamma }\right) C^{\mu \nu \rho \lambda }  \notag \\
&&+\left( 3C_{\alpha \beta }^{\ast }C^{\alpha \beta \mu }-2H_{\alpha }^{\ast
}C^{\alpha \mu }\right) \eta ^{\nu \rho \lambda }+3H_{\alpha }^{\ast
}C^{\alpha \mu \nu }B^{\rho \lambda },  \label{ide2}
\end{eqnarray}%
\begin{eqnarray}
&&T_{2}^{\mu \nu \rho \lambda }=H_{\alpha }^{\ast }\left( \left( H_{\beta
}^{\ast }B^{\alpha \beta }-3C_{\beta \gamma }^{\ast }\eta ^{\alpha \beta
\gamma }\right) C^{\mu \nu \rho \lambda }+3H_{\beta }^{\ast }C^{\alpha \beta
\mu }\eta ^{\nu \rho \lambda }\right)  \notag \\
&&+\left( \left( 4H_{\alpha }^{\ast }C_{\beta \gamma \delta }^{\ast
}+3C_{\alpha \beta }^{\ast }C_{\gamma \delta }^{\ast }\right) C^{\alpha
\beta \gamma \delta }\right.  \notag \\
&&\left. +H_{\alpha }^{\ast }\left( 3C_{\beta \gamma }^{\ast }C^{\alpha
\beta \gamma }+H_{\beta }^{\ast }C^{\alpha \beta }\right) \right) \eta ^{\mu
\nu \rho \lambda },  \label{ide3}
\end{eqnarray}%
\begin{equation}
T_{3}^{\mu \nu \rho \lambda }=H_{\alpha }^{\ast }H_{\beta }^{\ast }\left(
\left( H_{\gamma }^{\ast }C^{\alpha \beta \gamma }+3C_{\gamma \delta }^{\ast
}C^{\alpha \beta \gamma \delta }\right) \eta ^{\mu \nu \rho \lambda
}-H_{\gamma }^{\ast }\eta ^{\alpha \beta \gamma }C^{\mu \nu \rho \lambda
}\right) ,  \label{ide4}
\end{equation}%
\begin{equation}
T_{4}^{\mu \nu \rho \lambda }=H_{\alpha }^{\ast }H_{\beta }^{\ast }H_{\gamma
}^{\ast }H_{\delta }^{\ast }C^{\alpha \beta \gamma \delta }\eta ^{\mu \nu
\rho \lambda },  \label{ide5}
\end{equation}%
\begin{eqnarray}
&&U_{0}^{\mu \nu \rho \lambda }=\left( \frac{1}{2}\eta _{\alpha \beta \gamma
\delta }^{\ast }\eta ^{\alpha \beta \gamma \delta }+\eta _{\alpha \beta
\gamma }^{\ast }\eta ^{\alpha \beta \gamma }+B_{\alpha \beta }^{\ast
}B^{\alpha \beta }-6A_{\alpha }^{\ast }A^{\alpha }\right) \eta ^{\mu \nu
\rho \lambda }  \notag \\
&&+\left( A^{\ast \mu }\eta +\frac{3}{2}B_{\alpha \beta }^{\ast }\eta
^{\alpha \beta \mu }-A_{\alpha }B^{\alpha \mu }\right) \eta ^{\nu \rho
\lambda }+\frac{1}{2}B^{\mu \nu }B^{\rho \lambda }\eta ,  \label{ide6}
\end{eqnarray}%
\begin{eqnarray}
&&U_{1}^{\mu \nu \rho \lambda }=\left( \frac{1}{4}\eta C^{\ast \mu \nu \rho
\lambda }-A^{\mu }C^{\ast \nu \rho \lambda }+3B^{\ast \mu \nu }C^{\ast \rho
\lambda }-2\eta ^{\ast \mu \nu \rho }H^{\ast \lambda }\right) \eta _{\alpha
\beta \gamma \delta }\eta ^{\alpha \beta \gamma \delta }  \notag \\
&&+\left( \left( \frac{1}{2}C_{\alpha \beta \gamma }^{\ast }\eta +\frac{3}{2}%
C_{\alpha \beta }^{\ast }A_{\gamma }-3B_{\alpha \beta }^{\ast }H_{\gamma
}^{\ast }\right) \eta ^{\alpha \beta \gamma }+\left( \frac{1}{2}C_{\alpha
\beta }^{\ast }B^{\alpha \beta }+A^{\ast \alpha }H_{\alpha }^{\ast }\right)
\eta \right.  \notag \\
&&\left. +H_{\alpha }^{\ast }A_{\beta }B^{\alpha \beta }\right) \eta ^{\mu
\nu \rho \lambda }+\left( \frac{3}{2}\left( \frac{1}{2}C_{\alpha \beta
}^{\ast }\eta +H_{\alpha }^{\ast }A_{\beta }\right) \eta ^{\alpha \beta \mu
}-H_{\alpha }^{\ast }B^{\alpha \mu }\eta \right) \eta ^{\nu \rho \lambda },
\label{ide7}
\end{eqnarray}%
\begin{eqnarray}
&&U_{2}^{\mu \nu \rho \lambda }=H_{\alpha }^{\ast }\left( \frac{3}{2}\left(
C_{\beta \gamma }^{\ast }\eta +H_{\beta }^{\ast }A_{\gamma }\right) \eta
^{\alpha \beta \gamma }+H_{\beta }^{\ast }B^{\alpha \beta }\eta \right) \eta
^{\mu \nu \rho \lambda }  \notag \\
&&+\frac{3}{4}H_{\alpha }^{\ast }H_{\beta }^{\ast }\eta \eta ^{\alpha \beta
\mu }\eta ^{\nu \lambda \rho }+\left( H^{\ast \mu }\left( C^{\ast \nu \rho
\lambda }\eta +3H^{\ast \nu }B^{\ast \rho \lambda }\right) \right.  \notag \\
&&\left. +3\left( \frac{1}{4}C^{\ast \mu \nu }\eta +H^{\ast \mu }A^{\nu
}\right) C^{\ast \rho \lambda }\right) \eta _{\alpha \beta \gamma \delta
}\eta ^{\alpha \beta \gamma \delta },  \label{ide8}
\end{eqnarray}%
\begin{eqnarray}
&&U_{3}^{\mu \nu \rho \lambda }=\frac{1}{2}H^{\ast \mu }H^{\ast \nu }\left(
3C^{\ast \rho \lambda }\eta +2H^{\ast \rho }A^{\lambda }\right) \eta
_{\alpha \beta \gamma \delta }\eta ^{\alpha \beta \gamma \delta }  \notag \\
&&+\frac{1}{2}H_{\alpha }^{\ast }H_{\beta }^{\ast }H_{\gamma }^{\ast }\eta
\eta ^{\alpha \beta \gamma }\eta ^{\mu \nu \rho \lambda },  \label{ide9}
\end{eqnarray}%
\begin{equation}
U_{4}^{\mu \nu \rho \lambda }=\frac{1}{2}\eta H^{\ast \mu }H^{\ast \nu
}H^{\ast \rho }H^{\ast \lambda }\eta _{\alpha \beta \gamma \delta }\eta
^{\alpha \beta \gamma \delta },  \label{ide10}
\end{equation}%
\begin{equation}
Q_{0}^{(1)}=-\frac{1}{2}\left( \psi ^{\ast \mu }\gamma _{\mu }\gamma _{5}\xi
+\mathrm{i}\bar{\psi}_{\mu }\gamma ^{\mu \nu }\gamma _{5}\psi _{\nu }\right)
\eta +\frac{\mathrm{i}}{2}\bar{\xi}\gamma ^{\mu \nu }\gamma _{5}\left( 2\psi
_{\nu }A_{\mu }+\xi B_{\mu \nu }^{\ast }\right) ,  \label{q1}
\end{equation}%
\begin{equation}
Q_{1}^{(1)}=\frac{\mathrm{i}}{2}H_{\mu }^{\ast }\bar{\xi}\gamma ^{\mu \nu
}\gamma _{5}\left( \xi A_{\nu }-2\psi _{\nu }\eta \right) +\frac{\mathrm{i}}{%
4}C_{\mu \nu }^{\ast }\bar{\xi}\gamma ^{\mu \nu }\gamma _{5}\xi \eta ,
\label{q2}
\end{equation}%
\begin{equation}
Q_{2}^{(1)}=\frac{\mathrm{i}}{4}H_{\mu }^{\ast }H_{\nu }^{\ast }\bar{\xi}%
\gamma ^{\mu \nu }\gamma _{5}\xi \eta ,  \label{q3}
\end{equation}%
\begin{equation}
Q_{0}^{(2)}=-\frac{\mathrm{i}}{2}\left( \psi ^{\ast \mu }\gamma _{\mu }\xi +%
\mathrm{i}\bar{\psi}_{\mu }\gamma ^{\mu \nu }\psi _{\nu }\right) \eta -\frac{%
1}{2}\bar{\xi}\gamma ^{\mu \nu }\left( 2\psi _{\nu }A_{\mu }+\xi B_{\mu \nu
}^{\ast }\right) ,  \label{q4}
\end{equation}%
\begin{equation}
Q_{1}^{(2)}=-\frac{1}{2}H_{\mu }^{\ast }\bar{\xi}\gamma ^{\mu \nu }\left(
\xi A_{\nu }-2\psi _{\nu }\eta \right) -\frac{1}{4}C_{\mu \nu }^{\ast }\bar{%
\xi}\gamma ^{\mu \nu }\xi \eta ,  \label{q5}
\end{equation}%
\begin{equation}
Q_{2}^{(2)}=-\frac{1}{4}H_{\mu }^{\ast }H_{\nu }^{\ast }\bar{\xi}\gamma
^{\mu \nu }\xi \eta ,  \label{q6}
\end{equation}%
\begin{equation}
Q_{0}^{(3)}=\frac{3\mathrm{i}}{2}\bar{\psi}_{\mu }\gamma ^{\mu }\xi ,
\label{q7}
\end{equation}%
\begin{equation}
Q_{1}^{(3)}=-\frac{3\mathrm{i}}{4}H_{\mu }^{\ast }\bar{\xi}\gamma ^{\mu }\xi
,  \label{q8}
\end{equation}%
\begin{eqnarray}
Q_{0}^{(4)} &=&2\mathrm{i}\left( \xi ^{\ast }\gamma _{5}\xi +\psi ^{\ast \mu
}\gamma _{5}\psi _{\mu }\right) \varepsilon _{\alpha \beta \gamma \delta
}\eta ^{\alpha \beta \gamma \delta }+2\mathrm{i}A_{\mu }^{\ast }\bar{\xi}%
\gamma ^{\mu }\xi  \notag \\
&&-\left( 2\mathrm{i}\psi ^{\ast \rho }\gamma _{5}\xi +\bar{\psi}_{\mu
}\gamma ^{\mu \nu \rho }\gamma _{5}\psi _{\nu }\right) \varepsilon _{\rho
\beta \gamma \delta }\eta ^{\beta \gamma \delta }  \notag \\
&&+\bar{\xi}\gamma ^{\mu \nu \rho }\gamma _{5}\psi _{\rho }\varepsilon _{\mu
\nu \alpha \beta }B^{\alpha \beta },  \label{q9}
\end{eqnarray}%
\begin{eqnarray}
Q_{1}^{(4)} &=&\left[ -\frac{1}{6}C_{\mu \nu \rho }^{\ast }\bar{\xi}\gamma
^{\mu \nu \rho }\gamma _{5}\xi +C_{\mu \nu }^{\ast }\bar{\xi}\gamma ^{\mu
\nu \rho }\gamma _{5}\psi _{\rho }+H_{\mu }^{\ast }\left( \bar{\psi}_{\nu
}\gamma ^{\mu \nu \rho }\gamma _{5}\psi _{\rho }\right. \right.  \notag \\
&&\left. \left. -2\mathrm{i}\psi ^{\ast \mu }\gamma _{5}\xi \right) \right]
\varepsilon _{\alpha \beta \gamma \delta }\eta ^{\alpha \beta \gamma \delta
}+\left( 2H_{\mu }^{\ast }\bar{\xi}\gamma ^{\mu \nu \rho }\gamma _{5}\psi
_{\nu }-\frac{1}{2}C_{\mu \nu }^{\ast }\bar{\xi}\gamma ^{\mu \nu \rho
}\gamma _{5}\xi \right) \times  \notag \\
&&\times \varepsilon _{\rho \beta \gamma \delta }\eta ^{\beta \gamma \delta
}-2\mathrm{i}H_{\mu }^{\ast }\bar{\xi}\gamma _{\nu }\xi B^{\mu \nu },
\label{q10}
\end{eqnarray}%
\begin{eqnarray}
Q_{2}^{(4)} &=&H_{\mu }^{\ast }H_{\nu }^{\ast }\left( \bar{\xi}\gamma ^{\mu
\nu \rho }\gamma _{5}\psi _{\rho }\varepsilon _{\alpha \beta \gamma \delta
}\eta ^{\alpha \beta \gamma \delta }-\frac{1}{2}\bar{\xi}\gamma ^{\mu \nu
\rho }\gamma _{5}\xi \varepsilon _{\rho \beta \gamma \delta }\eta ^{\beta
\gamma \delta }\right)  \notag \\
&&-\frac{1}{2}H_{\mu }^{\ast }C_{\nu \rho }^{\ast }\bar{\xi}\gamma ^{\mu \nu
\rho }\gamma _{5}\xi \varepsilon _{\alpha \beta \gamma \delta }\eta ^{\alpha
\beta \gamma \delta },  \label{q11}
\end{eqnarray}%
\begin{equation}
Q_{3}^{(4)}=-\frac{1}{6}H_{\mu }^{\ast }H_{\nu }^{\ast }H_{\rho }^{\ast }%
\bar{\xi}\gamma ^{\mu \nu \rho }\gamma _{5}\xi \varepsilon _{\alpha \beta
\gamma \delta }\eta ^{\alpha \beta \gamma \delta }  \label{q12}
\end{equation}%
together with
\begin{eqnarray}
X\left( \varphi \right) &=&W\left( \varphi \right) M\left( \varphi \right)
,\qquad Y\left( \varphi \right) =W\left( \varphi \right) \frac{dM\left(
\varphi \right) }{d\varphi },  \label{ide11} \\
Z_{1}\left( \varphi \right) &=&W\left( \varphi \right) \frac{dU_{2}\left(
\varphi \right) }{d\varphi }+2U_{1}\left( \varphi \right) U_{3}\left(
\varphi \right) ,  \label{ide11a} \\
Z_{2}\left( \varphi \right) &=&W\left( \varphi \right) \frac{dU_{3}\left(
\varphi \right) }{d\varphi }-2U_{1}\left( \varphi \right) U_{2}\left(
\varphi \right) ,  \label{ide11b} \\
Z_{3}\left( \varphi \right) &=&\left( U_{2}\left( \varphi \right) \right)
^{2}+\left( U_{3}\left( \varphi \right) \right) ^{2},\qquad Z_{4}\left(
\varphi \right) =U_{1}\left( \varphi \right) M\left( \varphi \right) .
\label{ide11c}
\end{eqnarray}%
It is clear that none of the terms involving any of the functions $X$, $Y$, $%
Z_{1}$, $Z_{2}$, $Z_{3}$, $Z_{4}$ or their derivatives with respect to the
scalar field can be written as it is required by equation (\ref{bfa2.6}),
namely, like the $s$-variation of some local functional, and therefore they
must vanish. In other words, the consistency of the first-order deformation
at order two in the coupling constant, namely the existence of a local $%
S_{2} $ as solution to equation (\ref{bfa2.6}), restricts the five functions
of the undifferentiated scalar field that parameterize $S_{1}$ to satisfy
the equations%
\begin{equation}
X\left( \varphi \right) =0,\qquad Y\left( \varphi \right) =0,\qquad
Z_{i}\left( \varphi \right) =0,\qquad i=\overline{1,4}.  \label{consisteqs}
\end{equation}%
Due to (\ref{consisteqs}), from (\ref{bfa45}) it is obvious that $\left(
S_{1},S_{1}\right) =0$, and thus we can take
\begin{equation}
S_{2}=0  \label{s2int}
\end{equation}%
as solution to (\ref{bfa2.6}). By relying on (\ref{s2int}), it is easy to
show that one can safely put%
\begin{equation}
S_{k}=0,\qquad k>2.  \label{skint}
\end{equation}%
Collecting formulas (\ref{s2int}) and (\ref{skint}), we can state that the
complete deformed solution to the master equation for the model under study,
which complies with all the working hypotheses and is consistent to all
orders in the coupling constant, stops at order one in the coupling constant
and reads as%
\begin{equation}
S=\bar{S}+gS_{1},  \label{defsolmast}
\end{equation}%
where $\bar{S}$ is given in (\ref{bfa14}) and $S_{1}$ is expressed by (\ref%
{bfa44}). The fully deformed solution to the master equations comprises five
types of smooth functions that depend only on the undifferentiated scalar
field: $W$, $M$, and $\left( U_{i}\right) _{i=1,2,3}$. They are no longer
arbitrary, but satisfy equations (\ref{consisteqs}), imposed by the
consistency conditions.

There appear two complementary solutions to the above equations,
\begin{eqnarray}
U_{2}\left( \varphi \right) &=&U_{3}\left( \varphi \right) =M\left( \varphi
\right) =0,  \label{condcaseI} \\
U_{1}\left( \varphi \right) &=&U_{2}\left( \varphi \right) =U_{3}\left(
\varphi \right) =W\left( \varphi \right) =0,  \label{condcaseII}
\end{eqnarray}%
which will be analyzed separately below. They yield two types of deformed
models, whose Lagrangian formulation will be discussed in the following.

\section{Lagrangian formulation of the coupled model\label{lag}}

The first case corresponds to solution (\ref{condcaseI}) of equations (\ref%
{consisteqs}), such that the deformed solution to the master equation is
parameterized by the smooth, but otherwise arbitrary, functions $W\left(
\varphi \right) $ and $U_{1}\left( \varphi \right) $. It is given by (\ref%
{defsolmast}), with $S_{1}$ from (\ref{bfa44}) particularized to $%
U_{2}=U_{3}=M=0$, and reads as
\begin{eqnarray}
&&S^{\left( \mathrm{I}\right) }=\int d^{4}x\left\{ H^{\mu }\left( \partial
_{\mu }\varphi -\lambda WA_{\mu }\right) -\frac{\mathrm{i}}{2}\left( \bar{%
\psi}_{\mu }\gamma ^{\mu \nu \rho }\left( \partial _{\nu }+\mathrm{i}\gamma
_{5}A_{\nu }U_{1}\right) \psi _{\rho }\right) \right.  \notag \\
&&+\frac{1}{2}B^{\mu \nu }\partial _{\lbrack \mu }A_{\nu ]}+A_{\mu }^{\ast
}\partial ^{\mu }\eta +\lambda W\varphi ^{\ast }\eta +H^{\ast \mu }\left[
\lambda \frac{dW}{d\varphi }\left( -\eta H_{\mu }+2A^{\nu }C_{\mu \nu
}\right) \right.  \notag \\
&&\left. +2\partial ^{\nu }C_{\mu \nu }-\frac{\lambda }{2}\frac{dU_{1}}{%
d\varphi }H_{\mu }^{\ast }\left( 2\bar{\xi}\gamma ^{\mu \nu \rho }\gamma
_{5}\psi _{\nu }A_{\rho }+\bar{\psi}_{\nu }\gamma ^{\mu \nu \rho }\gamma
_{5}\psi _{\rho }\eta \right) \right]  \notag \\
&&+B_{\mu \nu }^{\ast }\left( -3\partial _{\rho }\eta ^{\mu \nu \rho
}+2\lambda WC^{\mu \nu }+\lambda U_{1}\bar{\xi}\gamma ^{\mu \nu \rho }\gamma
_{5}\psi _{\rho }\right) +\psi ^{\ast \mu }\left[ \partial _{\mu }\xi \right.
\notag \\
&&\left. -\mathrm{i}\lambda U_{1}\left( \gamma _{5}\psi _{\mu }\eta -\gamma
_{5}\xi A_{\mu }\right) \right] +\lambda H^{\ast \mu }H^{\ast \nu }\left[
\frac{1}{4}\frac{d^{2}U_{1}}{d\varphi ^{2}}\bar{\xi}\gamma _{\mu \nu \rho
}\gamma _{5}\left( \xi A^{\rho }-2\psi ^{\rho }\eta \right) \right.  \notag
\\
&&\left. +\frac{d^{2}W}{d\varphi ^{2}}\eta C_{\mu \nu }\right] +\lambda
\frac{dU_{1}}{d\varphi }H_{\mu }^{\ast }\left( \frac{1}{2}\bar{\xi}\gamma
^{\mu \nu \rho }\gamma _{5}\xi B_{\nu \rho }^{\ast }+\mathrm{i}\psi ^{\ast
\mu }\gamma _{5}\xi \eta \right)  \notag \\
&&-3C^{\ast \mu \nu }\left( \partial ^{\rho }C_{\mu \nu \rho }+\lambda \frac{%
dW}{d\varphi }A^{\rho }C_{\mu \nu \rho }\right) +\eta _{\mu \nu \rho }^{\ast
}\left( 4\partial _{\lambda }\eta ^{\mu \nu \rho \lambda }-2\lambda WC^{\mu
\nu \rho }\right)  \notag \\
&&+\lambda C^{\ast \mu \nu }\left[ \frac{dW}{d\varphi }\eta C_{\mu \nu }+%
\frac{1}{4}\frac{dU_{1}}{d\varphi }\bar{\xi}\gamma _{\mu \nu \rho }\gamma
_{5}\left( \xi A^{\rho }-2\psi ^{\rho }\eta \right) \right] -\lambda
U_{1}\left( \mathrm{i}\xi ^{\ast }\gamma _{5}\xi \eta \right.  \notag \\
&&\left. -\frac{1}{6}\eta _{\mu \nu \rho }^{\ast }\bar{\xi}\gamma ^{\mu \nu
\rho }\gamma _{5}\xi \right) -3\lambda H^{\ast \mu }\left( \frac{d^{2}W}{%
d\varphi ^{2}}H^{\ast \nu }A^{\rho }+2\frac{dW}{d\varphi }B^{\ast \nu \rho
}\right) C_{\mu \nu \rho }  \notag \\
&&+4C^{\ast \mu \nu \rho }\left( \partial ^{\lambda }C_{\mu \nu \rho \lambda
}+\lambda \frac{dW}{d\varphi }A^{\lambda }\right) +2\lambda W\eta ^{\ast \mu
\nu \rho \lambda }C_{\mu \nu \rho \lambda }  \notag \\
&&-\lambda C^{\ast \mu \nu \rho }\left( \frac{dW}{d\varphi }\eta C_{\mu \nu
\rho }-\frac{1}{12}\frac{dU_{1}}{d\varphi }\bar{\xi}\gamma ^{\mu \nu \rho
}\gamma _{5}\xi \eta \right)  \notag \\
&&+4\lambda \left[ 3\frac{d^{2}W}{d\varphi ^{2}}H^{\ast \mu }C^{\ast \nu
\rho }A^{\lambda }+\frac{dW}{d\varphi }\left( 3B^{\ast \mu \nu }C^{\ast \rho
\lambda }+2H^{\ast \mu }\eta ^{\ast \nu \rho \lambda }\right) \right] C_{\mu
\nu \rho \lambda }  \notag \\
&&+4\lambda H^{\ast \mu }H^{\ast \nu }\left( 3\frac{d^{2}W}{d\varphi ^{2}}%
B^{\ast \rho \lambda }+\frac{d^{3}W}{d\varphi ^{3}}H^{\ast \rho }A^{\lambda
}\right) C_{\mu \nu \rho \lambda }  \notag \\
&&-\lambda H^{\ast \mu }\left[ 3C^{\ast \nu \rho }\left( \frac{d^{2}W}{%
d\varphi ^{2}}C_{\mu \nu \rho }-\frac{1}{12}\frac{d^{2}U_{1}}{d\varphi ^{2}}%
\bar{\xi}\gamma _{\mu \nu \rho }\gamma _{5}\xi \right) \right.  \notag \\
&&\left. +H^{\ast \nu }H^{\ast \rho }\left( \frac{d^{3}W}{d\varphi ^{3}}%
C_{\mu \nu \rho }-\frac{1}{12}\frac{d^{3}U_{1}}{d\varphi ^{3}}\bar{\xi}%
\gamma _{\mu \nu \rho }\gamma _{5}\xi \right) \right] \eta  \notag \\
&&+\lambda \left[ \frac{dW}{d\varphi }C^{\ast \mu \nu \rho \lambda }+\frac{%
d^{2}W}{d\varphi ^{2}}\left( H^{\ast \lbrack \mu }C^{\ast \nu \rho \lambda
]}+C^{\ast \lbrack \mu \nu }C^{\ast \rho \lambda ]}\right) \right.  \notag \\
&&\left. \left. +\frac{d^{3}W}{d\varphi ^{3}}H^{\ast \lbrack \mu }H^{\ast
\nu }C^{\ast \rho \lambda ]}+\frac{d^{4}W}{d\varphi ^{4}}H^{\ast \mu
}H^{\ast \nu }H^{\ast \rho }H^{\ast \lambda }\right] \eta C_{\mu \nu \rho
\lambda }\right\} .  \label{sdef1}
\end{eqnarray}

By virtue of the discussion from the end of Section \ref{free} on the
significance of terms with various antighost numbers from the solution to
the master equation, at this stage we can extract all the information on the
gauge structure of the coupled model. From the antifield-independent piece
in (\ref{sdef1}) we read that the overall Lagrangian action of the
interacting gauge theory has the expression
\begin{eqnarray}
&&\tilde{S}^{\left( \mathrm{I}\right) }[A^{\mu },H^{\mu },\varphi ,B^{\mu
\nu },\psi _{\mu }]=\int d^{4}x\left[ H_{\mu }\left( \partial ^{\mu }\varphi
-\lambda W\left( \varphi \right) A^{\mu }\right) \right.  \notag \\
&&\left. +\frac{1}{2}B^{\mu \nu }\partial _{\lbrack \mu }A_{\nu ]}-\frac{%
\mathrm{i}}{2}\left( \bar{\psi}_{\mu }\gamma ^{\mu \nu \rho }\left( \partial
_{\nu }+\lambda \mathrm{i}\gamma _{5}A_{\nu }U_{1}\right) \psi _{\rho
}\right) \right] ,  \label{ac1}
\end{eqnarray}%
while from the components linear in the antighost number one antifields we
conclude that it is invariant under the gauge transformations
\begin{equation}
\bar{\delta}_{\epsilon }^{\left( \mathrm{I}\right) }\varphi =\lambda W\left(
\varphi \right) \epsilon ,  \label{1}
\end{equation}%
\begin{equation}
\bar{\delta}_{\epsilon }^{\left( \mathrm{I}\right) }H^{\mu }=2\bar{D}_{\nu
}\epsilon ^{\mu \nu }+\lambda \left[ \left( \frac{1}{2}\frac{dU_{1}}{%
d\varphi }\bar{\psi}_{\nu }\gamma ^{\mu \nu \rho }\gamma _{5}\psi _{\rho }-%
\frac{dW}{d\varphi }H^{\mu }\right) \epsilon +\frac{dU_{1}}{d\varphi }%
A_{\rho }\bar{\psi}_{\nu }\gamma ^{\mu \nu \rho }\gamma _{5}\chi \right] ,
\label{2}
\end{equation}%
\begin{equation}
\bar{\delta}_{\epsilon }^{\left( \mathrm{I}\right) }A^{\mu }=\partial ^{\mu
}\epsilon ,  \label{3}
\end{equation}%
\begin{equation}
\bar{\delta}_{\epsilon }^{\left( \mathrm{I}\right) }B^{\mu \nu }=-3\partial
_{\rho }\epsilon ^{\mu \nu \rho }+2\lambda W\left( \varphi \right) \epsilon
^{\mu \nu }-\lambda U_{1}\left( \varphi \right) \bar{\psi}_{\rho }\gamma
^{\mu \nu \rho }\gamma _{5}\chi ,  \label{4}
\end{equation}%
\begin{equation}
\bar{\delta}_{\epsilon }^{\left( \mathrm{I}\right) }\psi _{\mu }=\partial
_{\mu }\chi -\mathrm{i}\lambda U_{1}\left( \varphi \right) \left( \gamma
_{5}\psi _{\mu }\epsilon -\gamma _{5}\chi A_{\mu }\right) ,  \label{5}
\end{equation}%
where we employed the notation
\begin{equation}
\bar{D}_{\nu }=\partial _{\nu }+\lambda \frac{dW}{d\varphi }A_{\nu }.
\label{dercov}
\end{equation}%
We observe that (\ref{sdef1}) contains two kinds of pieces quadratic
in the ghosts of pure ghost number one: ones are linear in their
antifields, and the others are quadratic in the antifields of the
original fields, which indicates the open behavior of the deformed
gauge algebra. This is translated into the fact that the commutators
among the deformed gauge generators only close on-shell, where
on-shell means on the stationary surface of field equations for
action (\ref{ac1}). Also, we notice the presence of terms linear in
the ghosts with pure ghost number two and three in (\ref{sdef1}),
which shows that the gauge generators of the coupled model are also
second-order reducible, but some of the reducibility functions are
modified and, moreover, some of the reducibility relations only hold
on-shell. The remaining elements in (\ref{sdef1}) give us
information on the higher-order gauge structure of the interacting
model. All these ingredients of the Lagrangian gauge structure of
the deformed theory are listed in Appendix \ref{appendix}.

It is interesting to mention certain similarities between the model under
study and $N=1$, $D=4$ conformal SUGRA \cite{fratse}. First, there are
common fields in both theories, namely a gravitino $\psi _{\mu }$ with
undeformed gauge symmetries described by ordinary local $Q$-supersymmetry
transformation $\chi $ (see the latter formula from (\ref{bfa2.1b})) and a
vector field $A^{\mu }$ with initial $U\left( 1\right) $ gauge
transformation $\epsilon $ (see the first relation in (\ref{bfa2.1a})). It
is suggestive to make the notations%
\begin{equation}
\chi \equiv \epsilon _{Q},\qquad \epsilon \equiv \epsilon _{U},\qquad D_{\mu
}\equiv \partial _{\mu }+\mathrm{i}\lambda U_{1}\left( \varphi \right)
\gamma _{5}A_{\mu },  \label{covferm}
\end{equation}%
in terms of which the deformed gauge transformations of the Rarita-Schwinger
spinors, (\ref{5}), become%
\begin{equation}
\bar{\delta}_{\epsilon }^{\left( \mathrm{I}\right) }\psi _{\mu }=D_{\mu
}\epsilon _{Q}-\mathrm{i}\lambda U_{1}\left( \varphi \right) \gamma
_{5}\epsilon _{U}\psi _{\mu }.  \label{varpsinou}
\end{equation}%
Second, if we make the choices%
\begin{equation}
U_{1}\left( \varphi \right) =-\frac{3}{4},\qquad \lambda =1,  \label{partU1}
\end{equation}%
then the deformed gauge transformations of gravitini, (\ref{varpsinou}),
take the form%
\begin{equation}
\bar{\delta}_{\epsilon }^{\left( \mathrm{I}\right) }\psi _{\mu }=D_{\mu
}\epsilon _{Q}+\frac{3}{4}\mathrm{i}\gamma _{5}\epsilon _{U}\psi _{\mu },
\label{varpsiconf}
\end{equation}%
in terms of the covariant derivative%
\begin{equation}
D_{\mu }\equiv \partial _{\mu }-\frac{3}{4}\mathrm{i}\gamma _{5}A_{\mu }.
\label{covderconf}
\end{equation}%
We observe that (\ref{varpsiconf}) are nothing but the standard $N=1$, $D=4$
conformal SUGRA gauge transformations of the spin-$3/2$ field in the absence
of local dilatational $D$-transformation, special conformal $S$%
-supersymmetry, local spacetime translations and local Lorentz rotations,
i.e. solely in the presence of local $Q$-supersymmetry and $U\left( 1\right)
$-symmetry. Indeed, it coincides with formula (2.24) for $\psi _{\mu }$
given in \cite{fratse} if one sets $\epsilon _{D}=\epsilon _{S}=\epsilon
_{P}^{a}=\epsilon _{M}^{ab}=0$. By contrast to conformal SUGRA, where the
gauge transformation of the vector field $A_{\mu }$ gains $Q$-supersymmetric
contributions, here it remains $U\left( 1\right) $ also for the coupled
model (see formula (\ref{3})). This is mainly because the BF model under
discussion does not include a Maxwell Lagrangian with respect to $A_{\mu }$,
but the term $(1/2)B^{\mu \nu }\partial _{\lbrack \mu }A_{\nu ]}$ (see (\ref%
{ac1})). For this topological model it is precisely the two-form
$B^{\mu \nu }$ whose gauge transformations (\ref{4}) gain
$Q$-supersymmetric
contributions%
\begin{equation}
\bar{\delta}_{\epsilon }^{\left( \mathrm{I}\right) }B^{\mu \nu }=\mathrm{%
something}+\frac{3}{4}\bar{\psi}_{\rho }\gamma ^{\mu \nu \rho }\gamma
_{5}\epsilon _{Q}  \label{varBnou}
\end{equation}%
that compensate the gauge variation of the last term from the right-hand
side of (\ref{ac1}), namely, $\left( \mathrm{i}/2\right) \bar{\psi}_{\mu
}\gamma ^{\mu \nu \rho }D_{\nu }\psi _{\rho }$. In conclusion, we can state
that (\ref{5}) describes a certain generalization of gravitino gauge
transformations from conformal SUGRA in the sense that puts the standard $Q$%
-supersymmetry and $U\left( 1\right) $-symmetry parts in a `background'
potential $U_{1}\left( \varphi \right) $.

\subsection{Case II. No couplings to the Rarita-Schwinger fields \label%
{caseII}}

The second solution to equations (\ref{consisteqs}) is (\ref{condcaseII}),
so the deformed solution of the master equation is uniquely parameterized in
this situation by the arbitrary function $M\left( \varphi \right) $ (of the
undifferentiated scalar field). This case is less interesting from the point
of view of interactions as the Rarita-Schwinger spinors are no longer
coupled to the BF fields. From (\ref{bfa45}) and the higher-order
deformation equations, (\ref{bfa2.7}), etc., it follows that we can safely
take all the deformations (of order two or higher) to vanish
\begin{equation}
S_{k}=0,\qquad k>1.  \label{SkII}
\end{equation}%
Therefore, the overall deformed solution to the master equation that is
consistent to all orders in the coupling constant is equal to the sum
between the free\ solution, (\ref{bfa14}), and the first-order deformation, (%
\ref{bfa44}), where we set (\ref{condcaseII}), $S^{\left( \mathrm{II}\right)
}=\bar{S}+\lambda S_{1}$, being expressed by
\begin{eqnarray}
&&S^{\left( \mathrm{II}\right) }=\int d^{4}x\left[ H_{\mu }\partial ^{\mu
}\varphi +\frac{1}{2}B^{\mu \nu }\left( \partial _{\lbrack \mu }A_{\nu
]}+\lambda \varepsilon _{\mu \nu \rho \lambda }MB^{\rho \lambda }\right)
+\right.  \notag \\
&&-\frac{\mathrm{i}}{2}\bar{\psi}_{\mu }\gamma ^{\mu \nu \rho }\partial
_{\nu }\psi _{\rho }+A^{\ast \mu }\left( \partial _{\mu }\eta -2\lambda
M\varepsilon _{\mu \alpha \beta \gamma }\eta ^{\alpha \beta \gamma }\right)
\notag \\
&&+2H_{\mu }^{\ast }\left( \partial _{\nu }C^{\mu \nu }+\lambda \frac{dM}{%
d\varphi }B^{\mu \alpha }\varepsilon _{\alpha \beta \gamma \delta }\eta
^{\beta \gamma \delta }\right) -3B_{\mu \nu }^{\ast }\partial _{\rho }\eta
^{\mu \nu \rho }  \notag \\
&&-3C_{\mu \nu }^{\ast }\partial _{\rho }C^{\mu \nu \rho }+4\eta _{\mu \nu
\rho }^{\ast }\partial _{\lambda }\eta ^{\mu \nu \rho \lambda }  \notag \\
&&+\lambda \left[ \left( \frac{dM}{d\varphi }C_{\rho \lambda }^{\ast }+\frac{%
d^{2}M}{d\varphi ^{2}}H_{\rho }^{\ast }H_{\lambda }^{\ast }\right) B^{\rho
\lambda }+2\left( \frac{dM}{d\varphi }H_{\mu }^{\ast }A^{\ast \mu }-M\eta
^{\ast }\right) \right] \varepsilon _{\alpha \beta \gamma \delta }\eta
^{\alpha \beta \gamma \delta }  \notag \\
&&-\frac{9}{4}\lambda \left( \frac{dM}{d\varphi }C_{\rho \lambda }^{\ast }+%
\frac{d^{2}M}{d\varphi ^{2}}H_{\rho }^{\ast }H_{\lambda }^{\ast }\right)
\varepsilon _{\alpha \beta \gamma \delta }\eta ^{\rho \alpha \beta }\eta
^{\lambda \gamma \delta }+4C_{\mu \nu \rho }^{\ast }\partial _{\lambda
}C^{\mu \nu \rho \lambda }  \notag \\
&&-\lambda \left( \frac{dM}{d\varphi }C_{\nu \rho \lambda }^{\ast }+\frac{%
d^{2}M}{d\varphi ^{2}}H_{\left[ \nu \right. }^{\ast }C_{\left. \rho \lambda %
\right] }^{\ast }+\frac{d^{3}M}{d\varphi ^{3}}H_{\nu }^{\ast }H_{\rho
}^{\ast }H_{\lambda }^{\ast }\right) \eta ^{\nu \rho \lambda }\varepsilon
_{\alpha \beta \gamma \delta }\eta ^{\alpha \beta \gamma \delta }  \notag \\
&&+\frac{\lambda }{2}\left( \frac{dM}{d\varphi }C_{\mu \nu \rho \lambda
}^{\ast }+\frac{d^{2}M}{d\varphi ^{2}}H_{\left[ \mu \right. }^{\ast
}C_{\left. \nu \rho \lambda \right] }^{\ast }+\frac{d^{2}M}{d\varphi ^{2}}C_{%
\left[ \mu \nu \right. }^{\ast }C_{\left. \rho \lambda \right] }^{\ast
}\right.  \notag \\
&&\left. \left. +\frac{d^{3}M}{d\varphi ^{3}}H_{\left[ \mu \right. }^{\ast
}H_{\nu }^{\ast }C_{\left. \rho \lambda \right] }^{\ast }+\frac{d^{4}M}{%
d\varphi ^{4}}H_{\mu }^{\ast }H_{\nu }^{\ast }H_{\rho }^{\ast }H_{\lambda
}^{\ast }\right) \eta ^{\mu \nu \rho \lambda }\varepsilon _{\alpha \beta
\gamma \delta }\eta ^{\alpha \beta \gamma \delta }\right] .  \label{sdef2}
\end{eqnarray}

Its antighost number zero part emphasizes the Lagrangian action of the
deformed theory
\begin{eqnarray}
&&\tilde{S}^{\left( \mathrm{II}\right) }[A^{\mu },H^{\mu },\varphi ,B^{\mu
\nu },y^{i}]=\int d^{4}x\left[ H_{\mu }\partial ^{\mu }\varphi -\frac{%
\mathrm{i}}{2}\bar{\psi}_{\mu }\gamma ^{\mu \nu \rho }\partial _{\nu }\psi
_{\rho }\right.  \notag \\
&&+\left. \frac{1}{2}B^{\mu \nu }\left( \partial _{\lbrack \mu }A_{\nu
]}+\lambda \varepsilon _{\mu \nu \rho \lambda }MB^{\rho \lambda }\right) %
\right] ,  \label{ac2}
\end{eqnarray}%
while the antighost number one components provide the gauge transformations
of action (\ref{ac2})
\begin{equation}
\bar{\delta}_{\epsilon }^{\left( \mathrm{II}\right) }A_{\mu }=\partial _{\mu
}\epsilon -2\lambda M\left( \varphi \right) \varepsilon _{\mu \alpha \beta
\gamma }\epsilon ^{\alpha \beta \gamma }\equiv (\tilde{Z}_{(A)\mu }^{\left(
\mathrm{II}\right) })\epsilon +(\tilde{Z}_{(A)\mu }^{\left( \mathrm{II}%
\right) })_{\alpha \beta \gamma }\epsilon ^{\alpha \beta \gamma },
\label{i27}
\end{equation}%
\begin{equation}
\bar{\delta}_{\epsilon }^{\left( \mathrm{II}\right) }H^{\mu }=2\left(
\partial _{\nu }\epsilon ^{\mu \nu }-\lambda \frac{dM}{d\varphi }B^{\mu
\alpha }\varepsilon _{\alpha \beta \gamma \delta }\epsilon ^{\beta \gamma
\delta }\right) \equiv (\tilde{Z}_{(H)}^{\left( \mathrm{II}\right) \mu
})_{\alpha \beta }\epsilon ^{\alpha \beta }+(\tilde{Z}_{(H)}^{\left( \mathrm{%
II}\right) \mu })_{\alpha \beta \gamma }\epsilon ^{\alpha \beta \gamma },
\label{i28}
\end{equation}%
\begin{equation}
\bar{\delta}_{\epsilon }^{\left( \mathrm{II}\right) }\varphi =0,\;\bar{\delta%
}_{\epsilon }^{\left( \mathrm{II}\right) }B^{\mu \nu }=-3\partial _{\rho
}\epsilon ^{\mu \nu \rho }\equiv (\tilde{Z}_{(B)}^{\left( \mathrm{II}\right)
\mu \nu })_{\alpha \beta \gamma }\epsilon ^{\alpha \beta \gamma },
\label{i29}
\end{equation}%
\begin{equation}
\bar{\delta}_{\epsilon }^{\left( \mathrm{II}\right) }\psi ^{A\mu }=\partial
^{\mu }\chi ^{A}\equiv (\tilde{Z}_{(\psi )}^{\left( \mathrm{II}\right) A\mu
})_{B}\chi ^{B}.  \label{i30}
\end{equation}%
We observe that in this case the Rarita-Schwinger fields remain uncoupled to
the BF field sector. From (\ref{i27})--(\ref{i30}), we notice that the
nonvanishing gauge generators are
\begin{equation}
(\tilde{Z}_{(A)\mu }^{\left( \mathrm{II}\right) })(x,x^{\prime })=(Z_{(A)\mu
})(x,x^{\prime })=\partial _{\mu }^{x}\delta ^{4}(x-x^{\prime }),
\label{i31}
\end{equation}%
\begin{equation}
(\tilde{Z}_{(A)\mu }^{\left( \mathrm{II}\right) })_{\alpha \beta \gamma
}(x,x^{\prime })=-2\lambda M\left( \varphi \left( x\right) \right)
\varepsilon _{\mu \alpha \beta \gamma }\delta ^{4}(x-x^{\prime }),
\label{i32}
\end{equation}%
\begin{equation}
(\tilde{Z}_{(H)}^{\left( \mathrm{II}\right) \mu })_{\alpha \beta
}(x,x^{\prime })=(Z_{(H)}^{\mu })_{\alpha \beta }(x,x^{\prime })=-\partial _{%
\left[ \alpha \right. }^{x}\delta _{\left. \beta \right] }^{\mu }\delta
^{4}(x-x^{\prime }),  \label{i33}
\end{equation}%
\begin{equation}
(\tilde{Z}_{(H)}^{\left( \mathrm{II}\right) \mu })_{\alpha \beta \gamma
}(x,x^{\prime })=-2\lambda \frac{dM}{d\varphi }\left( x\right) B^{\mu \nu
}\left( x\right) \varepsilon _{\nu \alpha \beta \gamma }\delta
^{4}(x-x^{\prime }),  \label{i34}
\end{equation}%
\begin{equation}
(\tilde{Z}_{(B)}^{\left( \mathrm{II}\right) \mu \nu })_{\alpha \beta \gamma
}(x,x^{\prime })=(Z_{(B)}^{\mu \nu })_{\alpha \beta \gamma }(x,x^{\prime })=-%
\frac{1}{2}\partial _{\left[ \alpha \right. }^{x}\delta _{\beta }^{\mu
}\delta _{\left. \gamma \right] }^{\nu }\delta ^{4}(x-x^{\prime }),
\label{i35}
\end{equation}%
\begin{equation}
(\tilde{Z}_{(\psi )}^{\left( \mathrm{II}\right) A\mu })_{B}=(Z_{(\psi
)}^{A\mu })_{B}=\delta _{B}^{A}\partial _{x}^{\mu }\delta ^{4}(x-x^{\prime
}).  \label{i36}
\end{equation}%
The deformed gauge algebra (corresponding to the generating set (\ref{i27}%
)--(\ref{i30})) is open, as can be seen from the elements of antighost
number two in (\ref{sdef2}) that are quadratic in the ghosts of pure ghost
number one. The only non-Abelian commutators among the new gauge
transformations are expressed by
\begin{eqnarray}
&&\left( \tilde{Z}_{(B)}^{\left( \mathrm{II}\right) \rho \lambda }\right)
_{\alpha \beta \gamma }\frac{\delta (\tilde{Z}_{(H)}^{\left( \mathrm{II}%
\right) \mu })_{\alpha ^{\prime }\beta ^{\prime }\gamma ^{\prime }}}{\delta
B^{\rho \lambda }}-\left( \tilde{Z}_{(B)}^{\left( \mathrm{II}\right) \rho
\lambda }\right) _{\alpha ^{\prime }\beta ^{\prime }\gamma ^{\prime }}\frac{%
\delta (\tilde{Z}_{(H)}^{\left( \mathrm{II}\right) \mu })_{\alpha \beta
\gamma }}{\delta B^{\rho \lambda }}=  \notag \\
&&-\frac{\lambda }{4}\frac{dM}{d\varphi }(\tilde{Z}_{(H)}^{\left( \mathrm{II}%
\right) \mu })_{\rho \lambda }\delta _{\left[ \alpha \right. }^{\left[ \rho
\right. }\left( \varepsilon _{\left. \beta \gamma \right] \left[ \alpha
^{\prime }\beta ^{\prime }\right. }\right) \delta _{\left. \gamma ^{\prime }%
\right] }^{\left. \lambda \right] }  \notag \\
&&+\lambda \frac{d^{2}M}{d\varphi ^{2}}\delta _{\left[ \alpha \right. }^{%
\left[ \mu \right. }\left( \varepsilon _{\left. \beta \gamma \right] \left[
\alpha ^{\prime }\beta ^{\prime }\right. }\right) \delta _{\left. \gamma
^{\prime }\right] }^{\left. \nu \right] }\frac{\delta \tilde{S}^{\left(
\mathrm{II}\right) }}{\delta H^{\nu }}.  \label{i37}
\end{eqnarray}%
Looking at the remaining terms of antighost number two from (\ref{sdef2}),
we can state that, besides the original first-order reducibility relations (%
\ref{bfa5a}), there appear some new ones
\begin{equation}
(Z_{(H)}^{\left( \mathrm{II}\right) \mu })_{\rho \lambda }\left( \tilde{Z}%
_{1}^{\left( \mathrm{II}\right) \rho \lambda }\right) _{\alpha \beta \gamma
\delta }=2\lambda \varepsilon _{\alpha \beta \gamma \delta }\left( \frac{%
d^{2}M}{d\varphi ^{2}}B^{\mu \nu }\frac{\delta \tilde{S}^{\left( \mathrm{II}%
\right) }}{\delta H^{\nu }}+\frac{dM}{d\varphi }\sigma ^{\mu \nu }\frac{%
\delta \tilde{S}^{\left( \mathrm{II}\right) }}{\delta A^{\nu }}\right) ,
\label{i38}
\end{equation}%
\begin{equation}
(Z_{(A)}^{\left( \mathrm{II}\right) \mu })\left( \tilde{Z}_{1}^{\left(
\mathrm{II}\right) }\right) _{\alpha \beta \gamma \delta }=-2\lambda
\varepsilon _{\alpha \beta \gamma \delta }\frac{dM}{d\varphi }\sigma ^{\mu
\nu }\frac{\delta \tilde{S}^{\left( \mathrm{II}\right) }}{\delta H^{\nu }},
\label{i39}
\end{equation}%
which only close on-shell (i.e. on the stationary surface of field equations
resulting from action (\ref{sdef2})), where the accompanying first-order
reducibility functions are of the form
\begin{equation}
\left( \tilde{Z}_{1}^{\left( \mathrm{II}\right) \rho \lambda }\right)
_{\alpha \beta \gamma \delta }(x,x^{\prime })=\lambda \frac{dM}{d\varphi }%
\left( x\right) B^{\rho \lambda }\left( x\right) \varepsilon _{\alpha \beta
\gamma \delta }\delta ^{4}(x-x^{\prime }),  \label{i40}
\end{equation}%
\begin{equation}
\left( \tilde{Z}_{1}^{\left( \mathrm{II}\right) }\right) _{\alpha \beta
\gamma \delta }(x,x^{\prime })=-2\lambda M\left( \varphi \left( x\right)
\right) \varepsilon _{\alpha \beta \gamma \delta }\delta ^{4}(x-x^{\prime }).
\label{i41}
\end{equation}%
The second-order reducibility is not modified (it continues to be expressed
by (\ref{bfa5b})). The presence (in (\ref{sdef2})) of elements with
antighost number strictly greater than two that are proportional with the
coupling constant $\lambda $ signifies a higher-order gauge tensor structure
of the deformed model, due to the open character of the gauge algebra, as
well as to the field dependence of the deformed reducibility functions. The
case (II) appears thus to be less important from the perspective of
constructing effective couplings among the BF fields and the spin-vector,
since no nontrivial interactions among them are allowed.

\section{Conclusion\label{conc}}

To conclude with, in this paper we have investigated the consistent
interactions that can be introduced between a topological BF theory and a
massless Rarita-Schwinger field. Starting with the BRST differential for the
free theory, we give the consistent first-order deformation of the solution
to the master equation, and obtain that it is parameterized by five kinds of
functions depending only on the undifferentiated scalar fields. Next, we
analyze the consistency of the first-order deformation, which imposes
certain restrictions on these functions. Based on these restrictions, we
show that we can take all the remaining deformations, of order two or
higher, to vanish. As a consequence of our procedure, we are led to two
classes of interacting gauge theories. Only one is interesting, being
endowed with deformed gauge transformations, a non-Abelian gauge algebra
that only closes on-shell, and on-shell, second-order reducibility
relations. This coupled model emphasizes some contributions to the gauge
transformations of the spin-$3/2$ field that generalize the local $Q$%
-supersymmetry and $U\left( 1\right) $ gauge symmetry contributions from $%
N=1 $, $D=4$ conformal SUGRA in the sense of multiplying them with an
arbitrary function that depends only on the scalar field from the BF
spectrum.

\section*{Acknowledgments}

The authors are partially supported by the type A grant 581/2007 with the
Romanian National Council for Academic Scientific Research (C.N.C.S.I.S.)
and the Romanian Ministry of Education, Research and Youth (M.E.C.T.) and by
the European Commission FP6 program MRTN-CT-2004-005104.

\appendix

\section{Proof of formula (\protect\ref{om159})\label{formula1}}

The proof of formula (\ref{om159}) requires the derivative order assumption.
We start from the general solution to (\ref{om1}), which is of the form (\ref%
{3.8}) for $I=1$, being thus written as%
\begin{equation}
\tilde{a}_{1}^{\mathrm{int}}=H_{\mu }^{\ast }N^{\mu }\xi +\varphi ^{\ast
}N\xi +A_{\mu }^{\ast }\bar{N}^{\mu }\xi +B_{\mu \nu }^{\ast }\bar{N}^{\mu
\nu }\xi +\psi ^{\ast \mu }M_{\mu }\xi +\bar{M}^{\mu }\bar{\psi}_{\mu
}^{\ast }\eta ,  \label{om11}
\end{equation}%
where $N^{\mu }$, $N$, $\bar{N}^{\mu }$, $\bar{N}^{\mu \nu }$, and $\bar{M}%
^{\mu }$ are some fermionic, gauge-invariant, spinor-like functions and $%
M_{\mu }$ is a matrix $4\times 4$ with spinor indices and bosonic,
gauge-invariant elements. In addition, $\bar{N}^{\mu \nu }$ is antisymmetric
in its Lorentz indices. Because $\tilde{c}_{0}$ produced by (\ref{om11}) via
(\ref{ai12}) must contain at most one spacetime derivative, the objects $%
N^{\mu }$, $N$, $\bar{N}^{\mu }$, $\bar{N}^{\mu \nu }$, $\bar{M}_{\mu }$,
and $M_{\mu }$ are also constrained to include at most one derivative. Given
their properties, exposed above, the objects $N^{\mu }$, $N$, $\bar{N}^{\mu }
$, $\bar{N}^{\mu \nu }$, and $\bar{M}^{\mu }$ can be represented like%
\begin{eqnarray}
N^{\mu } &=&\partial _{\lbrack \alpha }\bar{\psi}_{\beta ]}N^{\mu \mid
\alpha \beta },\qquad N=\partial _{\lbrack \alpha }\bar{\psi}_{\beta
]}N^{\alpha \beta },\qquad \bar{N}^{\mu }=\partial _{\lbrack \alpha }\bar{%
\psi}_{\beta ]}\bar{N}^{\mu \mid \alpha \beta },  \label{om12a} \\
\bar{N}^{\mu \nu } &=&\partial _{\lbrack \alpha }\bar{\psi}_{\beta ]}\bar{N}%
^{\mu \nu \mid \alpha \beta },\qquad \bar{M}^{\mu }=\partial _{\lbrack
\alpha }\bar{\psi}_{\beta ]}\bar{M}^{\mu \mid \alpha \beta },  \label{om12b}
\end{eqnarray}%
in terms of some $4\times 4$ matrices $N^{\mu \mid \alpha \beta }$, $%
N^{\alpha \beta }$, $\bar{N}^{\mu \mid \alpha \beta }$, $\bar{N}^{\mu \nu
\mid \alpha \beta }$, and $\bar{M}^{\mu \mid \alpha \beta }$ with
spinor-like indices, whose elements are smooth functions depending at most
on the undifferentiated scalar field. Since $M_{\mu }$ is $4\times 4$,
bosonic matrix with, whose elements are gauge-invariant spinor functions
with at most one derivative, it can be represented like%
\begin{equation}
M_{\mu }=M_{\mu }^{(0)}+M_{\mu \mid \alpha }^{(1)}\partial ^{\alpha }\varphi
+M_{\mu }^{(2)}\partial _{\alpha }H^{\alpha }+M_{\mu \mid \alpha \beta
}^{(3)}\partial ^{\lbrack \alpha }A^{\beta ]}+M_{\mu \mid \alpha
}^{(4)}\partial _{\beta }B^{\alpha \beta },  \label{om13}
\end{equation}%
where $M_{\mu }^{(0)}$, $M_{\mu \mid \alpha }^{(1)}$, $M_{\mu }^{(2)}$, $%
M_{\mu \mid \alpha \beta }^{(3)}$, and $M_{\mu \mid \alpha }^{(4)}$ are $%
4\times 4$ spinor matrices, whose elements may depend at most on the
undifferentiated scalar field $\varphi $, with $M_{\mu \mid \alpha \beta
}^{(3)}$ also antisymmetric in $\alpha $ and $\beta $. Let us show initially
that (\ref{om13}) can always be reduced to its first term via some trivial
redefinitions (of the matrices from (\ref{om12a})--(\ref{om12b})) performed
in $\tilde{a}_{1}^{\mathrm{int}}$. In view of this, we introduce (\ref{om12a}%
)--(\ref{om13}) in (\ref{om11}) and deduce
\begin{eqnarray}
\tilde{a}_{1}^{\mathrm{int}} &=&H_{\mu }^{\ast }\partial _{\lbrack \alpha }%
\bar{\psi}_{\beta ]}\left( N^{\mu \mid \alpha \beta }-\frac{\mathrm{i}}{2}%
\sigma ^{\mu \nu }\gamma ^{\alpha \beta \rho }M_{\rho \mid \nu
}^{(1)}\right) \xi   \notag \\
&&+\varphi ^{\ast }\partial _{\lbrack \alpha }\bar{\psi}_{\beta ]}\left(
N^{\alpha \beta }+\frac{\mathrm{i}}{2}\gamma ^{\alpha \beta \rho }M_{\rho
}^{(2)}\right) \xi   \notag \\
&&+A_{\mu }^{\ast }\partial _{\lbrack \alpha }\bar{\psi}_{\beta ]}\left(
\bar{N}^{\mu \mid \alpha \beta }-\frac{\mathrm{i}}{2}\sigma ^{\mu \nu
}\gamma ^{\alpha \beta \rho }M_{\rho \mid \nu }^{(4)}\right) \xi   \notag \\
&&+B_{\mu \nu }^{\ast }\partial _{\lbrack \alpha }\bar{\psi}_{\beta ]}\left(
\bar{N}^{\mu \nu \mid \alpha \beta }-\frac{\mathrm{i}}{2}\sigma ^{\alpha
^{\prime }[\mu }\sigma ^{\nu ]\beta ^{\prime }}\gamma ^{\alpha \beta \rho
}M_{\rho \mid \alpha ^{\prime }\beta ^{\prime }}^{(3)}\right) \xi   \notag \\
&&+\psi ^{\ast \mu }M_{\mu }^{(0)}\xi +\partial _{\lbrack \alpha }\bar{\psi}%
_{\beta ]}\bar{M}^{\mu \mid \alpha \beta }\bar{\psi}_{\mu }^{\ast }\eta
\notag \\
&&+s\left( -H^{\ast \alpha }\psi ^{\ast \mu }M_{\mu \mid \alpha }^{(1)}\xi
+\varphi ^{\ast }\psi ^{\ast \mu }M_{\mu }^{(2)}\xi \right.   \notag \\
&&\left. -2B^{\ast \alpha \beta }\psi ^{\ast \mu }M_{\mu \mid \alpha \beta
}^{(3)}\xi -A^{\ast \alpha }\psi ^{\ast \mu }M_{\mu \mid \alpha }^{(4)}\xi
\right) .  \label{om14}
\end{eqnarray}%
Formula (\ref{om14}) allows us to take the solution to (\ref{om1}) under the
form%
\begin{eqnarray}
\tilde{a}_{1}^{\mathrm{int}} &=&H_{\mu }^{\ast }\partial _{\lbrack \alpha }%
\bar{\psi}_{\beta ]}N^{\mu \mid \alpha \beta }\xi +\varphi ^{\ast }\partial
_{\lbrack \alpha }\bar{\psi}_{\beta ]}N^{\alpha \beta }\xi +A_{\mu }^{\ast
}\partial _{\lbrack \alpha }\bar{\psi}_{\beta ]}\bar{N}^{\mu \mid \alpha
\beta }\xi   \notag \\
&&+B_{\mu \nu }^{\ast }\partial _{\lbrack \alpha }\bar{\psi}_{\beta ]}\bar{N}%
^{\mu \nu \mid \alpha \beta }\xi +\psi ^{\ast \mu }M_{\mu }^{(0)}\xi
+\partial _{\lbrack \alpha }\bar{\psi}_{\beta ]}\bar{M}^{\mu \mid \alpha
\beta }\bar{\psi}_{\mu }^{\ast }\eta ,  \label{om15}
\end{eqnarray}%
where $N^{\mu \mid \alpha \beta }$, $N^{\alpha \beta }$, $\bar{N}^{\mu \mid
\alpha \beta }$, $\bar{N}^{\mu \nu \mid \alpha \beta }$, and $M_{\mu }^{(0)}$
are $4\times 4$ matrices with spinor indices, whose elements are smooth
functions depending at most on $\varphi $ ($M_{\mu }^{(0)}$ is imposed to
effectively depend on the scalar field since otherwise the corresponding
term from $\tilde{a}_{1}^{\mathrm{int}}$ cannot describe cross-interactions,
i.e. mix the BF sector with the Rarita-Schwinger one). We ask that (\ref%
{om15}) satisfies (\ref{ai12}). With the help of definitions (\ref{bfa15})--(%
\ref{bfa17}), by direct computation we infer%
\begin{eqnarray}
\delta \tilde{a}_{1}^{\mathrm{int}} &=&\partial _{\lbrack \alpha }\bar{\psi}%
_{\beta ]}N^{\mu \mid \alpha \beta }\xi \partial _{\mu }\varphi -\partial
_{\lbrack \alpha }\bar{\psi}_{\beta ]}N^{\alpha \beta }\xi \partial _{\mu
}H^{\mu }+\partial _{\lbrack \alpha }\bar{\psi}_{\beta ]}\bar{N}^{\mu \mid
\alpha \beta }\xi \partial ^{\nu }B_{\mu \nu }  \notag \\
&&+\partial _{\lbrack \alpha }\bar{\psi}_{\beta ]}\bar{N}^{\mu \nu \mid
\alpha \beta }\xi \partial _{\mu }A_{\nu }-\mathrm{i}\partial _{\mu }\bar{%
\psi}_{\nu }\gamma ^{\mu \nu \rho }M_{\rho }^{(0)}\xi   \notag \\
&&-\frac{\mathrm{i}}{2}\partial _{\lbrack \alpha }\bar{\psi}_{\beta ]}\bar{M}%
^{\mu \mid \alpha \beta }\gamma _{\mu \nu \rho }\left( \partial ^{\lbrack
\nu }\psi ^{\rho ]}\right) \eta .  \label{om16}
\end{eqnarray}%
It is simpler to analyze $\delta \tilde{a}_{1}^{\mathrm{int}}$ via
decomposing it along the number of derivatives into%
\begin{equation}
\delta \tilde{a}_{1}^{\mathrm{int}}=\omega _{1}+\omega _{2},  \label{om17}
\end{equation}%
where
\begin{equation}
\omega _{1}=-\mathrm{i}\partial _{\mu }\bar{\psi}_{\nu }\gamma ^{\mu \nu
\rho }M_{\rho }^{(0)}\xi ,  \label{om18a}
\end{equation}%
\begin{eqnarray}
\omega _{2} &=&\partial _{\lbrack \alpha }\bar{\psi}_{\beta ]}N^{\mu \mid
\alpha \beta }\xi \partial _{\mu }\varphi -\partial _{\lbrack \alpha }\bar{%
\psi}_{\beta ]}N^{\alpha \beta }\xi \partial _{\mu }H^{\mu }+\partial
_{\lbrack \alpha }\bar{\psi}_{\beta ]}\bar{N}^{\mu \mid \alpha \beta }\xi
\partial ^{\nu }B_{\mu \nu }  \notag \\
&&+\partial _{\lbrack \alpha }\bar{\psi}_{\beta ]}\bar{N}^{\mu \nu \mid
\alpha \beta }\xi \partial _{\mu }A_{\nu }-\frac{\mathrm{i}}{2}\partial
_{\lbrack \alpha }\bar{\psi}_{\beta ]}\bar{M}^{\mu \mid \alpha \beta }\gamma
_{\mu \nu \rho }\left( \partial ^{\lbrack \nu }\psi ^{\rho ]}\right) \eta .
\label{om18b}
\end{eqnarray}%
Using decomposition (\ref{om17}), it follows that equation (\ref{ai12})
becomes equivalent to two independent equations, one for each component:%
\begin{eqnarray}
\omega _{1} &=&\gamma d_{0}+\partial _{\mu }v_{0}^{\mu },  \label{om19a} \\
\omega _{2} &=&\gamma d_{1}+\partial _{\mu }v_{1}^{\mu },  \label{om19b}
\end{eqnarray}%
where%
\begin{equation}
\tilde{c}_{0}=-\left( d_{0}+d_{1}\right) ,\qquad \tilde{m}_{0}^{\mu
}=v_{0}^{\mu }+v_{1}^{\mu }.  \label{om120}
\end{equation}%
The objects denoted by $d_{0}$ or $v_{0}^{\mu }$ are derivative-free, while $%
d_{1}$ and $v_{1}^{\mu }$ comprise a single spacetime derivative. We will
approach equations (\ref{om19a}) and (\ref{om19b}) separately.

Related to (\ref{om19a}), from (\ref{om18a}) we find that%
\begin{eqnarray}
\omega _{1} &=&\gamma \left( -\mathrm{i}\bar{\psi}_{\mu }\gamma ^{\mu \nu
\rho }M_{\rho }^{(0)}\psi _{\nu }\right) +\partial _{\mu }\left\{ -\mathrm{i}%
\bar{\psi}_{\nu }\left[ \gamma ^{\mu \nu \rho }M_{\rho }^{(0)}+\gamma
^{0}\left( \gamma ^{0}\gamma ^{\mu \nu \rho }M_{\rho }^{(0)}\right) ^{\top }%
\right] \xi \right\}  \notag \\
&&-\frac{\mathrm{i}}{2}\bar{\psi}_{\mu }\left[ \gamma ^{\mu \nu \rho
}\partial _{\lbrack \nu }M_{\rho ]}^{(0)}+\gamma ^{0}\left( \gamma
^{0}\gamma ^{\mu \nu \rho }\partial _{\lbrack \nu }M_{\rho ]}^{(0)}\right)
^{\top }\right] \xi  \notag \\
&&+\mathrm{i}\partial _{\mu }\bar{\psi}_{\nu }\left[ \gamma ^{0}\left(
\gamma ^{0}\gamma ^{\mu \nu \rho }M_{\rho }^{(0)}\right) ^{\top }-\gamma
^{\mu \nu \rho }M_{\rho }^{(0)}\right] \xi -\omega _{1},  \label{om121}
\end{eqnarray}%
and hence equation (\ref{om19a}) is fulfilled if and only if
\begin{eqnarray}
\gamma ^{0}\left( \gamma ^{0}\gamma ^{\mu \nu \rho }M_{\rho }^{(0)}\right)
^{\top }-\gamma ^{\mu \nu \rho }M_{\rho }^{(0)} &=&0,  \label{om122a} \\
\gamma ^{0}\left( \gamma ^{0}\gamma ^{\mu \nu \rho }M_{\rho }^{(0)}\right)
^{\top }+\gamma ^{\mu \nu \rho }M_{\rho }^{(0)} &=&A,  \label{om122b}
\end{eqnarray}%
where $A$ is a constant $4\times 4$ matrix with spinor-like indices.
Relations (\ref{om122a}) and (\ref{om122b}) are equivalent with%
\begin{equation*}
\gamma ^{\mu \nu \rho }M_{\rho }^{(0)}=\gamma ^{0}\left( \gamma ^{0}\gamma
^{\mu \nu \rho }M_{\rho }^{(0)}\right) ^{\top }=\frac{1}{2}A,
\end{equation*}%
i.e. the matrix $M_{\rho }^{(0)}$ is purely constant, such that the
corresponding term from $\tilde{a}_{1}^{\mathrm{int}}$ cannot produce
cross-couplings, as required. As a consequence, we can safely work with%
\begin{equation}
M_{\rho }^{(0)}=0  \label{om123}
\end{equation}%
in $\tilde{a}_{1}^{\mathrm{int}}$, which further leads to
\begin{equation}
\omega _{1}=d_{0}=v_{0}^{\mu }=0.  \label{partialomega1}
\end{equation}

Regarding (\ref{om19b}), we observe that $\omega _{2}$ is written as a sum
of four different types of terms%
\begin{equation}
\omega _{2}=\omega _{2}^{(1)}+\omega _{2}^{(2)}+\omega _{2}^{(3)}+\omega
_{2}^{(4)},  \label{om124}
\end{equation}%
where
\begin{eqnarray}
\omega _{2}^{(1)} &=&\partial _{\lbrack \alpha }\bar{\psi}_{\beta ]}N^{\mu
\mid \alpha \beta }\xi \partial _{\mu }\varphi ,  \label{om124a} \\
\omega _{2}^{(2)} &=&-\partial _{\lbrack \alpha }\bar{\psi}_{\beta
]}N^{\alpha \beta }\xi \partial _{\mu }H^{\mu },  \label{om124b} \\
\omega _{2}^{(3)} &=&\partial _{\lbrack \alpha }\bar{\psi}_{\beta ]}\bar{N}%
^{\mu \mid \alpha \beta }\xi \partial ^{\nu }B_{\mu \nu },  \label{om124c} \\
\omega _{2}^{(4)} &=&\partial _{\lbrack \alpha }\bar{\psi}_{\beta ]}\bar{N}%
^{\mu \nu \mid \alpha \beta }\xi \partial _{\mu }A_{\nu }-\frac{\mathrm{i}}{2%
}\partial _{\lbrack \alpha }\bar{\psi}_{\beta ]}\bar{M}^{\mu \mid \alpha
\beta }\gamma _{\mu \nu \rho }\left( \partial ^{\lbrack \nu }\psi ^{\rho
]}\right) \eta ,  \label{om124d}
\end{eqnarray}%
so each type mixes the Rarita-Schwinger spinor with various fields/ghosts
from the BF complex. Recalling decomposition (\ref{om124}), it is easy to
see that $\omega _{2}$ is $\gamma $-exact modulo $d$ if and only if each of
its components, $\omega _{2}^{(j)}$, $j=\overline{1,4}$, is so:%
\begin{equation}
\omega _{2}^{(j)}=\gamma d_{1}^{(j)}+\partial _{\mu }v_{1}^{(j)\mu },\qquad
j=\overline{1,4},  \label{om125}
\end{equation}%
where%
\begin{equation}
d_{1}=\sum\limits_{j=1}^{4}d_{1}^{(j)},\qquad v_{1}^{\mu
}=\sum\limits_{j=1}^{4}v_{1}^{(j)\mu }.  \label{om126}
\end{equation}%
We will analyze equations (\ref{om125}) separately as well. Firstly, we
consider (\ref{om125}) for $j=1$ and denote by $\hat{N}^{\mu \mid \alpha
\beta }$ some $4\times 4$ matrices with spinor-like indices, whose elements
are nothing but some indefinite integrals of the corresponding elements of $%
N^{\mu \mid \alpha \beta }$. In terms of this new matrices, (\ref{om124a})
becomes%
\begin{eqnarray}
\omega _{2}^{(1)} &=&\partial _{\lbrack \alpha }\bar{\psi}_{\beta ]}\left(
\partial _{\mu }\hat{N}^{\mu \mid \alpha \beta }\right) \xi  \notag \\
&=&\partial _{\mu }\left[ \left( \partial _{\lbrack \alpha }\bar{\psi}%
_{\beta ]}\right) \hat{N}^{\mu \mid \alpha \beta }\xi \right] +\gamma \left[
-\left( \partial _{\lbrack \alpha }\bar{\psi}_{\beta ]}\right) \hat{N}^{\mu
\mid \alpha \beta }\psi _{\mu }\right]  \notag \\
&&-\left( \partial _{\mu }\partial _{\lbrack \alpha }\bar{\psi}_{\beta
]}\right) \hat{N}^{\mu \mid \alpha \beta }\xi ,  \label{om127}
\end{eqnarray}%
such that $\omega _{2}^{(1)}$ is solution to (\ref{om125}) for $j=1$ if and
only if
\begin{equation}
\left( \partial _{\mu }\partial _{\lbrack \alpha }\bar{\psi}_{\beta
]}\right) \hat{N}^{\mu \mid \alpha \beta }=0,  \label{om128}
\end{equation}%
or, in other words, matrices $\hat{N}^{\mu \mid \alpha \beta }$ are
completely antisymmetric in their Lorentz indices. Since we work in $D=4$,
the general solution to (\ref{om128}) reads as%
\begin{equation}
\hat{N}^{\mu \mid \alpha \beta }=\gamma ^{\mu \alpha \beta }\left( \mathrm{i}%
\hat{U}_{7}+\hat{U}_{8}\gamma _{5}\right) ,  \label{om129}
\end{equation}%
where $\hat{U}_{7}$ and $\hat{U}_{8}$ are some real, smooth functions of $%
\varphi $. Replacing (\ref{om129}) into (\ref{om127}), we get%
\begin{eqnarray}
\omega _{2}^{(1)} &\equiv &\partial _{\lbrack \alpha }\bar{\psi}_{\beta
]}\gamma ^{\mu \alpha \beta }\left( \mathrm{i}\partial _{\mu }\hat{U}%
_{7}+\partial _{\mu }\hat{U}_{8}\gamma _{5}\right) \xi =\partial _{\mu }%
\left[ \left( \partial _{\lbrack \alpha }\bar{\psi}_{\beta ]}\right) \gamma
^{\mu \alpha \beta }\left( \mathrm{i}\hat{U}_{7}+\hat{U}_{8}\gamma
_{5}\right) \xi \right]  \notag \\
&&+\gamma \left[ -\left( \partial _{\lbrack \alpha }\bar{\psi}_{\beta
]}\right) \gamma ^{\mu \alpha \beta }\left( \mathrm{i}\hat{U}_{7}+\hat{U}%
_{8}\gamma _{5}\right) \psi _{\mu }\right] ,  \label{om130}
\end{eqnarray}%
which enables us to make the identifications
\begin{eqnarray}
d_{1}^{(1)} &=&-\left( \partial _{\lbrack \alpha }\bar{\psi}_{\beta
]}\right) \gamma ^{\mu \alpha \beta }\left( \mathrm{i}\hat{U}_{7}+\hat{U}%
_{8}\gamma _{5}\right) \psi _{\mu }  \label{om131a} \\
v_{1}^{(1)\mu } &=&\left( \partial _{\lbrack \alpha }\bar{\psi}_{\beta
]}\right) \gamma ^{\mu \alpha \beta }\left( \mathrm{i}\hat{U}_{7}+\hat{U}%
_{8}\gamma _{5}\right) \xi .  \label{om131b}
\end{eqnarray}%
Due to the relationship between $\hat{N}^{\mu \mid \alpha \beta }$ and $%
N^{\mu \mid \alpha \beta }$, we can write
\begin{equation}
N^{\mu \mid \alpha \beta }=\gamma ^{\mu \alpha \beta }\left( \mathrm{i}\frac{%
d\hat{U}_{7}}{d\varphi }+\frac{d\hat{U}_{8}}{d\varphi }\gamma _{5}\right) .
\label{om129a}
\end{equation}%
Next, we approach equation (\ref{om125}) for $j=2$. By integrating by parts,
we arrive at%
\begin{eqnarray}
\omega _{2}^{(2)} &=&\partial _{\mu }\left( -H^{\mu }\partial _{\lbrack
\alpha }\bar{\psi}_{\beta ]}N^{\alpha \beta }\xi \right) +\gamma \left(
H^{\mu }\partial _{\lbrack \alpha }\bar{\psi}_{\beta ]}N^{\alpha \beta }\psi
_{\mu }\right)  \notag \\
&&-2\partial _{\lbrack \alpha }\bar{\psi}_{\beta ]}N^{\alpha \beta }\psi
_{\mu }\partial _{\nu }C^{\mu \nu }+H^{\mu }\partial _{\mu }\left( \partial
_{\lbrack \alpha }\bar{\psi}_{\beta ]}N^{\alpha \beta }\right) \xi ,
\label{om132}
\end{eqnarray}%
so (\ref{om125}) for $j=2$ imposes the following restrictions (to be
satisfied simultaneously):%
\begin{eqnarray}
\partial _{\lbrack \alpha }\bar{\psi}_{\beta ]}N^{\alpha \beta }\psi _{\mu }
&=&\partial _{\mu }l,  \label{om133a} \\
\partial _{\mu }\left( \partial _{\lbrack \alpha }\bar{\psi}_{\beta
]}N^{\alpha \beta }\right) &=&0.  \label{om133b}
\end{eqnarray}%
The former equality, (\ref{om133a}), cannot take place. (The Euler-Lagrange
derivatives of the right-hand of (\ref{om133a}) with respect to $\psi _{\rho
}$ vanishes obviously, while that of the left-hand side is nonvanishing.)
Neither does equation (\ref{om133b}) because in the opposite situation the
quantity $\partial _{\lbrack \alpha }\bar{\psi}_{\beta ]}N^{\alpha \beta }$
should be constant. The requirement that $\omega _{2}^{(2)}$ is of the form (%
\ref{om125}) for $j=2$ implies thus the condition%
\begin{equation}
N^{\alpha \beta }=0.  \label{om134}
\end{equation}%
Integrating now by parts equation (\ref{om125}) for $j=3$, it results%
\begin{eqnarray}
\omega _{2}^{(3)} &=&\partial ^{\nu }\left[ \left( \partial _{\lbrack \alpha
}\bar{\psi}_{\beta ]}\right) \bar{N}^{\mu \mid \alpha \beta }\xi B_{\mu \nu }%
\right] +\gamma \left[ -\left( \partial _{\lbrack \alpha }\bar{\psi}_{\beta
]}\right) \bar{N}^{\mu \mid \alpha \beta }\psi ^{\nu }B_{\mu \nu }\right]
\notag \\
&&-3\left( \partial _{\lbrack \alpha }\bar{\psi}_{\beta ]}\right) \bar{N}%
^{\mu \mid \alpha \beta }\psi ^{\nu }\partial ^{\rho }\eta _{\mu \nu \rho
}-\partial ^{\nu }\left( \partial _{\lbrack \alpha }\bar{\psi}_{\beta ]}\bar{%
N}^{\mu \mid \alpha \beta }\right) \xi B_{\mu \nu },  \label{om135}
\end{eqnarray}%
so $\omega _{2}^{(3)}$ reads as in (\ref{om125}) for $j=3$ if and only if
the next properties are verified simultaneously:%
\begin{eqnarray}
\left( \partial _{\lbrack \alpha }\bar{\psi}_{\beta ]}\right) \bar{N}^{\mu
\mid \alpha \beta }\psi ^{\nu }-\left( \partial _{\lbrack \alpha }\bar{\psi}%
_{\beta ]}\right) \bar{N}^{\nu \mid \alpha \beta }\psi ^{\mu } &=&\partial
^{\lbrack \mu }k^{\nu ]},  \label{om136a} \\
\partial _{\lbrack \alpha }\bar{\psi}_{\beta ]}\bar{N}^{\mu \mid \alpha
\beta } &=&\partial ^{\mu }k,  \label{om136b}
\end{eqnarray}%
with $k^{\mu }$ and $k$ local quantities and $k$ having in addition a
spinor-like behavior. Equations (\ref{om136a}) and (\ref{om136b}) are
incompatible, so the requirement that $\omega _{2}^{(3)}$ is of the form (%
\ref{om125}) for $j=3$ reveals the condition%
\begin{equation}
\bar{N}^{\mu \mid \alpha \beta }=0.  \label{om137}
\end{equation}%
(The incompatibility between conditions (\ref{om136a}) and (\ref{om136b})
can be emphasized for instance by taking $\bar{N}^{\mu \mid \alpha \beta }$
to be a $4\times 4$ matrix with spinor-like indices that is solution to (\ref%
{om136b}). Since $k$ is a derivative-free spinor, this solution inserted in
the left-hand side of (\ref{om136b}) gives%
\begin{eqnarray}
\left( \partial _{\lbrack \alpha }\bar{\psi}_{\beta ]}\right) \bar{N}^{\mu
\mid \alpha \beta }\psi ^{\nu }-\left( \partial _{\lbrack \alpha }\bar{\psi}%
_{\beta ]}\right) \bar{N}^{\nu \mid \alpha \beta }\psi ^{\mu } &=&\left(
\partial ^{\mu }k\right) \psi ^{\nu }-\left( \partial ^{\nu }k\right) \psi
^{\mu }  \notag \\
&=&\partial ^{\lbrack \mu }\left( k\psi ^{\nu ]}\right) -k\left( \partial
^{\lbrack \mu }\psi ^{\nu ]}\right)  \notag \\
&\neq &\partial ^{\lbrack \mu }k^{\nu ]},  \label{om138}
\end{eqnarray}%
which shows that (\ref{om136a}) cannot be satisfied and proves thus our
assertion.) We are left now with equation (\ref{om125}) for $j=4$.
Integrating it by parts, we obtain%
\begin{eqnarray}
\omega _{2}^{(4)} &=&\partial _{\mu }\left[ A_{\nu }\left( \partial
_{\lbrack \alpha }\bar{\psi}_{\beta ]}\right) \bar{N}^{\mu \nu \mid \alpha
\beta }\xi -\eta \left( \partial _{\lbrack \alpha }\bar{\psi}_{\beta
]}\right) \bar{N}^{\mu \nu \mid \alpha \beta }\psi _{\nu }\right]  \notag \\
&&+\gamma \left[ A_{\mu }\left( \partial _{\lbrack \alpha }\bar{\psi}_{\beta
]}\right) \bar{N}^{\mu \nu \mid \alpha \beta }\psi _{\nu }\right]  \notag \\
&&+\partial _{\mu }\left[ \left( \partial _{\lbrack \alpha }\bar{\psi}%
_{\beta ]}\right) \bar{N}^{\mu \nu \mid \alpha \beta }\right] \left( \psi
_{\nu }\eta -\xi A_{\nu }\right)  \notag \\
&&+\frac{1}{2}\left( \partial _{\lbrack \alpha }\bar{\psi}_{\beta ]}\right)
\left( \bar{N}^{\mu \nu \mid \alpha \beta }-\mathrm{i}\sigma _{\rho \lambda }%
\bar{M}^{\rho \mid \alpha \beta }\gamma ^{\lambda \mu \nu }\right) \left(
\partial _{\lbrack \mu }\psi _{\nu ]}\right) \eta .  \label{om139}
\end{eqnarray}%
From (\ref{om139}) we notice that $\omega _{2}^{(4)}$ can be written like in
(\ref{om125}) for $j=4$ if and only if the next formulas hold simultaneously:%
\begin{eqnarray}
\left( \partial _{\lbrack \alpha }\bar{\psi}_{\beta ]}\right) \bar{N}^{\mu
\nu \mid \alpha \beta } &=&\partial _{\rho }K^{\mu \nu \rho },
\label{om140a} \\
\left( \gamma ^{0}\hat{N}^{\mu \nu \mid \alpha \beta }\right) ^{\top }
&=&\gamma ^{0}\hat{N}^{\alpha \beta \mid \mu \nu }.  \label{om140b}
\end{eqnarray}%
In the above $K^{\mu \nu \rho }$ has a spinor behavior, is derivative-free,
and completely antisymmetric in its Lorentz indices, while $\hat{N}^{\mu \nu
\mid \alpha \beta }$ is defined via
\begin{equation}
\hat{N}^{\mu \nu \mid \alpha \beta }=\bar{N}^{\mu \nu \mid \alpha \beta }-%
\mathrm{i}\sigma _{\rho \lambda }\bar{M}^{\rho \mid \alpha \beta }\gamma
^{\lambda \mu \nu }.  \label{om141}
\end{equation}%
Equation (\ref{om140a}) shows that $\bar{N}^{\mu \nu \mid \alpha \beta }$
cannot depend on the scalar field $\varphi $, being therefore some constant,
$4\times 4$ matrices with spinor-like indices. Under these circumstances, (%
\ref{om140a}) becomes equivalent to
\begin{equation}
\partial _{\rho }\left( 2\bar{\psi}_{\lambda }\bar{N}^{\mu \nu \mid \rho
\lambda }-K^{\mu \nu \rho }\right) =0  \label{om142}
\end{equation}%
and, since the quantity under derivative is derivative-free, (\ref{om142})
produces
\begin{equation}
2\bar{\psi}_{\lambda }\bar{N}^{\mu \nu \mid \rho \lambda }=K^{\mu \nu \rho }.
\label{om143}
\end{equation}%
Taking into account the complete antisymmetry of $K^{\mu \nu \rho }$, from
the last formula it follows that $\bar{N}^{\mu \nu \mid \rho \lambda }$ must
be also antisymmetric with respect to its first three indices. On the other
hand, $\bar{N}^{\mu \nu \mid \rho \lambda }$ is known to be antisymmetric in
its last two indices, so they must be fully antisymmetric with respect to
their Lorentz indices. In $D=4$ there is a single possibility for the
solution of equation (\ref{om140a}), namely%
\begin{equation}
\bar{N}^{\mu \nu \mid \rho \lambda }=\varepsilon ^{\mu \nu \rho \lambda
}\left( k_{1}+\mathrm{i}k_{2}\gamma _{5}\right) ,  \label{om144}
\end{equation}%
with $k_{1}$ and $k_{2}$ some arbitrary, real constants. In order to solve
the last condition, (\ref{om140b}), we decompose the matrices $\hat{N}^{\mu
\nu \mid \alpha \beta }$ according to the basis (\ref{b2})
\begin{equation}
\hat{N}^{\mu \nu \mid \alpha \beta }=\hat{n}^{\mu \nu \mid \alpha \beta }+%
\hat{n}_{\rho }^{\mu \nu \mid \alpha \beta }\gamma ^{\rho }+\hat{n}_{\rho
\lambda }^{\mu \nu \mid \alpha \beta }\gamma ^{\rho \lambda }+\hat{n}_{\rho
\lambda \sigma }^{\mu \nu \mid \alpha \beta }\gamma ^{\rho \lambda \sigma }+%
\bar{n}^{\mu \nu \mid \alpha \beta }\gamma _{5},  \label{om145}
\end{equation}%
where $\hat{n}^{\mu \nu \mid \alpha \beta }$, $\hat{n}_{\rho }^{\mu \nu \mid
\alpha \beta }$, $\hat{n}_{\rho \lambda }^{\mu \nu \mid \alpha \beta }$, $%
\hat{n}_{\rho \lambda \sigma }^{\mu \nu \mid \alpha \beta }$, and $\bar{n}%
^{\mu \nu \mid \alpha \beta }$ are some Lorentz tensors, completely
antisymmetric in their lower indices, displaying the same
symmetry/antisymmetry properties like $\hat{N}^{\mu \nu \mid \alpha \beta }$
with respect to their upper indices, and depending only on the
undifferentiated scalar field. Since in $D=4$ the only constant Lorentz
tensors have an even number of indices, the decomposition (\ref{om145})
reduces to%
\begin{equation}
\hat{N}^{\mu \nu \mid \alpha \beta }=\hat{n}^{\mu \nu \mid \alpha \beta }+%
\hat{n}_{\rho \lambda }^{\mu \nu \mid \alpha \beta }\gamma ^{\rho \lambda }+%
\bar{n}^{\mu \nu \mid \alpha \beta }\gamma _{5}.  \label{om146}
\end{equation}%
Asking that (\ref{om146}) satisfies (\ref{om140b}) and using properties (\ref%
{sym1})--(\ref{sym2}), it follows that $\hat{n}^{\mu \nu \mid \alpha \beta }$%
, $\hat{n}_{\rho \lambda }^{\mu \nu \mid \alpha \beta }$, and $\bar{n}^{\mu
\nu \mid \alpha \beta }$ consequently exhibit the symmetry/antisymmetry
properties%
\begin{eqnarray}
\hat{n}^{\mu \nu \mid \alpha \beta } &=&-\hat{n}^{\alpha \beta \mid \mu \nu
},  \label{om147a} \\
\hat{n}_{\rho \lambda }^{\mu \nu \mid \alpha \beta } &=&\hat{n}_{\rho
\lambda }^{\alpha \beta \mid \mu \nu },  \label{om147b} \\
\bar{n}^{\mu \nu \mid \alpha \beta } &=&-\bar{n}^{\alpha \beta \mid \mu \nu
}.  \label{om147c}
\end{eqnarray}%
Because all the constant Lorentz tensors can be constructed out of the flat
metric and the Levi-Civita symbol and we work in $D=4$, it is easy to see
that there are no such tensors that fulfill the properties (\ref{om147a})--(%
\ref{om147c}) and hence the solution to (\ref{om140b}) is purely trivial%
\begin{equation}
\hat{N}^{\mu \nu \mid \alpha \beta }=0.  \label{trivNhat}
\end{equation}%
Replacing now (\ref{om144}) and (\ref{trivNhat}) into (\ref{om141}), we
conclude that%
\begin{equation}
\mathrm{i}\sigma _{\rho \lambda }\bar{M}^{\rho \mid \alpha \beta }\gamma
^{\lambda \mu \nu }=\varepsilon ^{\mu \nu \rho \lambda }\left( k_{1}+\mathrm{%
i}k_{2}\gamma _{5}\right) .  \label{om148}
\end{equation}%
The last relations allow us to identify the concrete expression of the
matrices $\bar{M}^{\rho \mid \alpha \beta }$, which are restricted to be
constant. This can be done by decomposing these matrices along the basis (%
\ref{b2})
\begin{equation}
\bar{M}^{\rho \mid \alpha \beta }=\bar{m}^{\rho \mid \alpha \beta }+\bar{m}%
_{\sigma }^{\rho \mid \alpha \beta }\gamma ^{\sigma }+\bar{m}_{\sigma
\varepsilon }^{\rho \mid \alpha \beta }\gamma ^{\sigma \varepsilon }+\bar{m}%
_{\sigma \varepsilon \gamma }^{\rho \mid \alpha \beta }\gamma ^{\sigma
\varepsilon \gamma }+\hat{m}^{\rho \mid \alpha \beta }\gamma _{5},
\label{om149}
\end{equation}%
where $\bar{m}^{\rho \mid \alpha \beta }$, $\bar{m}_{\sigma }^{\rho \mid
\alpha \beta }$, $\bar{m}_{\sigma \varepsilon }^{\rho \mid \alpha \beta }$, $%
\bar{m}_{\sigma \varepsilon \gamma }^{\rho \mid \alpha \beta }$, and $\hat{m}%
^{\rho \mid \alpha \beta }$ are some constant, nonderivative Lorentz
tensors, antisymmetric in their lower indices and with the symmetry
properties of $\bar{M}^{\rho \mid \alpha \beta }$ with respect to their
upper indices. Invoking one more time the even number of Lorentz indices for
any constant tensor in $D=4$, only the second and the fourth terms from
decomposition (\ref{om149}) will survive,%
\begin{equation}
\bar{M}^{\rho \mid \alpha \beta }=\bar{m}_{\sigma }^{\rho \mid \alpha \beta
}\gamma ^{\sigma }+\bar{m}_{\sigma \varepsilon \gamma }^{\rho \mid \alpha
\beta }\gamma ^{\sigma \varepsilon \gamma }.  \label{om150}
\end{equation}%
Their general expressions are%
\begin{eqnarray}
\bar{m}_{\sigma }^{\rho \mid \alpha \beta } &=&\bar{k}_{1}\varepsilon
_{\qquad \sigma }^{\rho \alpha \beta }+\bar{k}_{2}\sigma ^{\rho \lbrack
\alpha }\delta _{\sigma }^{\beta ]},  \label{om151a} \\
\bar{m}_{\sigma \varepsilon \gamma }^{\rho \mid \alpha \beta } &=&-\frac{1}{6%
}\bar{k}_{3}\varepsilon _{\sigma \varepsilon \gamma }^{\qquad \lbrack \alpha
}\sigma ^{\beta ]\rho }+\frac{1}{6}\bar{k}_{4}\delta _{\sigma }^{[\rho
}\delta _{\varepsilon }^{\alpha }\delta _{\gamma }^{\beta ]},  \label{om151b}
\end{eqnarray}%
with $\bar{k}$s some constants. From the Fierz identities%
\begin{eqnarray}
\gamma _{\alpha }\gamma ^{\mu \nu \rho } &=&\delta _{\alpha }^{[\mu }\gamma
^{\nu \rho ]}+\gamma _{\alpha }^{\ \ \mu \nu \rho },  \label{f1} \\
\gamma _{\alpha \beta \gamma }\gamma ^{\mu \nu \rho } &=&-\delta _{\alpha
}^{[\mu }\delta _{\beta }^{\nu }\delta _{\gamma }^{\rho ]}-\delta _{\lbrack
\alpha }^{[\mu }\delta _{\beta }^{\nu }\gamma _{\gamma ]}^{\ \ \rho ]},
\label{f2}
\end{eqnarray}%
the duality relations (\ref{d1}), and formulas (\ref{om150})--(\ref{om151b}%
), the left-hand side of equation (\ref{om148}) becomes%
\begin{eqnarray}
\mathrm{i}\sigma _{\rho \lambda }\bar{M}^{\rho \mid \alpha \beta }\gamma
^{\lambda \mu \nu } &=&2\varepsilon ^{\mu \nu \alpha \beta }\left( -\mathrm{i%
}\bar{k}_{3}+\bar{k}_{2}\gamma _{5}\right) -2\sigma ^{\mu \lbrack \alpha
}\sigma ^{\beta ]\nu }\left( \mathrm{i}\bar{k}_{4}+\bar{k}_{1}\gamma
_{5}\right)  \notag \\
&&+\sigma ^{\mu \lbrack \alpha }\gamma ^{\beta ]\nu }\left[ \mathrm{i}\left(
\bar{k}_{2}-\bar{k}_{4}\right) +\left( -\bar{k}_{1}+\bar{k}_{3}\right)
\gamma _{5}\right]  \notag \\
&&-\sigma ^{\nu \lbrack \alpha }\gamma ^{\beta ]\mu }\left[ \mathrm{i}\left(
\bar{k}_{2}-\bar{k}_{4}\right) +\left( -\bar{k}_{1}+\bar{k}_{3}\right)
\gamma _{5}\right] ,  \label{om152}
\end{eqnarray}%
such that (\ref{om148}) projected on the elements of the basis (\ref{b2})
amounts to three independent equations%
\begin{equation}
-2\mathrm{i}\varepsilon ^{\mu \nu \alpha \beta }\bar{k}_{3}-2\mathrm{i}%
\sigma ^{\mu \lbrack \alpha }\sigma ^{\beta ]\nu }\bar{k}_{4}=\varepsilon
^{\mu \nu \alpha \beta }k_{1},  \label{om153a}
\end{equation}%
\begin{equation}
2\varepsilon ^{\mu \nu \alpha \beta }\bar{k}_{2}-2\sigma ^{\mu \lbrack
\alpha }\sigma ^{\beta ]\nu }\bar{k}_{1}=\mathrm{i}\varepsilon ^{\mu \nu
\alpha \beta }k_{2},  \label{om153b}
\end{equation}%
\begin{equation}
\left( \sigma ^{\mu \lbrack \alpha }\delta _{\lbrack \rho }^{\beta ]}\delta
_{\lambda ]}^{\nu }-\sigma ^{\nu \lbrack \alpha }\delta _{\lbrack \rho
}^{\beta ]}\delta _{\lambda ]}^{\mu }\right) \left( \bar{k}_{2}-\bar{k}%
_{4}\right) -\left( \sigma ^{\mu \lbrack \alpha }\varepsilon _{\quad \rho
\lambda }^{\beta ]\nu }-\sigma ^{\nu \lbrack \alpha }\varepsilon _{\quad
\rho \lambda }^{\beta ]\mu }\right) \left( -\bar{k}_{1}+\bar{k}_{3}\right)
=0.  \label{om153c}
\end{equation}%
The previous relations are nothing but an algebraic system with the unknowns
$k_{1}$, $k_{2}$, $\bar{k}_{1}$, $\bar{k}_{2}$, $\bar{k}_{3}$, and $\bar{k}%
_{4}$:
\begin{eqnarray}
2\mathrm{i}\bar{k}_{3}+k_{1} &=&0,\qquad \bar{k}_{4}=0,  \label{om154a} \\
2\bar{k}_{2}-\mathrm{i}k_{2} &=&0,\qquad \bar{k}_{1}=0,  \label{om154b} \\
\bar{k}_{2}-\bar{k}_{4} &=&0,\qquad -\bar{k}_{1}+\bar{k}_{3}=0,
\label{om154c}
\end{eqnarray}%
whose solution is purely trivial%
\begin{equation}
k_{1}=k_{2}=\bar{k}_{1}=\bar{k}_{2}=\bar{k}_{3}=\bar{k}_{4}=0.  \label{om155}
\end{equation}%
So far, we have shown that $\omega _{2}^{(4)}$ satisfies equation (\ref%
{om125}) for $j=4$ if and only if%
\begin{equation}
\bar{M}^{\rho \mid \alpha \beta }=\bar{N}^{\mu \nu \mid \rho \lambda }=0.
\label{om156}
\end{equation}

We are now in a position to prove the assertion that $\tilde{a}_{1}^{\mathrm{%
int}}$ can be taken to be trivial. Substituting results (\ref{om129a}), (\ref%
{om134}), (\ref{om137}), and (\ref{om156}) into (\ref{om15}), we identify
the component $\tilde{a}_{1}^{\mathrm{int}}$ of (\ref{ai11}), i.e. the
solution to (\ref{om1}) that fulfills (\ref{ai12}), as%
\begin{equation}
\tilde{a}_{1}^{\mathrm{int}}=H_{\mu }^{\ast }\left( \partial _{\lbrack
\alpha }\bar{\psi}_{\beta ]}\right) \gamma ^{\mu \alpha \beta }\left(
\mathrm{i}\frac{d\hat{U}_{7}}{d\varphi }+\frac{d\hat{U}_{8}}{d\varphi }%
\gamma _{5}\right) \xi .  \label{om157}
\end{equation}%
From the actions of the differentials $\delta $, $\gamma $, and $s$, it is
simple to see that (\ref{om157}) reads as%
\begin{eqnarray}
\tilde{a}_{1}^{\mathrm{int}} &=&\partial _{\mu }\left[ 2\mathrm{i}\psi
^{\ast \mu }\left( \mathrm{i}\hat{U}_{7}+\hat{U}_{8}\gamma _{5}\right) \xi %
\right] +s\left[ 2\mathrm{i}H_{\mu }^{\ast }\psi ^{\ast \mu }\left( \mathrm{i%
}\frac{d\hat{U}_{7}}{d\varphi }+\frac{d\hat{U}_{8}}{d\varphi }\gamma
_{5}\right) \xi \right.   \notag \\
&&\left. -2\mathrm{i}\xi ^{\ast }\left( \mathrm{i}\hat{U}_{7}+\hat{U}%
_{8}\gamma _{5}\right) \xi \right] +\gamma \left[ -2\mathrm{i}\psi ^{\ast
\mu }\left( \mathrm{i}\hat{U}_{7}+\hat{U}_{8}\gamma _{5}\right) \psi _{\mu }%
\right] ,  \label{om158}
\end{eqnarray}%
so, in agreement with (\ref{3.1a}) and (\ref{r68}), it will produce only
trivial deformations, so it can indeed be made to vanish.

\section{Proof of formula (\protect\ref{om034})\label{formula2}}

In the following we solve equation (\ref{om02b}) and hence proof formula (%
\ref{om034}). It is advantageous to decompose $\bar{a}_{0}^{\prime \prime
\mathrm{int}}$ with respect to the number of derivatives into%
\begin{equation}
\bar{a}_{0}^{\prime \prime \mathrm{int}}=\overset{(0)}{\pi }+\overset{(1)}{%
\pi },  \label{om05}
\end{equation}%
where $\overset{(0)}{\pi }$ is derivative-free and $\overset{(1)}{\pi }$
comprises a single spacetime derivative of the fields. Again, $\overset{(0)}{%
\pi }$ and $\overset{(1)}{\pi }$ mix the Rarita-Schwinger field with the BF
field sector. Due to (\ref{om05}), equation (\ref{om02b}) is clearly
equivalent to%
\begin{eqnarray}
\gamma \overset{(0)}{\pi } &=&\partial _{\mu }\overset{(0)}{u}_{0}^{\mu },
\label{om06a} \\
\gamma \overset{(1)}{\pi } &=&\partial _{\mu }\overset{(1)}{u}_{0}^{\mu }.
\label{om06b}
\end{eqnarray}%
Making use of definitions (\ref{bfa21}) and (\ref{bfa21.1}), we infer%
\begin{eqnarray}
\gamma \overset{(0)}{\pi } &=&\frac{\partial ^{R}\overset{(0)}{\pi }}{%
\partial \psi _{\mu }}\partial _{\mu }\xi +2\frac{\partial \overset{(0)}{\pi
}}{\partial H^{\mu }}\partial _{\nu }C^{\mu \nu }+\frac{\partial \overset{(0)%
}{\pi }}{\partial A_{\mu }}\partial _{\mu }\eta -3\frac{\partial \overset{(0)%
}{\pi }}{\partial B^{\mu \nu }}\partial _{\rho }\eta ^{\mu \nu \rho }  \notag
\\
&=&\partial _{\mu }\left( \frac{\partial ^{R}\overset{(0)}{\pi }}{\partial
\psi _{\mu }}\xi -2\frac{\partial \overset{(0)}{\pi }}{\partial H^{\nu }}%
C^{\mu \nu }+\frac{\partial \overset{(0)}{\pi }}{\partial A_{\mu }}\eta -3%
\frac{\partial \overset{(0)}{\pi }}{\partial B^{\nu \rho }}\eta ^{\mu \nu
\rho }\right)   \notag \\
&&-\left( \partial _{\mu }\frac{\partial ^{R}\overset{(0)}{\pi }}{\partial
\psi _{\mu }}\right) \xi +\left( \partial _{\lbrack \mu }\frac{\partial
\overset{(0)}{\pi }}{\partial H^{\nu ]}}\right) C^{\mu \nu }-\left( \partial
_{\mu }\frac{\partial \overset{(0)}{\pi }}{\partial A_{\mu }}\right) \eta
\notag \\
&&+\left( \partial _{\lbrack \mu }\frac{\partial \overset{(0)}{\pi }}{%
\partial B^{\nu \rho ]}}\right) \eta ^{\mu \nu \rho },  \label{om07}
\end{eqnarray}%
so $\overset{(0)}{\pi }$ satisfies (\ref{om06a}) if%
\begin{eqnarray}
\partial _{\mu }\frac{\partial ^{R}\overset{(0)}{\pi }}{\partial \psi _{\mu }%
} &=&0,\qquad \partial _{\lbrack \mu }\frac{\partial \overset{(0)}{\pi }}{%
\partial H^{\nu ]}}=0,  \label{om08a} \\
\partial _{\mu }\frac{\partial \overset{(0)}{\pi }}{\partial A_{\mu }}
&=&0,\qquad \partial _{\lbrack \mu }\frac{\partial \overset{(0)}{\pi }}{%
\partial B^{\nu \rho ]}}=0.  \label{om08b}
\end{eqnarray}%
The solutions to the latter equations from (\ref{om08a}) and also to (\ref%
{om08b}) may be trivial, but the former equation from (\ref{om08a}) must
necessarily provide a nontrivial solution in order to produce
cross-interactions. But $\overset{(0)}{\pi }$ is derivative-free, such that
the solutions to (\ref{om08a}) and (\ref{om08b}) read as
\begin{eqnarray}
\frac{\partial ^{R}\overset{(0)}{\pi }}{\partial \psi _{\mu }} &=&\Psi ^{\mu
},\qquad \frac{\partial \overset{(0)}{\pi }}{\partial H^{\mu }}=h_{\mu },
\label{om09a} \\
\frac{\partial \overset{(0)}{\pi }}{\partial A_{\mu }} &=&a^{\mu },\qquad
\frac{\partial \overset{(0)}{\pi }}{\partial B^{\mu \nu }}=b_{\mu \nu },
\label{om09b}
\end{eqnarray}%
where $\Psi ^{\mu }$, $h_{\mu }$, $a^{\mu }$, and $b_{\mu \nu }$ are all
constants. In addition, $\Psi ^{\mu }$ is a spinor and $b_{\mu \nu }$ is
antisymmetric, $b_{\mu \nu }=-b_{\nu \mu }$. There are no such constants and
hence (\ref{om08a}) and (\ref{om08b}) possess only the trivial solution,
which further implies that the solution to (\ref{om06a}) is also trivial,%
\begin{equation}
\overset{(0)}{\pi }=0.  \label{om010}
\end{equation}%
Regarding equation (\ref{om06b}), from definitions (\ref{bfa21}) and (\ref%
{bfa21.1}) we can write%
\begin{eqnarray}
\gamma \overset{(1)}{\pi } &=&\frac{\partial ^{R}\overset{(1)}{\pi }}{%
\partial \psi _{\mu }}\partial _{\mu }\xi +\frac{\partial ^{R}\overset{(1)}{%
\pi }}{\partial \left( \partial _{\lambda }\psi _{\mu }\right) }\partial
_{\lambda }\partial _{\mu }\xi +2\frac{\partial \overset{(1)}{\pi }}{%
\partial H^{\mu }}\partial _{\nu }C^{\mu \nu }+2\frac{\partial \overset{(1)}{%
\pi }}{\partial \left( \partial _{\lambda }H^{\mu }\right) }\partial
_{\lambda }\partial _{\nu }C^{\mu \nu }  \notag \\
&&+\frac{\partial \overset{(1)}{\pi }}{\partial A_{\mu }}\partial _{\mu
}\eta +\frac{\partial \overset{(1)}{\pi }}{\partial \left( \partial
_{\lambda }A_{\mu }\right) }\partial _{\lambda }\partial _{\mu }\eta -3\frac{%
\partial \overset{(1)}{\pi }}{\partial B^{\mu \nu }}\partial _{\rho }\eta
^{\mu \nu \rho }-3\frac{\partial \overset{(1)}{\pi }}{\partial \left(
\partial _{\lambda }B^{\mu \nu }\right) }\partial _{\lambda }\partial _{\rho
}\eta ^{\mu \nu \rho }  \notag \\
&=&\partial _{\mu }\left[ \left( \frac{\partial ^{R}\overset{(1)}{\pi }}{%
\partial \psi _{\mu }}-\partial _{\lambda }\frac{\partial ^{R}\overset{(1)}{%
\pi }}{\partial \left( \partial _{\mu }\psi _{\lambda }\right) }\right) \xi +%
\frac{\partial ^{R}\overset{(1)}{\pi }}{\partial \left( \partial _{\lambda
}\psi _{\mu }\right) }\partial _{\lambda }\xi -2\frac{\partial \overset{(1)}{%
\pi }}{\partial \left( \partial _{\lambda }H^{\nu }\right) }\partial
_{\lambda }C^{\mu \nu }\right.   \notag \\
&&-2\frac{\partial \overset{(1)}{\pi }}{\partial H^{\nu }}C^{\mu \nu
}+2\left( \partial _{\rho }\frac{\partial \overset{(1)}{\pi }}{\partial
\left( \partial _{\mu }H^{\lambda }\right) }\right) C^{\rho \lambda }+\left(
\frac{\partial \overset{(1)}{\pi }}{\partial A_{\mu }}-\partial _{\lambda }%
\frac{\partial \overset{(1)}{\pi }}{\partial \left( \partial _{\mu
}A_{\lambda }\right) }\right) \eta   \notag \\
&&+\frac{\partial \overset{(1)}{\pi }}{\partial \left( \partial _{\lambda
}A_{\mu }\right) }\partial _{\lambda }\eta -3\left( \frac{\partial \overset{%
(1)}{\pi }}{\partial B^{\nu \rho }}\eta ^{\mu \nu \rho }+\frac{\partial
\overset{(1)}{\pi }}{\partial \left( \partial _{\lambda }B^{\nu \rho
}\right) }\partial _{\lambda }\eta ^{\mu \nu \rho }\right)   \notag \\
&&\left. +3\left( \partial _{\nu }\frac{\partial \overset{(1)}{\pi }}{%
\partial \left( \partial _{\mu }B^{\rho \lambda }\right) }\right) \eta ^{\nu
\rho \lambda }\right] -\left( \partial _{\mu }\frac{\delta ^{R}\overset{(1)}{%
\pi }}{\delta \psi _{\mu }}\right) \xi +\left( \partial _{\lbrack \mu }\frac{%
\delta ^{R}\overset{(1)}{\pi }}{\delta H^{\nu ]}}\right) C^{\mu \nu }  \notag
\\
&&-\left( \partial _{\mu }\frac{\delta \overset{(1)}{\pi }}{\delta A_{\mu }}%
\right) \eta +\left( \partial _{\lbrack \mu }\frac{\delta \overset{(1)}{\pi }%
}{\delta B^{\nu \rho ]}}\right) \eta ^{\mu \nu \rho },  \label{om011}
\end{eqnarray}%
so we conclude that $\overset{(1)}{\pi }$ fulfills (\ref{om06b}) if the
following equations take place:%
\begin{eqnarray}
\partial _{\mu }\frac{\delta ^{R}\overset{(1)}{\pi }}{\delta \psi _{\mu }}
&=&0,\qquad \partial _{\lbrack \mu }\frac{\delta \overset{(1)}{\pi }}{\delta
H^{\nu ]}}=0,  \label{om012a} \\
\partial _{\mu }\frac{\delta \overset{(1)}{\pi }}{\delta A_{\mu }}
&=&0,\qquad \partial _{\lbrack \mu }\frac{\delta \overset{(1)}{\pi }}{\delta
B^{\nu \rho ]}}=0.  \label{om012b}
\end{eqnarray}%
Their solutions are given by%
\begin{eqnarray}
\frac{\delta ^{R}\overset{(1)}{\pi }}{\delta \psi _{\mu }} &=&\partial _{\nu
}\Psi ^{\mu \nu },\qquad \frac{\delta \overset{(1)}{\pi }}{\delta H^{\mu }}%
=\partial _{\mu }H,  \label{om013a} \\
\frac{\delta \overset{(1)}{\pi }}{\delta A_{\mu }} &=&\partial _{\nu }A^{\mu
\nu },\qquad \frac{\delta \overset{(1)}{\pi }}{\delta B^{\mu \nu }}=\partial
_{\lbrack \mu }B_{\nu ]},  \label{om013b}
\end{eqnarray}%
where the objects $\Psi ^{\mu \nu }$, $H$, $A^{\mu \nu }$, and $B_{\mu }$
are functions (with $\Psi ^{\mu \nu }$ fermionic and spinor-like and the
remaining ones bosonic) depending only on the undifferentiated fields, with
both $\Psi ^{\mu \nu }$ and $A^{\mu \nu }$ antisymmetric%
\begin{equation}
\Psi ^{\mu \nu }=-\Psi ^{\nu \mu },\qquad A^{\mu \nu }=-A^{\nu \mu }.
\label{om014}
\end{equation}%
Before solving equations (\ref{om013a}) and (\ref{om013b}), we will show
that the dependence of $\overset{(1)}{\pi }$ on $H^{\mu }$, $A_{\mu }$, and $%
B^{\mu \nu }$ can be eliminated through some trivial deformations.

Let $N$ be a derivation in the algebra of the fields $\psi _{\mu }$, $H^{\mu
}$, $A_{\mu }$, and $B^{\mu \nu }$ and of their derivatives that counts the
powers of the fields and their derivatives%
\begin{eqnarray}
N &=&\sum\limits_{k\geq 0}\left( \left( \partial _{\mu _{1}\cdots \mu
_{k}}\psi _{\mu }\right) \frac{\partial ^{L}}{\partial \left( \partial _{\mu
_{1}\cdots \mu _{k}}\psi _{\mu }\right) }+\left( \partial _{\mu _{1}\cdots
\mu _{k}}H^{\mu }\right) \frac{\partial }{\partial \left( \partial _{\mu
_{1}\cdots \mu _{k}}H^{\mu }\right) }\right.  \notag \\
&&\left. +\left( \partial _{\mu _{1}\cdots \mu _{k}}A_{\mu }\right) \frac{%
\partial }{\partial \left( \partial _{\mu _{1}\cdots \mu _{k}}A_{\mu
}\right) }+\left( \partial _{\mu _{1}\cdots \mu _{k}}B_{\mu \nu }\right)
\frac{\partial }{\partial \left( \partial _{\mu _{1}\cdots \mu _{k}}B_{\mu
\nu }\right) }\right) .  \label{om015}
\end{eqnarray}%
Then, it is easy to see that for every nonintegrated density $\phi $ we have%
\begin{equation}
N\phi =\frac{\delta ^{R}\phi }{\delta \psi _{\mu }}\psi _{\mu }+\frac{\delta
\phi }{\delta H^{\mu }}H^{\mu }+\frac{\delta \phi }{\delta A_{\mu }}A_{\mu }+%
\frac{\delta \phi }{\delta B^{\mu \nu }}B^{\mu \nu }+\partial _{\mu }s^{\mu
},  \label{om016}
\end{equation}%
where $\delta ^{(R)}\phi /\delta \Phi ^{\alpha _{0}}$ denotes the (right)
variational derivative of $\phi $ with respect to the field $\Phi ^{\alpha
_{0}}$. If $\phi ^{(l)}$ is a homogeneous polynomial of degree $l$ in the
fields $\psi _{\mu }$, $H^{\mu }$, $A_{\mu }$, $B^{\mu \nu }$ and their
derivatives, then $N\phi ^{(l)}=l\phi ^{(l)}$. Inserting (\ref{om013a}) and (%
\ref{om013b}) in (\ref{om016}) for $\phi =\overset{(1)}{\pi }$ and moving
the derivatives such to act on the fields, we infer
\begin{equation}
N\overset{(1)}{\pi }=\frac{1}{2}\Psi ^{\mu \nu }\partial _{\lbrack \mu }\psi
_{\nu ]}-H\partial _{\mu }H^{\mu }+\frac{1}{2}A^{\mu \nu }\partial _{\lbrack
\mu }A_{\nu ]}+2B^{\mu }\partial ^{\nu }B_{\mu \nu }+\partial _{\mu }t^{\mu
}.  \label{om017}
\end{equation}%
Decomposing $\overset{(1)}{\pi }$ as a sum of homogeneous polynomials of
various degrees in the fields and their derivatives%
\begin{equation}
\overset{(1)}{\pi }=\sum\limits_{k\geq 2}\overset{(1)}{\pi }^{(k)},
\label{om018}
\end{equation}%
such that $N\overset{(1)}{\pi }^{(k)}=k\overset{(1)}{\pi }^{(k)}$ (with $%
k\geq 2$ since $\overset{(1)}{\pi }$ contains at least two spinors), we
deduce%
\begin{equation}
N\overset{(1)}{\pi }=\sum\limits_{k\geq 2}k\overset{(1)}{\pi }^{(k)}.
\label{om019}
\end{equation}%
Comparing (\ref{om019}) with (\ref{om017}), we conclude that decomposition (%
\ref{om018}) induces a similar decomposition at the level of the functions $%
\Psi ^{\mu \nu }$, $H$, $A^{\mu \nu }$, and $B_{\mu }$, i.e.
\begin{eqnarray}
\Psi ^{\mu \nu } &=&\sum\limits_{k\geq 2}\Psi _{(k-1)}^{\mu \nu },\qquad
H=\sum\limits_{k\geq 2}H_{(k-1)},  \label{om020a} \\
A^{\mu \nu } &=&\sum\limits_{k\geq 2}A_{(k-1)}^{\mu \nu },\qquad B^{\mu
}=\sum\limits_{k\geq 2}B_{(k-1)}^{\mu }.  \label{om020b}
\end{eqnarray}%
Substituting (\ref{om020a}) and (\ref{om020b}) in (\ref{om017}) and then
comparing the result with (\ref{om019}), we find that
\begin{eqnarray}
\overset{(1)}{\pi }^{(k)} &=&\frac{1}{2k}\Psi _{(k-1)}^{\mu \nu }\partial
_{\lbrack \mu }\psi _{\nu ]}-\frac{1}{k}H_{(k-1)}\partial _{\mu }H^{\mu }
\notag \\
&&+\frac{1}{2k}A_{(k-1)}^{\mu \nu }\partial _{\lbrack \mu }A_{\nu ]}+\frac{2%
}{k}B_{(k-1)}^{\mu }\partial ^{\nu }B_{\mu \nu }+\partial _{\mu }\hat{t}%
_{(k)}^{\mu },  \label{om021}
\end{eqnarray}%
which further inserted in (\ref{om018}) provides $\overset{(1)}{\pi }$ as%
\begin{equation}
\overset{(1)}{\pi }=\frac{1}{2}\hat{\Psi}^{\mu \nu }\partial _{\lbrack \mu
}\psi _{\nu ]}-\hat{H}\partial _{\mu }H^{\mu }+\frac{1}{2}\hat{A}^{\mu \nu
}\partial _{\lbrack \mu }A_{\nu ]}+2\hat{B}^{\mu }\partial ^{\nu }B_{\mu \nu
}+\partial _{\mu }\hat{t}^{\mu },  \label{om022}
\end{equation}%
where we made the notations%
\begin{eqnarray}
\hat{\Psi}^{\mu \nu } &=&\sum\limits_{k\geq 21}\frac{1}{k}\Psi _{(k-1)}^{\mu
\nu },\qquad \hat{H}=\sum\limits_{k\geq 2}\frac{1}{k}H_{(k-1)},
\label{om023a} \\
\hat{A}^{\mu \nu } &=&\sum\limits_{k\geq 2}\frac{1}{k}A_{(k-1)}^{\mu \nu
},\qquad \hat{B}^{\mu }=\sum\limits_{k\geq 2}\frac{1}{k}B_{(k-1)}^{\mu }.
\label{om023b}
\end{eqnarray}%
As a consequence, we succeeded in bringing the solution $\overset{(1)}{\pi }$
of equation (\ref{om06b}) to the form (\ref{om022}). On the one hand, it is
direct to see (from the former definition in (\ref{bfa16}), definitions (\ref%
{bfa16.1}), and formula (\ref{om022})) that all the terms from $\overset{(1)}%
{\pi }$ but the first vanish on-shell modulo $d$, and therefore they can be
written in a $\delta $-exact modulo $d$ form%
\begin{equation}
\overset{(1)}{\pi }=\frac{1}{2}\hat{\Psi}^{\mu \nu }\partial _{\lbrack \mu
}\psi _{\nu ]}+\delta \left( -\varphi ^{\ast }\hat{H}-B_{\mu \nu }^{\ast }%
\hat{A}^{\mu \nu }-2A_{\mu }^{\ast }\hat{B}^{\mu }\right) +\partial _{\mu }%
\hat{t}^{\mu }.  \label{om024}
\end{equation}%
On the other hand, equation (\ref{om06b}) shows that $\overset{(1)}{\pi }$
belongs to $H^{0}\left( \gamma |d\right) $ in antighost number zero. Using
the general result \cite{gen1} according to which the elements of $H\left(
\gamma |d\right) $ independent of antifields are nontrivial elements of $%
H\left( s|d\right) $ if and only if they do not vanish on-shell modulo $d$,
we can state that all the terms from (\ref{om022}) excepting the first one
are trivial elements of $H\left( s|d\right) $, so they can be eliminated by
trivial redefinitions of the fields. In conclusion, the entire dependence of
$\overset{(1)}{\pi }$ on $H^{\mu }$, $A_{\mu }$, and $B^{\mu \nu }$ can be
removed, which proves our statement from the end of the previous paragraph.

Now, we come back to completing $\overset{(1)}{\pi }$ as solution to (\ref%
{om06b}). By virtue of the above discussion, we can state that the general
form of the nontrivial candidate to the solution of equation (\ref{om06b})
can be chosen under the form
\begin{equation}
\overset{(1)}{\pi }=\frac{1}{2}\hat{\Psi}^{\mu \nu }\partial _{\lbrack \mu
}\psi _{\nu ]}+\partial _{\mu }\bar{t}^{\mu },  \label{om025}
\end{equation}%
where $\hat{\Psi}^{\mu \nu }$ is antisymmetric, $\hat{\Psi}^{\mu \nu }=-\hat{%
\Psi}^{\nu \mu }$, spinor-like, derivative-free (due to the derivative order
assumption) and depends effectively on the undifferentiated scalar field $%
\varphi $ (because otherwise $\overset{(1)}{\pi }$ would not describe
cross-couplings, but only self-interactions in the Rarita-Schwinger sector).
It is convenient to represent $\hat{\Psi}^{\mu \nu }$ like%
\begin{equation}
\hat{\Psi}^{\mu \nu }=\bar{\psi}_{\rho }\hat{\Psi}^{\mu \nu \mid \rho },
\label{om026}
\end{equation}%
where $\hat{\Psi}^{\mu \nu \mid \rho }$ are $4\times 4$ matrices with
spinor-like indices, whose elements are bosonic functions depending on the
undifferentiated fields $\psi _{\lambda }$ and $\varphi $. According to the
chosen basis (\ref{b2}), $\hat{\Psi}^{\mu \nu \mid \rho }$ decomposes as%
\begin{equation}
\hat{\Psi}^{\mu \nu \mid \rho }=\tilde{\Psi}^{\mu \nu \mid \rho }+\tilde{\Psi%
}_{\alpha }^{\mu \nu \mid \rho }\gamma ^{\alpha }+\tilde{\Psi}_{\alpha \beta
}^{\mu \nu \mid \rho }\gamma ^{\alpha \beta }+\tilde{\Psi}_{\alpha \beta
\gamma }^{\mu \nu \mid \rho }\gamma ^{\alpha \beta \gamma }+\bar{\Psi}^{\mu
\nu \mid \rho }\gamma _{5},  \label{om027}
\end{equation}%
where $\tilde{\Psi}^{\mu \nu \mid \rho }$, $\tilde{\Psi}_{\alpha }^{\mu \nu
\mid \rho }$, $\tilde{\Psi}_{\alpha \beta }^{\mu \nu \mid \rho }$, $\tilde{%
\Psi}_{\alpha \beta \gamma }^{\mu \nu \mid \rho }$, and $\bar{\Psi}^{\mu \nu
\mid \rho }$ are some bosonic, Lorentz tensors constructed out of $\psi
_{\lambda }$ and $\varphi $, separately antisymmetric in their lower indices
and in the upper pair $\{\mu ,\nu \}$ respectively. From (\ref{om027}) and
properties (\ref{sym1})--(\ref{sym2}), (\ref{om025}) can be expressed as%
\begin{eqnarray}
\overset{(1)}{\pi } &=&\frac{1}{2}\left[ \left( \partial _{\lbrack \mu }\bar{%
\psi}_{\nu ]}\right) \psi _{\rho }\tilde{\Psi}^{\mu \nu \mid \rho }-\left(
\partial _{\lbrack \mu }\bar{\psi}_{\nu ]}\right) \gamma ^{\alpha }\psi
_{\rho }\tilde{\Psi}_{\alpha }^{\mu \nu \mid \rho }-\left( \partial
_{\lbrack \mu }\bar{\psi}_{\nu ]}\right) \gamma ^{\alpha \beta }\psi _{\rho }%
\tilde{\Psi}_{\alpha \beta }^{\mu \nu \mid \rho }\right.   \notag \\
&&\left. +\left( \partial _{\lbrack \mu }\bar{\psi}_{\nu ]}\right) \gamma
^{\alpha \beta \gamma }\psi _{\rho }\tilde{\Psi}_{\alpha \beta \gamma }^{\mu
\nu \mid \rho }+\left( \partial _{\lbrack \mu }\bar{\psi}_{\nu ]}\right)
\gamma _{5}\psi _{\rho }\bar{\Psi}^{\mu \nu \mid \rho }\right] +\partial
_{\mu }\bar{t}^{\mu }.  \label{om028}
\end{eqnarray}%
By applying the differential $\gamma $ on (\ref{om028}), we get
\begin{eqnarray}
\gamma \overset{(1)}{\pi } &=&\frac{1}{2}\left[ \tilde{\Psi}^{\mu \nu \mid
\rho }\left( \partial _{\lbrack \mu }\bar{\psi}_{\nu ]}\right) -\tilde{\Psi}%
_{\alpha }^{\mu \nu \mid \rho }\left( \partial _{\lbrack \mu }\bar{\psi}%
_{\nu ]}\right) \gamma ^{\alpha }-\tilde{\Psi}_{\alpha \beta }^{\mu \nu \mid
\rho }\left( \partial _{\lbrack \mu }\bar{\psi}_{\nu ]}\right) \gamma
^{\alpha \beta }\right.   \notag \\
&&+\tilde{\Psi}_{\alpha \beta \gamma }^{\mu \nu \mid \rho }\left( \partial
_{\lbrack \mu }\bar{\psi}_{\nu ]}\right) \gamma ^{\alpha \beta \gamma }+\bar{%
\Psi}^{\mu \nu \mid \rho }\left( \partial _{\lbrack \mu }\bar{\psi}_{\nu
]}\right) \gamma _{5}+\left( \partial _{\lbrack \mu }\bar{\psi}_{\nu
]}\right) \psi _{\lambda }\frac{\partial ^{R}\tilde{\Psi}^{\mu \nu \mid
\lambda }}{\partial \psi _{\rho }}  \notag \\
&&-\left( \partial _{\lbrack \mu }\bar{\psi}_{\nu ]}\right) \gamma ^{\alpha
}\psi _{\lambda }\frac{\partial ^{R}\tilde{\Psi}_{\alpha }^{\mu \nu \mid
\lambda }}{\partial \psi _{\rho }}-\left( \partial _{\lbrack \mu }\bar{\psi}%
_{\nu ]}\right) \gamma ^{\alpha \beta }\psi _{\lambda }\frac{\partial ^{R}%
\tilde{\Psi}_{\alpha \beta }^{\mu \nu \mid \lambda }}{\partial \psi _{\rho }}
\notag \\
&&\left. +\left( \partial _{\lbrack \mu }\bar{\psi}_{\nu ]}\right) \gamma
^{\alpha \beta \gamma }\psi _{\lambda }\frac{\partial ^{R}\tilde{\Psi}%
_{\alpha \beta \gamma }^{\mu \nu \mid \lambda }}{\partial \psi _{\rho }}%
+\left( \partial _{\lbrack \mu }\bar{\psi}_{\nu ]}\right) \gamma _{5}\psi
_{\lambda }\frac{\partial ^{R}\bar{\Psi}^{\mu \nu \mid \lambda }}{\partial
\psi _{\rho }}\right] \partial _{\rho }\xi +\partial _{\mu }\left( \gamma
\bar{t}^{\mu }\right) ,  \label{om029}
\end{eqnarray}%
such that equation (\ref{om06b}) is fulfilled if
\begin{eqnarray}
&&\partial _{\rho }\left[ \tilde{\Psi}^{\mu \nu \mid \rho }\left( \partial
_{\lbrack \mu }\bar{\psi}_{\nu ]}\right) -\tilde{\Psi}_{\alpha }^{\mu \nu
\mid \rho }\left( \partial _{\lbrack \mu }\bar{\psi}_{\nu ]}\right) \gamma
^{\alpha }-\tilde{\Psi}_{\alpha \beta }^{\mu \nu \mid \rho }\left( \partial
_{\lbrack \mu }\bar{\psi}_{\nu ]}\right) \gamma ^{\alpha \beta }\right.
\notag \\
&&+\tilde{\Psi}_{\alpha \beta \gamma }^{\mu \nu \mid \rho }\left( \partial
_{\lbrack \mu }\bar{\psi}_{\nu ]}\right) \gamma ^{\alpha \beta \gamma }+\bar{%
\Psi}^{\mu \nu \mid \rho }\left( \partial _{\lbrack \mu }\bar{\psi}_{\nu
]}\right) \gamma _{5}+\left( \partial _{\lbrack \mu }\bar{\psi}_{\nu
]}\right) \psi _{\lambda }\frac{\partial ^{R}\tilde{\Psi}^{\mu \nu \mid
\lambda }}{\partial \psi _{\rho }}  \notag \\
&&-\left( \partial _{\lbrack \mu }\bar{\psi}_{\nu ]}\right) \gamma ^{\alpha
}\psi _{\lambda }\frac{\partial ^{R}\tilde{\Psi}_{\alpha }^{\mu \nu \mid
\lambda }}{\partial \psi _{\rho }}-\left( \partial _{\lbrack \mu }\bar{\psi}%
_{\nu ]}\right) \gamma ^{\alpha \beta }\psi _{\lambda }\frac{\partial ^{R}%
\tilde{\Psi}_{\alpha \beta }^{\mu \nu \mid \lambda }}{\partial \psi _{\rho }}
\notag \\
&&\left. +\left( \partial _{\lbrack \mu }\bar{\psi}_{\nu ]}\right) \gamma
^{\alpha \beta \gamma }\psi _{\lambda }\frac{\partial ^{R}\tilde{\Psi}%
_{\alpha \beta \gamma }^{\mu \nu \mid \lambda }}{\partial \psi _{\rho }}%
+\left( \partial _{\lbrack \mu }\bar{\psi}_{\nu ]}\right) \gamma _{5}\psi
_{\lambda }\frac{\partial ^{R}\bar{\Psi}^{\mu \nu \mid \lambda }}{\partial
\psi _{\rho }}\right] =0.  \label{om030}
\end{eqnarray}%
The above equation implies the existence of an antisymmetric spinor-tensor $%
\Phi ^{\rho \lambda }$, $\Phi ^{\rho \lambda }=-\Phi ^{\lambda \rho }$,
depending on $\varphi $ and $\psi _{\mu }$, in terms of which%
\begin{eqnarray}
&&\tilde{\Psi}^{\mu \nu \mid \rho }\left( \partial _{\lbrack \mu }\bar{\psi}%
_{\nu ]}\right) -\tilde{\Psi}_{\alpha }^{\mu \nu \mid \rho }\left( \partial
_{\lbrack \mu }\bar{\psi}_{\nu ]}\right) \gamma ^{\alpha }-\tilde{\Psi}%
_{\alpha \beta }^{\mu \nu \mid \rho }\left( \partial _{\lbrack \mu }\bar{\psi%
}_{\nu ]}\right) \gamma ^{\alpha \beta }  \notag \\
&&+\tilde{\Psi}_{\alpha \beta \gamma }^{\mu \nu \mid \rho }\left( \partial
_{\lbrack \mu }\bar{\psi}_{\nu ]}\right) \gamma ^{\alpha \beta \gamma }+\bar{%
\Psi}^{\mu \nu \mid \rho }\left( \partial _{\lbrack \mu }\bar{\psi}_{\nu
]}\right) \gamma _{5}+\left( \partial _{\lbrack \mu }\bar{\psi}_{\nu
]}\right) \psi _{\lambda }\frac{\partial ^{R}\tilde{\Psi}^{\mu \nu \mid
\lambda }}{\partial \psi _{\rho }}  \notag \\
&&-\left( \partial _{\lbrack \mu }\bar{\psi}_{\nu ]}\right) \gamma ^{\alpha
}\psi _{\lambda }\frac{\partial ^{R}\tilde{\Psi}_{\alpha }^{\mu \nu \mid
\lambda }}{\partial \psi _{\rho }}-\left( \partial _{\lbrack \mu }\bar{\psi}%
_{\nu ]}\right) \gamma ^{\alpha \beta }\psi _{\lambda }\frac{\partial ^{R}%
\tilde{\Psi}_{\alpha \beta }^{\mu \nu \mid \lambda }}{\partial \psi _{\rho }}
\notag \\
&&+\left( \partial _{\lbrack \mu }\bar{\psi}_{\nu ]}\right) \gamma ^{\alpha
\beta \gamma }\psi _{\lambda }\frac{\partial ^{R}\tilde{\Psi}_{\alpha \beta
\gamma }^{\mu \nu \mid \lambda }}{\partial \psi _{\rho }}+\left( \partial
_{\lbrack \mu }\bar{\psi}_{\nu ]}\right) \gamma _{5}\psi _{\lambda }\frac{%
\partial ^{R}\bar{\Psi}^{\mu \nu \mid \lambda }}{\partial \psi _{\rho }}%
=\partial _{\lambda }\Phi ^{\rho \lambda }.  \label{om031}
\end{eqnarray}%
In order to analyze equation (\ref{om031}), we compute its Euler-Lagrange
derivative with respect to $\varphi $ and deduce the necessary condition%
\begin{eqnarray}
&&\frac{\delta }{\delta \varphi }\left[ \tilde{\Psi}^{\mu \nu \mid \rho
}\left( \partial _{\lbrack \mu }\bar{\psi}_{\nu ]}\right) -\tilde{\Psi}%
_{\alpha }^{\mu \nu \mid \rho }\left( \partial _{\lbrack \mu }\bar{\psi}%
_{\nu ]}\right) \gamma ^{\alpha }-\tilde{\Psi}_{\alpha \beta }^{\mu \nu \mid
\rho }\left( \partial _{\lbrack \mu }\bar{\psi}_{\nu ]}\right) \gamma
^{\alpha \beta }\right.   \notag \\
&&+\tilde{\Psi}_{\alpha \beta \gamma }^{\mu \nu \mid \rho }\left( \partial
_{\lbrack \mu }\bar{\psi}_{\nu ]}\right) \gamma ^{\alpha \beta \gamma }+\bar{%
\Psi}^{\mu \nu \mid \rho }\left( \partial _{\lbrack \mu }\bar{\psi}_{\nu
]}\right) \gamma _{5}+\left( \partial _{\lbrack \mu }\bar{\psi}_{\nu
]}\right) \psi _{\lambda }\frac{\partial ^{R}\tilde{\Psi}^{\mu \nu \mid
\lambda }}{\partial \psi _{\rho }}  \notag \\
&&-\left( \partial _{\lbrack \mu }\bar{\psi}_{\nu ]}\right) \gamma ^{\alpha
}\psi _{\lambda }\frac{\partial ^{R}\tilde{\Psi}_{\alpha }^{\mu \nu \mid
\lambda }}{\partial \psi _{\rho }}-\left( \partial _{\lbrack \mu }\bar{\psi}%
_{\nu ]}\right) \gamma ^{\alpha \beta }\psi _{\lambda }\frac{\partial ^{R}%
\tilde{\Psi}_{\alpha \beta }^{\mu \nu \mid \lambda }}{\partial \psi _{\rho }}
\notag \\
&&\left. +\left( \partial _{\lbrack \mu }\bar{\psi}_{\nu ]}\right) \gamma
^{\alpha \beta \gamma }\psi _{\lambda }\frac{\partial ^{R}\tilde{\Psi}%
_{\alpha \beta \gamma }^{\mu \nu \mid \lambda }}{\partial \psi _{\rho }}%
+\left( \partial _{\lbrack \mu }\bar{\psi}_{\nu ]}\right) \gamma _{5}\psi
_{\lambda }\frac{\partial ^{R}\bar{\Psi}^{\mu \nu \mid \lambda }}{\partial
\psi _{\rho }}\right] =0.  \label{om032}
\end{eqnarray}%
Since $\hat{\Psi}^{\mu \nu \mid \rho }$s are bosonic, they can be decomposed
as sums of homogeneous polynomials of various, even degrees in the
Rarita-Schwinger spinors, so it is enough to analyze equation (\ref{om032})
for a fixed value of this degree, say $2p$. Consequently, all the components
from (\ref{om027}) (namely $\tilde{\Psi}^{\mu \nu \mid \rho }$, $\tilde{\Psi}%
_{\alpha }^{\mu \nu \mid \rho }$, $\tilde{\Psi}_{\alpha \beta }^{\mu \nu
\mid \rho }$, $\tilde{\Psi}_{\alpha \beta \gamma }^{\mu \nu \mid \rho }$,
and $\bar{\Psi}^{\mu \nu \mid \rho }$) will display the same degree with
respect to $\psi _{\rho }$. Multiplying (\ref{om032}) at the right with $%
\psi _{\rho }$ and using (\ref{om026}), (\ref{om027}), and the homogeneity
assumption, we finally find%
\begin{equation}
(2p+1)\frac{\delta }{\delta \varphi }\left( \hat{\Psi}^{\mu \nu }\partial
_{\lbrack \mu }\psi _{\nu ]}\right) =0,  \label{homnophi}
\end{equation}%
which, according to (\ref{om025}), indicates that $\overset{(1)}{\pi }$
cannot depend nontrivially on the scalar field, and therefore it describes
only self-interactions in the Rarita-Schwinger sector, so it can be made to
vanish%
\begin{equation}
\overset{(1)}{\pi }=0.  \label{om033}
\end{equation}%
Based on results (\ref{om010}) and (\ref{om033}), from (\ref{om05}) we
obtain precisely (\ref{om034}).

\section{Gauge generators, commutators, and reducibility of the coupled
model in case I\label{appendix}}

From the terms of antighost number one present in (\ref{sdef1}), we read the
nonvanishing gauge generators of the coupled model (written in De Witt
condensed notations) as
\begin{eqnarray}
(\tilde{Z}_{(A)}^{\left( \mathrm{I}\right) \mu }) &=&(Z_{(A)}^{\mu
})=\partial ^{\mu },\qquad (\tilde{Z}_{(H)}^{\left( \mathrm{I}\right) \mu
})_{\alpha \beta }=-D_{[\alpha }\delta _{\beta ]}^{\mu },  \label{i1a} \\
(\tilde{Z}_{(H)}^{\left( \mathrm{I}\right) \mu }) &=&\lambda \left( \frac{1}{%
2}\frac{dU_{1}}{d\varphi }\bar{\psi}_{\nu }\gamma ^{\mu \nu \rho }\gamma
_{5}\psi _{\rho }-\frac{dW}{d\varphi }H^{\mu }\right) ,  \label{i1b} \\
(\tilde{Z}_{(H)}^{\left( \mathrm{I}\right) \mu })_{A} &=&\lambda \frac{dU_{1}%
}{d\varphi }A_{\rho }\left( \bar{\psi}_{\nu }\gamma ^{\mu \nu \rho }\gamma
_{5}\right) _{A},\qquad (\tilde{Z}_{(\varphi )}^{\left( \mathrm{I}\right)
})=\lambda W,  \label{i1c} \\
(\tilde{Z}_{(B)}^{\left( \mathrm{I}\right) \mu \nu })_{\alpha \beta \gamma }
&=&-\frac{1}{2}\partial _{\lbrack \alpha }\delta _{\beta }^{\mu }\delta
_{\gamma ]}^{\nu },\qquad (\tilde{Z}_{(B)}^{\left( \mathrm{I}\right) \mu \nu
})_{\alpha \beta }=\lambda W\delta _{\lbrack \alpha }^{\mu }\delta _{\beta
]}^{\nu },  \label{i1d} \\
(\tilde{Z}_{(B)}^{\left( \mathrm{I}\right) \mu \nu })_{A} &=&-\lambda
U_{1}\left( \bar{\psi}_{\rho }\gamma ^{\mu \nu \rho }\gamma _{5}\right) _{A},
\label{i1e} \\
(\tilde{Z}_{(\psi )}^{\left( \mathrm{I}\right) A\mu }) &=&-\mathrm{i}\lambda
U_{1}\left( \gamma _{5}\right) _{\quad B}^{A}\psi ^{B\mu },  \label{i1g} \\
(\tilde{Z}_{(\psi )}^{\left( \mathrm{I}\right) A\mu })_{B} &=&\left( \delta
_{B}^{A}\partial ^{\mu }+\mathrm{i}\lambda U_{1}\left( \gamma _{5}\right)
_{\quad B}^{A}A^{\mu }\right) .  \label{i1f}
\end{eqnarray}%
The nonvanishing commutators among the gauge transformations of the coupled
model result from the pieces in (\ref{sdef1}) that are quadratic in the
ghosts of pure ghost number one and take the form
\begin{eqnarray}
&&(\tilde{Z}_{(\varphi )}^{\left( \mathrm{I}\right) })\frac{\delta (\tilde{Z}%
_{(H)}^{\left( \mathrm{I}\right) \mu })_{\alpha \beta }}{\delta \varphi }+(%
\tilde{Z}_{(A)}^{\left( \mathrm{I}\right) \rho })\frac{\delta (\tilde{Z}%
_{(H)}^{\left( \mathrm{I}\right) \mu })_{\alpha \beta }}{\delta A^{\rho }}-(%
\tilde{Z}_{(H)}^{\left( \mathrm{I}\right) \rho })_{\alpha \beta }\frac{%
\delta (\tilde{Z}_{(H)}^{\left( \mathrm{I}\right) \mu })}{\delta H^{\rho }}
\notag \\
&=&\lambda \frac{dW}{d\varphi }(\tilde{Z}_{(H)}^{\left( \mathrm{I}\right)
\mu })_{\alpha \beta }+\lambda \frac{d^{2}W}{d\varphi ^{2}}\delta _{\left[
\alpha \right. }^{\mu }\frac{\delta \tilde{S}^{\left( \mathrm{I}\right) }}{%
\delta H^{\left. \beta \right] }},  \label{i21}
\end{eqnarray}%
\begin{equation}
(\tilde{Z}_{(\varphi )}^{\left( \mathrm{I}\right) })\frac{\delta (\tilde{Z}%
_{(B)}^{\left( \mathrm{I}\right) \mu \nu })_{\alpha \beta }}{\delta \varphi }%
=\lambda \frac{dW}{d\varphi }(\tilde{Z}_{(B)}^{\left( \mathrm{I}\right) \mu
\nu })_{\alpha \beta },  \label{i22}
\end{equation}%
\begin{eqnarray}
&&(\tilde{Z}_{(\varphi )}^{\left( \mathrm{I}\right) })\frac{\delta (\tilde{Z}%
_{(H)}^{\left( \mathrm{I}\right) \mu })_{A}}{\delta \varphi }-(\tilde{Z}%
_{(H)}^{\left( \mathrm{I}\right) \rho })_{A}\frac{\delta (\tilde{Z}%
_{(H)}^{\left( \mathrm{I}\right) \mu })}{\delta H^{\rho }}+(\tilde{Z}%
_{(A)}^{\left( \mathrm{I}\right) \rho })\frac{\delta (\tilde{Z}%
_{(H)}^{\left( \mathrm{I}\right) \mu })_{A}}{\delta A^{\rho }}  \notag \\
&&+\frac{\delta ^{R}(\tilde{Z}_{(H)}^{\left( \mathrm{I}\right) \mu })_{A}}{%
\delta \psi ^{B\rho }}(\tilde{Z}_{(\psi )}^{\left( \mathrm{I}\right) B\rho
})-\frac{\delta ^{R}(\tilde{Z}_{(H)}^{\left( \mathrm{I}\right) \mu })}{%
\delta \psi ^{B\rho }}(\tilde{Z}_{(\psi )}^{\left( \mathrm{I}\right) B\rho
})_{A}  \notag \\
&=&\mathrm{i}\lambda U_{1}(\tilde{Z}_{(H)}^{\left( \mathrm{I}\right) \mu
})_{B}\left( \gamma _{5}\right) _{\quad A}^{B}-\frac{\lambda }{2}\frac{dU_{1}%
}{d\varphi }\left( \bar{\psi}_{\gamma }\gamma ^{\alpha \beta \gamma }\gamma
_{5}\right) _{A}(\tilde{Z}_{(H)}^{\left( \mathrm{I}\right) \mu })_{\alpha
\beta }  \notag \\
&&+\lambda \frac{d^{2}U_{1}}{d\varphi ^{2}}\frac{\delta \tilde{S}^{\left(
\mathrm{I}\right) }}{\delta H^{\nu }}\left( \bar{\psi}_{\rho }\gamma ^{\mu
\nu \rho }\gamma _{5}\right) _{A}+\mathrm{i}\lambda \frac{dU_{1}}{d\varphi }%
\frac{\delta ^{R}\tilde{S}^{\left( \mathrm{I}\right) }}{\delta \psi _{\mu
}^{B}}\left( \gamma _{5}\right) _{\quad A}^{B},  \label{i23}
\end{eqnarray}%
\begin{eqnarray}
&&\frac{\delta (\tilde{Z}_{(\psi )}^{\left( \mathrm{I}\right) A\mu })_{B}}{%
\delta \varphi }(\tilde{Z}_{(\varphi )}^{\left( \mathrm{I}\right) })+\frac{%
\delta (\tilde{Z}_{(\psi )}^{\left( \mathrm{I}\right) A\mu })_{B}}{\delta
A^{\rho }}(\tilde{Z}_{(A)}^{\left( \mathrm{I}\right) \rho })-\frac{\delta
^{R}(\tilde{Z}_{(\psi )}^{\left( \mathrm{I}\right) A\mu })}{\delta \psi
^{C\rho }}(\tilde{Z}_{(\psi )}^{\left( \mathrm{I}\right) C\rho })_{B}  \notag
\\
&=&\mathrm{i}\lambda U_{1}(\tilde{Z}_{(\psi )}^{\left( \mathrm{I}\right)
A\mu })_{C}\left( \gamma _{5}\right) _{\quad B}^{C}-\mathrm{i}\lambda \frac{%
dU_{1}}{d\varphi }\frac{\delta \tilde{S}^{\left( \mathrm{I}\right) }}{\delta
H_{\mu }}\left( \gamma _{5}\right) _{\quad B}^{A},  \label{i24}
\end{eqnarray}%
\begin{eqnarray}
&&\frac{\delta ^{R}(\tilde{Z}_{(H)}^{\left( \mathrm{I}\right) \mu })_{B}}{%
\delta \psi ^{C\rho }}(\tilde{Z}_{(\psi )}^{\left( \mathrm{I}\right) C\rho
})_{A}+\frac{\delta ^{R}(\tilde{Z}_{(H)}^{\left( \mathrm{I}\right) \mu })_{A}%
}{\delta \psi ^{C\rho }}(\tilde{Z}_{(\psi )}^{\left( \mathrm{I}\right) C\rho
})_{B}  \notag \\
&=&\frac{\lambda }{2}\frac{dU_{1}}{d\varphi }A_{\gamma }\left( \gamma
^{0}\gamma ^{\alpha \beta \gamma }\gamma _{5}\right) _{AB}(\tilde{Z}%
_{(H)}^{\left( \mathrm{I}\right) \mu })_{\alpha \beta } \\
&&-\lambda \frac{d^{2}U_{1}}{d\varphi ^{2}}\frac{\delta \tilde{S}^{\left(
\mathrm{I}\right) }}{\delta H^{\nu }}A_{\rho }\left( \gamma ^{0}\gamma ^{\mu
\nu \rho }\gamma _{5}\right) _{AB}-\lambda \frac{dU_{1}}{d\varphi }\frac{%
\delta \tilde{S}^{\left( \mathrm{I}\right) }}{\delta B^{\nu \rho }}\left(
\gamma ^{0}\gamma ^{\mu \nu \rho }\gamma _{5}\right) _{AB},  \label{i25}
\end{eqnarray}%
\begin{eqnarray}
&&\frac{\delta ^{R}(\tilde{Z}_{(B)}^{\left( \mathrm{I}\right) \mu \nu })_{B}%
}{\delta \psi ^{C\rho }}(\tilde{Z}_{(\psi )}^{\left( \mathrm{I}\right) C\rho
})_{A}+\frac{\delta ^{R}(\tilde{Z}_{(B)}^{\left( \mathrm{I}\right) \mu \nu
})_{A}}{\delta \psi ^{C\rho }}(\tilde{Z}_{(\psi )}^{\left( \mathrm{I}\right)
C\rho })_{B}  \notag \\
&=&\lambda \frac{dU_{1}}{d\varphi }\frac{\delta \tilde{S}^{\left( \mathrm{I}%
\right) }}{\delta H}\left( \gamma ^{0}\gamma ^{\mu \nu \rho }\gamma
_{5}\right) _{AB}+\frac{\lambda }{3}U_{1}(\tilde{Z}_{(B)}^{\left( \mathrm{I}%
\right) \mu \nu })_{\alpha \beta \gamma }\left( \gamma ^{0}\gamma ^{\alpha
\beta \gamma }\gamma _{5}\right) _{AB}.  \label{i26}
\end{eqnarray}

The structure of the terms linear in the ghosts with pure ghost number two
or three from (\ref{sdef1}) shows that some of the reducibility functions
are modified with respect to the free theory and, moreover, some of the
reducibility relations only hold on-shell. From the analysis of these terms
we infer the first-order reducibility functions
\begin{eqnarray}
(\tilde{Z}_{1}^{\left( \mathrm{I}\right) \alpha \beta })_{\mu \nu \rho } &=&-%
\frac{1}{2}D_{[\mu }\delta _{\nu }^{\alpha }\delta _{\rho ]}^{\beta },
\label{i11a} \\
(\tilde{Z}_{1}^{\left( \mathrm{I}\right) \alpha \beta \gamma })_{\mu \nu
\rho } &=&-\frac{1}{3}\lambda W\delta _{\lbrack \mu }^{\alpha }\delta _{\nu
}^{\beta }\delta _{\rho ]}^{\gamma },  \label{i11b} \\
(\tilde{Z}_{1}^{\left( \mathrm{I}\right) \alpha \beta \gamma })_{\mu \nu
\rho \lambda } &=&-\frac{1}{6}\partial _{\lbrack \mu }\delta _{\nu }^{\alpha
}\delta _{\rho }^{\beta }\delta _{\lambda ]}^{\gamma },  \label{i11c}
\end{eqnarray}%
and respectively the second-order ones
\begin{eqnarray}
\left( \tilde{Z}_{2}^{\left( \mathrm{I}\right) \mu \nu \rho }\right)
_{\alpha \beta \gamma \delta } &=&-\frac{1}{6}D_{[\alpha }\delta _{\beta
}^{\mu }\delta _{\gamma }^{\nu }\delta _{\delta ]}^{\rho },  \label{i12a} \\
(\tilde{Z}_{2}^{\left( \mathrm{I}\right) \mu \nu \rho \lambda })_{\alpha
\beta \gamma \delta } &=&\frac{1}{12}\lambda W\delta _{\lbrack \alpha }^{\mu
}\delta _{\beta }^{\nu }\delta _{\gamma }^{\rho }\delta _{\delta ]}^{\lambda
}.  \label{i12b}
\end{eqnarray}%
Finally, the associated first- and second-order reducibility relations are
listed below:
\begin{equation}
(\tilde{Z}_{(H)}^{\left( \mathrm{I}\right) \mu })_{\alpha \beta }(\tilde{Z}%
_{1}^{\left( \mathrm{I}\right) \alpha \beta })_{\nu \rho \lambda }=-\lambda
\frac{d^{2}W}{d\varphi ^{2}}A_{[\nu }\delta _{\rho }^{\mu }\frac{\delta
\tilde{S}^{\left( \mathrm{I}\right) }}{\delta H^{\lambda ]}}-2\lambda \frac{%
dW}{d\varphi }\delta _{\lbrack \nu }^{\mu }\frac{\delta \tilde{S}^{\left(
\mathrm{I}\right) }}{\delta B^{\rho \lambda ]}},  \label{i16}
\end{equation}%
\begin{equation}
(\tilde{Z}_{(B)}^{\left( \mathrm{I}\right) \mu \nu })_{\alpha \beta }(\tilde{%
Z}_{1}^{\left( \mathrm{I}\right) \alpha \beta })_{\rho \lambda \sigma }+(%
\tilde{Z}_{(B)}^{\left( \mathrm{I}\right) \mu \nu })_{\alpha \beta \gamma }(%
\tilde{Z}_{1}^{\left( \mathrm{I}\right) \alpha \beta \gamma })_{\rho \lambda
\sigma }=\lambda \frac{dW}{d\varphi }\delta _{\lbrack \rho }^{\mu }\delta
_{\lambda }^{\nu }\frac{\delta \tilde{S}^{\left( \mathrm{I}\right) }}{\delta
H^{\sigma ]}},  \label{i17}
\end{equation}%
\begin{equation}
(\tilde{Z}_{(B)}^{\left( \mathrm{I}\right) \mu \nu })_{\alpha \beta \gamma }(%
\tilde{Z}^{\left( \mathrm{I}\right) \alpha \beta \gamma })_{\rho \lambda
\sigma \xi }=0,  \label{i18}
\end{equation}%
\begin{equation}
(\tilde{Z}_{1}^{\left( \mathrm{I}\right) \alpha \beta })_{\mu \nu \rho
}\left( \tilde{Z}_{2}^{\left( \mathrm{I}\right) \mu \nu \rho }\right)
_{\gamma \delta \varepsilon \eta }=\frac{\lambda }{2}\frac{d^{2}W}{d\varphi
^{2}}A_{[\gamma }\delta _{\delta }^{\alpha }\delta _{\varepsilon }^{\beta }%
\frac{\delta \tilde{S}^{\left( \mathrm{I}\right) }}{\delta H^{\eta ]}}%
-\lambda \frac{dW}{d\varphi }\delta _{\lbrack \gamma }^{\alpha }\delta
_{\delta }^{\beta }\frac{\delta \tilde{S}^{\left( \mathrm{I}\right) }}{%
\delta B^{\varepsilon \eta ]}},  \label{i19}
\end{equation}%
\begin{equation}
(\tilde{Z}_{1}^{\left( \mathrm{I}\right) \alpha \beta \gamma })_{\mu \nu
\rho }\left( \tilde{Z}_{2}^{\left( \mathrm{I}\right) \mu \nu \rho }\right)
_{\delta \varepsilon \eta \kappa }+(\tilde{Z}_{1}^{\left( \mathrm{I}\right)
\alpha \beta \gamma })_{\mu \nu \rho \lambda }(\tilde{Z}_{2}^{\left( \mathrm{%
I}\right) \mu \nu \rho \lambda })_{\delta \varepsilon \eta \kappa }=\frac{%
\lambda }{3}\frac{dW}{d\varphi }\delta _{\lbrack \delta }^{\alpha }\delta
_{\varepsilon }^{\beta }\delta _{\eta }^{\gamma }\frac{\delta \tilde{S}%
^{\left( \mathrm{I}\right) }}{\delta H^{\xi ]}}.  \label{i20}
\end{equation}


\begin{thebibliography}{99}
\bibitem{birmingham91} D. Birmingham, M. Blau, M. Rakowski and G. Thompson,
Topological field theory, \textit{Phys. Rept.} \textbf{209} (1991) 129--340.

\bibitem{labastida97} J. M. F. Labastida and C. Losano, Lectures on
topological QFT, in \textit{Proceedings of La Plata-CERN-Santiago de
Compostela Meeting on Trends in Theoretical Physics, La Plata, Argentina,
April-May 1997,} eds. H. Falomir, R. E. Gamboa Sarav\'{\i}, F. A. Schaposnik
(AIP, New York 1998), AIP Conference Proceedings vol. \textbf{419}, 54.

\bibitem{stroblspec} P. Schaller and T. Strobl, Poisson structure induced
(topological) field theories, \textit{Mod. Phys. Lett.} \textbf{A9} (1994)
3129--3136 [arXiv:hep-th/9405110].

\bibitem{psmikeda94} N. Ikeda, Two-dimensional gravity and nonlinear gauge
theory, \textit{Annals Phys.} \textbf{235} (1994) 435--464
[arXiv:hep-th/9312059].

\bibitem{psmstrobl95} A. Yu. Alekseev, P. Schaller and T. Strobl,
Topological $G/G$ WZW model in the generalized momentum representation,
\textit{Phys. Rev.} \textbf{D52} (1995) 7146--7160 [arXiv:hep-th/9505012].

\bibitem{psmstroblCQG961} T. Kl\"{o}sch and T. Strobl, Classical and quantum
gravity in 1+1 dimensions: I. A unifying approach, \textit{Class. Quantum
Grav.} \textbf{13} (1996) 965--983 [arXiv:gr-qc/9508020]; Erratum-ibid.
\textbf{14} (1997) 825.

\bibitem{psmstroblCQG962} T. Kl\"{o}sch and T. Strobl, Classical and quantum
gravity in 1+1 dimensions. II: The universal coverings, \textit{Class.
Quantum Grav.} \textbf{13} (1996) 2395--2421 [arXiv:gr-qc/9511081].

\bibitem{psmstrobl97} T. Kl\"{o}sch and T. Strobl, Classical and quantum
gravity in 1+1 dimensions: III. Solutions of arbitrary topology, \textit{%
Class. Quantum Grav.} \textbf{14} (1997) 1689--1723 [arXiv:hep-th/9607226].

\bibitem{psmcattaneo2000} A. S. Cattaneo and G. Felder,  A path integral
approach to the Kontsevich quantization formula, \textit{Commun. Math. Phys.}
\textbf{212} (2000) 591--611 [arXiv:math/9902090].

\bibitem{psmcattaneo2001} A. S. Cattaneo and G. Felder, Poisson sigma models
and deformation quantization, \textit{Mod. Phys. Lett.} \textbf{A16} (2001)
179--189 [arXiv:hep-th/0102208].

\bibitem{grav2teit83} C. Teitelboim, Gravitation and hamiltonian structure
in two spacetime dimensions, \textit{Phys. Lett.} \textbf{B126} (1983)
41--45.

\bibitem{grav2jackiw85} R. Jackiw, Lower dimensional gravity, \textit{Nucl.
Phys.} \textbf{B252} (1985) 343--356.

\bibitem{grav2katanaev86} M. O. Katanayev and I. V. Volovich, String model
with dynamical geometry and torsion, \textit{Phys. Lett.} \textbf{B175}
(1986) 413--416.

\bibitem{grav2brown88} J. Brown, \textit{Lower Dimensional Gravity,} World
Scientific, Singapore 1988.

\bibitem{grav2katanaev90} M. O. Katanaev and I. V. Volovich, Two-dimensional
gravity with dynamical torsion and strings, \textit{Annals Phys.} \textbf{197%
} (1990) 1--32.

\bibitem{grav2schmidt} H.-J. Schmidt, Scale-invariant gravity in two
dimensions, \textit{J. Math. Phys.} \textbf{32} (1991) 1562.

\bibitem{grav2solod} S. N. Solodukhin, Topological $2D$ Riemann-Cartan-Weyl
gravity, \textit{Class. Quantum Grav.} \textbf{10} (1993) 1011--1021.

\bibitem{grav2ikedaizawa90} N. Ikeda and K. I. Izawa, General form of
dilaton gravity and nonlinear gauge theory, \textit{Prog. Theor. Phys.}
\textbf{90} (1993) 237--246 [arXiv:hep-th/9304012].

\bibitem{grav2strobl94} T. Strobl, Dirac quantization of gravity-Yang-Mills
systems in 1+1 dimensions, \textit{Phys. Rev.} \textbf{D50} (1994)
7346--7350 [arXiv:hep-th/9403121].

\bibitem{grav2grumvassil02} D. Grumiller, W. Kummer and D. V. Vassilevich,
Dilaton gravity in two dimensions, \textit{Phys. Rept.} \textbf{369} (2002)
327--430 [arXiv:hep-th/0204253].

\bibitem{grav2strobl00} T. Strobl, \textit{Gravity in two space-time
dimensions,} Habilitation thesis RWTH Aachen, May 1999, arXiv:hep-th/0011240.

\bibitem{ezawa} K. Ezawa, Ashtekar's Formulation for $N=1,2$ Supergravities
as \textquotedblleft constrained\textquotedblright\ BF theories, \textit{%
Prog. Theor. Phys. }\textbf{95} (1996) 863--882 [arXiv:hep-th/9511047].

\bibitem{freidel} L. Freidel, K. Krasnov and R. Puzio, BF description of
higher-dimensional gravity theories, \textit{Adv. Theor. Math. Phys.}
\textbf{3} (1999) 1289--1324 [arXiv:hep-th/9901069].

\bibitem{smolin} L. Smolin, Holographic formulation of quantum general
relativity, \textit{Phys. Rev.} \textbf{D61} (2000) 084007
[arXiv:hep-th/9808191].

\bibitem{ling} Y. Ling and L. Smolin, Holographic formulation of quantum
supergravity, \textit{Phys. Rev.} \textbf{D63} (2001) 064010
[arXiv:hep-th/0009018].

\bibitem{defBFizawa2000} K.-I. Izawa, On nonlinear gauge theory from a
deformation theory perspective, \textit{Prog. Theor. Phys.} \textbf{103}
(2000) 225--228 [arXiv:hep-th/9910133].

\bibitem{defBFmpla} C. Bizdadea, Note on two-dimensional nonlinear gauge
theories,  \textit{Mod. Phys. Lett.} \textbf{A15} (2000) 2047--2055
[arXiv:hep-th/0201059].

\bibitem{defBFikeda00} N. Ikeda, A deformation of three dimensional BF
theory, \textit{J.High Energy Phys.} JHEP\textbf{11}(2000)009
[arXiv:hep-th/0010096].

\bibitem{defBFikeda01} N. Ikeda, Deformation of BF theories, topological
open membrane and a generalization of the star deformation, \textit{J. High
Energy Phys.} JHEP\textbf{07}(2001)037 [arXiv:hep-th/0105286].

\bibitem{defBFijmpa} C. Bizdadea, E. M. Cioroianu and S. O. Saliu,
Hamiltonian cohomological derivation of four-dimensional nonlinear gauge
theories, \textit{Int. J. Mod. Phys.} \textbf{A17} (2002) 2191--2210
[arXiv:hep-th/0206186].

\bibitem{defBFjhep} C. Bizdadea, C. C. Ciob\^{\i}rc\u{a}, E. M. Cioroianu,
S. O. Saliu and S. C. S\u{a}raru, Hamiltonian BRST deformation of a class of
n-dimensional BF-type theories, \textit{J. High Energy Phys.} JHEP\textbf{01}%
(2003)049 [arXiv:hep-th/0302037].

\bibitem{defBFijmpajuvi06} E. M. Cioroianu and S. C. S\u{a}raru, PT-symmetry
breaking Hamiltonian interactions in BF models, \textit{Int. J. Mod. Phys.}
\textbf{A21} (2006) 2573--2599 [arXiv:hep-th/0606164].

\bibitem{defBFepjc} C. Bizdadea, E. M. Cioroianu, S. O. Saliu and S. C. S%
\u{a}raru, Couplings of a collection of BF models to matter theories,
\textit{Eur. Phys. J.} \textbf{C41} (2005) 401--420 [arXiv:hep-th/0508037].

\bibitem{defBFikeda03} N. Ikeda, Chern-Simons gauge theory coupled with BF
theory, \textit{Int. J. Mod. Phys.} \textbf{A18} (2003) 2689--2702
[arXiv:hep-th/0203043].

\bibitem{defBFijmpajuvi04} E. M. Cioroianu and S. C. S\u{a}raru,
Two-dimensional interactions between a BF-type theory and a collection of
vector fields, \textit{Int. J. Mod. Phys.} \textbf{A19} (2004) 4101--4125
[arXiv:hep-th/0501056].

\bibitem{defBFjhep06} C. Bizdadea, E. M. Cioroianu, I. Negru, S. O. Saliu
and S. C. S\u{a}raru, On the generalized Freedman-Townsend model, \textit{J.
High Energy Phys.} JHEP\textbf{10}(2006)004 [arXiv:0704.3407].

\bibitem{deflag} G. Barnich and M. Henneaux, Consistent couplings between
fields with a gauge freedom and deformations of the master equation, \textit{%
Phys. Lett.} \textbf{B311} (1993) 123--129 [arXiv:hep-th/9304057].

\bibitem{defham} C. Bizdadea, Consistent interactions in the Hamiltonian
BRST formalism, \textit{Acta Phys. Polon.} \textbf{B32} (2001) 2843--2862
[arXiv:hep-th/0003199].

\bibitem{otherBFikeda02} N. Ikeda, Topological field theories and geometry
of Batalin-Vilkovisky algebras, \textit{J. High Energy Phys.} JHEP\textbf{10}%
(2002)076 [arXiv:hep-th/0209042].

\bibitem{otherBFikedaizawa04} N. Ikeda and K.-I. Izawa, Dimensional
reduction of nonlinear gauge theories, \textit{J. High Energy Phys.} JHEP%
\textbf{09}(2004)030 [arXiv:hep-th/0407243].

\bibitem{17and5} M. Henneaux, Consistent interactions between gauge fields:
the cohomological approach, \textit{Contemp. Math.} \textbf{219} (1998) 93
[arXiv:hep-th/9712226].

\bibitem{gen1} G. Barnich, F. Brandt and M. Henneaux, Local BRST cohomology
in the antifield formalism: I. General theorems, \textit{Commun. Math. Phys.}
\textbf{174} (1995) 57--91 [arXiv:hep-th/9405109].

\bibitem{gen2} G. Barnich, F. Brandt and M. Henneaux, Local BRST cohomology
in gauge theories, \textit{Phys. Rept.} \textbf{338} (2000) 439--569
[arXiv:hep-th/0002245].

\bibitem{fratse} E. S. Fradkin and A. A. Tseytlin, Conformal supergravity,
\textit{Phys. Rept.} \textbf{119} (1985) 233--362.
\end{thebibliography}
\end{document}